\documentclass[12pt, a4paper]{article} 

\usepackage[includeheadfoot,
            marginratio={1:1,2:3}, 
            width=412pt, 
            height=688pt,]{geometry}

\usepackage{amsmath}
\usepackage{amsfonts}
\usepackage{amssymb}
\usepackage{mathtools}
\usepackage{mathalfa}
\usepackage{graphicx,caption,subfigure}
\usepackage{tikz} 
\usetikzlibrary{decorations.pathreplacing,calc,tikzmark}
\usepackage{booktabs}
\usepackage{adjustbox}
\usepackage[utf8]{inputenc}
 \usepackage[splitrule]{footmisc}

\usepackage{placeins}
\usepackage{empheq}
\usepackage{paralist}
\usepackage{cite}
\usepackage[normalem]{ulem}
\usepackage{soul}

\usepackage[colorlinks=false,urlbordercolor=red]{hyperref}
\usepackage{xcolor}
\usepackage{tensor}

\newcommand{\nc}{\newcommand}
\nc{\lb}{\llbracket}
\nc{\rb}{\rrbracket}
\nc{\gl}{\llbracket}
\nc{\gr}{\rrbracket}
\nc{\del}{\partial}
\nc{\tri}{\hspace{-3.5pt}\vartriangle\hspace{-3.5pt}}
\nc{\blacktri}{\blacktriangle}

\nc{\eq}[1]{\begin{equation}
                     \begin{split} #1 \end{split}
                     \end{equation}}
\nc{\ov}{\overline}

\nc{\fa}{\hat}
\nc{\fb}{\MakeUppercase}
\nc{\fc}{\tilde }
\nc{\Lie}{{\cal L}} 
\nc{\lambdabar}{{\mkern0.75mu\mathchar '26\mkern -9.75mu\lambda}}

\allowdisplaybreaks[2]
\numberwithin{equation}{section}

\DeclareUnicodeCharacter{2212}{-}
\begin{document}

\vspace*{-1.5cm}
\begin{flushright}
  {\small
  MPP-2022-95\\
  }
\end{flushright}

\vspace{1.5cm}
\begin{center}
  {\Large 
     Dimensional Reduction of Cobordism and K-theory} 
\vspace{0.4cm}

\end{center}

\vspace{0.35cm}
\begin{center}
{Ralph Blumenhagen, Niccol\`o Cribiori, Christian Knei\ss l, Andriana Makridou}
\end{center}

\

\vspace{0.1cm}
\begin{center} 
\emph{
Max-Planck-Institut f\"ur Physik (Werner-Heisenberg-Institut), \\[.1cm] 
   F\"ohringer Ring 6,  80805 M\"unchen, Germany } 
   \\[0.1cm] 
 \vspace{0.3cm} 
\end{center} 

\vspace{0.5cm}

\begin{abstract}
It has been proposed that cobordism and K-theory groups, which can be mathematically related in certain cases, are physically associated to generalised higher-form symmetries. 
As a consequence, they should be broken or gauged in any consistent theory of quantum gravity, in accordance with swampland conjectures.
We provide further support to this idea by showing that cobordism and K-theory groups of a general manifold $X$ reproduce the pattern of symmetries expected from the dimensional reduction of the theory on $X$, as well as their breaking and gauging.
To this end, we employ the  Atiyah--Hirzebruch spectral sequence to compute such groups for common choices of $X$ in string compactifications.
\end{abstract}

\thispagestyle{empty}
\clearpage

\setcounter{tocdepth}{2}

\tableofcontents

\newpage

\section{Introduction}

The absence of global symmetries is considered to be a fundamental property  of quantum gravity. 
It is also one of the central and most tested conjectures in the swampland program \cite{Palti:2019pca,vanBeest:2021lhn}. 
It is believed to hold true even when the definition of symmetry is extended, for example by including higher-form global symmetries \cite{Gaiotto:2014kfa}, see e.g.~\cite{Heidenreich:2020pkc,Heidenreich:2021xpr} for recent works. 
The cobordism conjecture \cite{McNamara:2019rup} is yet another instance of this general feature. 
It associates a generalised global charge to any non-trivial cobordism group and thus it postulates its cancellation. 
In this work, we apply the cobordism conjecture to setups with a given, fixed background $X$.

The cobordism charge is topological and should vanish on consistent quantum gravity backgrounds. 
These can, and in general have to, be endowed with defects and/or gauge fields, in order respectively to break or gauge the cobordism symmetry. 
As already discussed in \cite{McNamara:2019rup}, breaking the symmetry can lead to the prediction of new defects in quantum gravity. 
This idea has been further developed in several directions \cite{GarciaEtxebarria:2020xsr,Montero:2020icj,Dierigl:2020lai,Hamada:2021bbz,Debray:2021vob,McNamara:2021cuo,Bedroya:2021fbu}, in relation to the Hypothesis H \cite{Sati:2021uhj} (see also \cite{Fiorenza:2019usl,Sati:2019tqq,Sati:2020cml}), and it has been given an effective dynamical description \cite{Buratti:2021yia,Buratti:2021fiv,Angius:2022aeq,Blumenhagen:2022mqw,Angius:2022mgh}. 
Generally, the introduction of new ingredients into the setup modifies the structure (topological and geometric) of the given compact manifold. 
The aim is thus to get closer and closer to the postulated quantum gravity structure, even if a preferred path does not seem to be obvious. 
In this respect, in \cite{Andriot:2022mri} the Whitehead tower construction has been identified as a possible organising principle for topological structures entering the cobordism conjecture.

One can think of compact manifolds and defects in cobordism as being associated to closed and open strings respectively. 
Besides, it is well-known that open strings are related to K-theory \cite{Witten:1998cd}. 
At the mathematical level, a correspondence has been uncovered between certain versions of cobordism and K-theory. It is deeply rooted into the structure of such generalised (co)homology theories and relies on established theorems, such as the Conner--Floyd \cite{ConnerFloyd} and
the Hopkins--Hovey \cite{Hopkins1992} theorems. 
As proposed in \cite{Blumenhagen:2021nmi}, at the physical level it is tempting to interpret this correspondence as a manifestation of open-closed string duality. 
This would directly imply that if the cobordism charge must cancel, the K-theory charge should undergo the same fate. 
Indeed, cancellation of K-theory charge has already been proposed in the literature from different perspectives \cite{Uranga:2000xp,Blumenhagen:2019kqm,Damian:2019bkb,GarciaEtx2019}.

Elaborating on this intuition, in \cite{Blumenhagen:2021nmi} it has also been discussed how the structure behind certain tadpoles in string theory is precisely given by the interplay between cobordism and K-theory charges. 
This observation is pertinent to the case in which symmetries are gauged rather than broken, which has been little explored so far.

While cobordism and K-theory groups considered in \cite{McNamara:2019rup,Blumenhagen:2021nmi} are those of the point, e.g.~$\Omega^{\rm Spin^c}_n({\rm pt})$ and $K^{-n}({\rm pt})$, a more generic situation is to consider the case in which pt is replaced by a topological space $X$.
In fact, the Conner--Floyd and the Hopkins--Hovey theorems also hold in this more general case.
At the mathematical level this requires that, given a compact $n$-dimensional manifold $M$ representative of a cobordism class, there exists a continuous map $f:M\to X$ with certain properties at the boundary of the cobordism. 
As we are going to argue, at the physical level it means that one is fixing the topology of $X$ and studying global symmetries associated to fluctuations in topology along the unfixed, complementary directions (if any) within $M$. 
On the other hand, fixing a manifold $X$ is most natural from a string theory point of view, where one is typically forced to perform a background-dependent analysis. Thus, this is yet another motivation for looking at cobordism and K-theory groups of a given $X$.

It is the purpose of this paper to  work out in detail such cobordism and K-theory classes on $X$ and to see how the Hopkins--Hovey isomorphism continues to hold beyond the trivial case, $X={\rm pt}$. 
For simple examples of $X$, such  as spheres, tori and Calabi-Yau two- and three-folds, we can be very explicit and understand the results in terms of a straightforward dimensional reduction of global symmetries and their gauging and breaking.
More precisely, we are going to show that, when performing a dimensional reduction of a given effective theory on $X$, a lattice of charges arises which is reproduced by the evaluation of cobordism and K-theory groups $\Omega^{\rm Spin^c}_n(X)$ and $K^{-n}(X)$ (or $\Omega^{\rm Spin}_n(X)$ and $KO^{-n}(X)$).
In this way, by showing consistency upon compactification, our analysis gives further support to the proposal of \cite{Blumenhagen:2021nmi} that gauging of cobordism and K-theory symmetries related via the Hopkins--Hovey isomorphism generically occurs simultaneously, so that both charges appear in the same Bianchi identities and tadpole conditions.\footnote{As already noticed in \cite{Blumenhagen:2021nmi}, the presence of 2-torsion classes can lead to a decoupling of the two charges in the tadpole.}

The technical tool we employ to compute cobordism and K-theory groups of $X$ is the Atiyah--Hirzebruch spectral sequence. 
It is a standard technique in algebraic topology and we will refer mainly to \cite{davis2012lecture,Hatcher:478079}; see also \cite{Garcia-Etxebarria:2018ajm} for a recent review and application to particle physics. 
This spectral sequence has been applied in a string theoretic context already in \cite{Diaconescu:2000wy, Brunner:2001eg, Brunner:2001sk, Maldacena:2001xj} among others, and those works will be relevant for us.  
Aiming at giving a self-contained presentation for a physicist reader, we will introduce the aforementioned technique in a pedagogical way and collect known mathematical results, which are employed systematically in the computations.

After computing the cobordism and K-theory groups of interest, we explain how to extract the information on the dimensional reduction of global symmetries (either broken or gauged) contained in them. 
We discuss both cobordism and K-theory groups and show how they cooperate in reproducing known tadpoles in string theory for compactifications on $X$, hence extending the analysis of \cite{Blumenhagen:2021nmi}.

This work is organised as follows. 
In section \ref{sec_cobkth}, we review the main properties of cobordism and K-theory needed for our analysis. 
In particular, we explain the difference between groups of the point and of a manifold $X$, and we recall the physical consequences of the correspondence between cobordism and K-theory, leading to a bottom-up construction of tadpole cancellation conditions. 
We also review how continuous higher-form symmetries (either broken or gauged) behave under dimensional reduction, when this is performed in standard (co)homology. We expect that cobordism and K-theory should give a refined description.
In section \ref{sec_AHSS}, we compute cobordism and K-theory groups of $X$ via the Atiyah--Hirzebruch spectral sequence, for $X$ equal to spheres, tori, Calabi-Yau two- and three-folds. We explain in detail the main subtleties entering such computations and how they can be solved in our (simple) circumstances. 
In section \ref{sec_interpretation}, we propose a physical interpretation of these results in terms of the dimensional reduction of higher-form symmetries, when the theory is compactified on $X$. 
We show explicitly how the information of the reduction is encoded into cobordism and K-theory groups of $X$. This formalism is more precise than standard (co)homology, since it also takes into account quantum mechanical effects, such as cancellation of Freed--Witten anomalies.
The latter are however absent for the simple backgrounds we consider.
In section \ref{sec_conclusion}, we draw our conclusions and outline future research directions. Additional material is organised into three appendices: In appendix \ref{sec_app_math}, we summarise further useful notions of algebraic topology and we present a calculation of cobordism groups of spheres and tori by induction. In appendix \ref{sec_app_cobordism} and \ref{sec_app_Ktheory}, we collect tables with all the main results from section \ref{sec_AHSS}.

\section{Cobordism and K-theory}
\label{sec_cobkth}

In this section, we review elements of cobordism and K-theory which are relevant for our analysis. We concentrate in particular on the difference between groups of the point and of a given compact manifold $X$. We recall their interpretation in terms of global symmetries and the correspondence between cobordism and K-theory, which is relevant when gauging such symmetries.

\subsection{Cobordism groups of pt and $X$}
\label{sec:cobgroupsrev}

Cobordism is a generalised homology theory classifying compact manifolds of the same dimension.\footnote{We recall that a generalised (co)homology theory can be defined axiomatically. It satisfies the same axioms as ordinary (co)homology, but for the dimension axiom \cite{davis2012lecture, Hatcher:478079}: $H_n({\rm pt})=0$ for $n>0$.} 
Two compact $n$-dimensional manifolds $M$ and $N$ are cobordant, if there  exists an $(n+1)$-dimensional compact manifold $W$ called cobordism, such that 
$\partial W = M \sqcup N$. 
The notion of cobordism can be refined by considering tangential structures on $M$, $N$ and $W$. An elementary example is requiring $M$, $N$ to be oriented. The cobordism $W$ is then such that $\partial W = M \sqcup \bar N$, where the bar refers to reversed orientation.
In general, one can also consider manifolds with more refined tangential structures, as for example Spin or Spin$^c$.

Being cobordant is an equivalence relation. Under disjoint union, the set of equivalence classes of compact $n$-dimensional manifolds with $\xi$-structure is an abelian group, called the cobordism group $\Omega_n^\xi ({\rm pt})$. 
More precisely, this is the cobordism group of the point (pt). The cobordism conjecture of \cite{McNamara:2019rup} is  the statement that there exists a not necessarily unique quantum gravity structure QG such that the associated cobordism groups are trivial,
\begin{equation}
\Omega_n^{\rm QG}({\rm pt}) = 0.
\end{equation}

The physical interpretation is the following. One should think of a compact $n$-dimensional manifold $M$ representative of the cobordism class $[M]$ as the compact part of a $d$-dimensional background of a theory of quantum gravity, such as string theory.
A non-vanishing cobordism group, $\Omega_n^\xi ({\rm pt})\neq 0$, is then interpreted in \cite{McNamara:2019rup} as the presence of a $(d-n-1)$-form global symmetry in the $d$-dimensional effective field theory with a conserved global current, $d J_{n} =0$. 
Via gluing a non-trivial element $[M]\in  \Omega_n^\xi ({\rm pt})$ to $\mathbb{R}^n$,
one can construct a $(d-n)$ dimensional defect. This is a gravitational soliton  carrying charge
under the $(d-n-1)$-form global symmetry.\footnote{We thank the referee for pointing this out to us.}
In order to avoid global symmetries in quantum gravity, the cobordism charge has  either to be broken or gauged.

As explained in \cite{McNamara:2019rup}, breaking the symmetry requires the introduction of  $(d-n-1)$-dimensional defects, such that current is not conserved anymore,
\begin{equation}
0 \neq d J_{n} = \sum_{\text{def $j$}}\delta^{(n+1)}(\Delta_{d-n-1,j})\,,    
\end{equation}
where the $\delta$-functions are the Poincar\'e dual of the $(d-n-1)$-cycles wrapped by the defects.
The structure is modified, $\xi \to \xi +\text{defects}$, in such a way that the cobordism group with the refined structure is possibly vanishing, $\Omega^{\xi + \text{defects}}_n({\rm pt})=0$. 
Thus, the cobordism conjecture predicts additional objects in the effective field theory. 
Instead, if the background has vanishing charge $[M] = 0 \in \Omega_n^\xi({\rm pt})$, even if the group is still non-trivial, the symmetry can be gauged. 
In this case, there exist appropriate gauge fields such that 
\begin{equation}
\label{J=dF}
    J_{n} = d F_{n-1}\,.
\end{equation}
One can now refine the structure by introducing these gauge fields, in such a way that manifolds in the zero equivalence class, $[M]=0$, lie in the cokernel of the map $\Omega^{\xi +\text{gauge}}_n ({\rm pt}) \to \Omega^{\xi }_n ({\rm pt}) $.

\begin{figure}[h]
\begin{center}
\begin{tikzpicture}
\draw [line width=0.3mm] (0,0) arc (0:360:0.3 and 1.3);
\draw [line width=0.3mm,dashed] (7.3,0) arc (0:360:0.3 and 1.3); 
\draw [line width=0.3mm] (7.,1.3) arc (90:-90:0.3 and 1.3); 
\draw [line width=0.3mm](1.8,-3) arc (360:0:-1.5 and .3); 
\draw[rounded corners=50pt,line width=0.3mm](-.3,1.3)--(3.6,1.3)--(7,1.3);
\draw[rounded corners=50pt, line width=0.3mm](-.3,-1.3)--(3.6,-1.3)--(7,-1.3);
\draw   [line width=0.3mm](.85,-.0) arc (-80:0:1cm and 0.5cm);
\draw  [line width=0.3mm] (1,0.04) arc (194:92:.7cm and 0.3cm);
\draw  [line width=0.3mm](4.,0.12) arc (175:315:1cm and 0.5cm);
\draw  [line width=0.3mm](5.4,-0.38) arc (-31:180:.7cm and 0.4cm);
\node (a) at (-0.3,0) {$M$};
\node (b) at (7.,0) {$N$};
\node (c) at (3.4,0.4) {$W$};
\node (d) at (3.3,-3) {$X$};
\node (e) at (3.6,-2) {$h$};
\node (f) at (1.3,-2) {$f$};
\node (g) at (5.5,-2) {$g$};
\draw [->]  (-0.2,-1.5) -- (1.7,-2.8) ;
\draw [->]  (3.3,-1.5) -- (3.3,-2.5);
\draw [->]  (6.9,-1.5) -- (4.9,-2.8) ;
\end{tikzpicture}
\caption{Cobordism $(W,h)$ between $(M,f)$ and $(N,g)$.}
\label{refcob}
\end{center}
\end{figure}
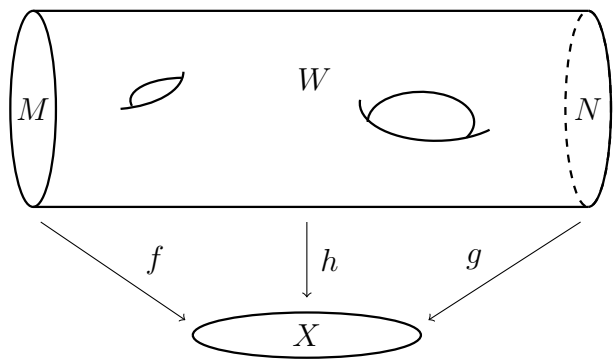

The discussion can be generalised by going from the cobordism group of the point to that of a topological space $X$, denoted $\Omega_n^\xi(X)$. Consider continuous maps $f:M\to X$ and $g: N \to X$, which can be understood as deformations of $M$ and $N$ into $X$, even if the construction is more general. 
In particular, the dimension of $X$ does not need to be the same as of $M$ and $N$. Two pairs $(M,f)$ and $(N,g)$ are cobordant if there is a cobordism $W$, such that 
$\partial W = M \sqcup N$, together with a map $h:W\to X$ appropriately restricting to $f$ and $g$ at the boundary $\partial W$. 
One can show that this is an equivalence relation and the cobordism group $\Omega_n^\xi(X)$ is the set of equivalence classes of pairs $(M,f)$. 
For $X={\rm pt}$ we recover the definition of cobordism group of the point.
We give a schematic representation in figure \ref{refcob}.

In the present work, we are interested in the case in which $X$ is a compact Spin or Spin$^c$ manifold without any additional structure, such as (higher-form) gauge fields.
In particular, we will consider $X=\{S^k, T^k, K3, CY_3\}$, but the analysis can be repeated in principle for any other space. Our choice is motivated by the fact that these manifolds are commonly employed in string compactifications and they turn out to be  simple enough to explicitly evaluate $\Omega_n^\xi(X)$.

When considering cobordism groups of the point, $\Omega^\xi_n$(pt), one is looking at global symmetries of the $d$-dimensional effective theory by scanning through all possible topologies of $n$-dimensional compact manifolds. 
We will see that, when going from ${\rm pt}$ to $X$, the cobordism group is generically enlarged.  In particular, the classes $\Omega_n^\xi({\rm pt})$ will also be present in $\Omega_n^\xi (X)$, but new classes can appear, depending on the topology of $X$. Intuitively, a non-trivial topology carries a charge which is detected by the cobordism group and, in turn, it increases its rank.

This intuitive picture on the relation between  $\Omega_n^\xi({\rm pt}) $ and  $\Omega_n^\xi (X)$ can be made precise by recalling a standard result which follows from the Splitting Lemma for abelian groups and which we will use systematically in our analysis. Consider the forgetful map
\begin{equation}
  \phi:\quad \Omega^\xi_n(X) \to \Omega^\xi_n({\rm pt})\,,
\end{equation}
acting as $\phi([M,f]) = [M]$. Its kernel is called the reduced cobordism group, denoted as
\begin{equation}
\ker \phi \equiv \tilde \Omega_n^\xi (X)\,,
\end{equation}
and one has $\tilde \Omega_n^\xi({\rm pt})=0$ by definition. One can show that this map is surjective and that the short exact sequence
\begin{equation}
0 \longrightarrow \tilde \Omega^\xi_n(X) \longrightarrow \Omega^\xi_n(X) \overset{\phi}{\longrightarrow} \Omega^\xi_n({\rm pt}) \longrightarrow 0
\end{equation}
is split. Therefore, one finds
\begin{equation}
\label{SplittLemmaCob}
\Omega^\xi_n(X) = \Omega^\xi_n({\rm pt})  \oplus \tilde \Omega^\xi_n(X),
\end{equation}
which is valid for any structure $\xi$. In accordance with our intuitive picture, when passing from $ \Omega^\xi_n ({\rm pt})$ to $ \Omega^\xi_n (X)$ the rank of the group is indeed increased. We interpret $\tilde \Omega_n^\xi(X)$ as the part of global symmetries genuinely stemming from having fixed a manifold $X$.

We will see explicitly in our examples that $\tilde \Omega_n^\xi(X)$ can be further decomposed into a direct sum of several pieces, but the details of the splitting depend on the topology of $X$. Each of these pieces will give rise to global symmetries. Therefore, for a given $X$, we will uncover a lattice of global symmetries organised according to the topology. 
These global symmetries are deemed pathological in a theory of quantum gravity, so one has to have
\begin{equation}
\Omega^{\rm QG}_n(X)=0,
\end{equation}
for $X$ a consistent on-shell background of quantum gravity. This is implied by the cobordism conjecture.

\subsection{K-theory groups of pt and $X$}

K-theory is a generalised cohomology theory classifying vector bundles over a space $X$. Consider two vector bundles $E$ and $F$ over $X$, which can be of different rank, and construct $(E,F) = E-F$. Then, one can introduce an equivalence relation
$(E\oplus H, F\oplus H)\sim (E,F)$, for any  bundle $H$.
The set of equivalence classes is a group, called the K-theory group $K(X)$. 
To be precise, $K(X)$ is the set of equivalence classes of complex vector bundles. We will also consider the set of equivalence classes of real vector bundles, denoted $KO(X)$.
Besides the simple choice $X={\rm pt}$, in this work we will focus on K-theory groups of $X=\{S^k,T^k,K3,CY_3\}$. 

Similarly to what we discussed for cobordism, the relation between K-theory groups of the point and of $X$ can be made mathematically precise by employing the Splitting Lemma. One can indeed consider the map
\begin{equation}
\varphi: \quad K(X) \to K({\rm pt}),    
\end{equation}
which can be identified with the (virtual) rank
\begin{equation}
\varphi[(E,F)] = \text{rank}(E)-\text{rank}(F), \qquad (E,F) \in K(X)\,.
\end{equation}
The reduced K-theory group is then defined as
\begin{equation}
\ker \varphi \equiv \widetilde K(X),
\end{equation}
namely it classifies vector bundles on $X$ with the same rank.
One can show that the map $\varphi$ is surjective and that the short exact sequence
\begin{equation}
    0\to \widetilde{K}(X) \to K(X) \overset{\varphi}{\to} K({\rm pt}) \to 0
\end{equation}
is split, $K(X) = K({\rm pt})\oplus \tilde K(X)$. 

By introducing the reduced suspension $\Sigma$, the higher reduced K-theory groups are then defined as
\begin{equation}
\begin{aligned}
\tilde K^{-n}(X) &=  \tilde K(\Sigma^n X)\,,\\
\widetilde{KO}^{-n}(X) &= \widetilde{KO}(\Sigma^n X),\,
\end{aligned}
\end{equation}
for $n \in \mathbb{Z}$, $n\geq 0$. By recalling the relations $K(X) = \widetilde K(X\sqcup {\rm pt})$ and $KO(X) = \widetilde{KO}(X\sqcup {\rm pt})$ and the properties of the reduced suspension (see appendix \ref{app_smash}), one has also
\begin{equation}
\begin{aligned}
\label{Knpt}
\widetilde K(S^n) &=\widetilde{K}(\Sigma^n S^0) = K^{-n}({\rm pt}),\\ 
\widetilde{KO}(S^n) &=\widetilde{KO}(\Sigma^n S^0)= KO^{-n}({\rm pt}).
\end{aligned}
\end{equation}
Then, the full generalised cohomology theories can be constructed as
\begin{equation}
\label{Ksplitting}
\begin{aligned}
K^{-n}(X) &= K^{-n}({\rm pt}) \oplus \widetilde{K}^{-n}(X),\\
KO^{-n}(X) &= KO^{-n}({\rm pt}) \oplus \widetilde{KO}^{-n}(X).
\end{aligned}
\end{equation}
They satisfy the important property known as Bott periodicity
\begin{equation}
\begin{aligned}
K^{-n}(X) &= K^{-n+2}(X),\\
KO^{-n}(X) &= KO^{-n+8}(X)\,.
\end{aligned}
\end{equation}

For $X={\rm pt}$, the physical interpretation of these groups is well-known: they classify D-branes in string theory \cite{Witten:1998cd} (see \cite{Olsen:1999xx,Evslin:2006cj} for a review). In particular, $KO^{-n}({\rm pt})$ classify $p=9-n$ branes in type I, while $K^{-n}({\rm pt})$ classify $p=9-n$ and $p=10-n$ branes in type IIB and IIA \cite{Witten:1998cd, Horava:1998jy}  respectively. This can be understood directly from \eqref{Knpt}, meaning that
K-theory groups of the point at degree $n$ classify D-branes in flat space which are point-like with respect to $S^n$. Note that this includes the type II and type I BPS D-branes charged with respect to the R-R forms, but it can also give stable non-BPS torsion branes.

As for cobordism, one can argue that a global symmetry is associated to each non-vanishing K-theory group, which must cancel in quantum gravity \cite{Uranga:2000xp,Blumenhagen:2019kqm}. It is known that K-theory charges are gauge charges \cite{Moore:1999gb,Freed:2000tt}, so that their global  symmetries will be  gauged rather than broken. 
When passing from K-theory groups of the point to groups of $X$ we are in fact introducing additional global symmetries. 
In particular, the reduced K-theory groups contain the information on symmetries genuinely associated to the choice of $X$.  We will see explicitly in our examples that $\widetilde K^{-n}(X)$ and  $\widetilde{KO}^{-n}(X)$ can be further split into a direct sum of several pieces, but the details of the splitting depend on the topology of $X$. 
In terms of D-branes, the first term on the right hand side of \eqref{Ksplitting}
is a universal contribution, i.e.~branes not depending on the choice of $X$ and wrapping it completely. The second term contains branes wrapping internal cycles of $X$ (and directions orthogonal to $X$).

Let us finally note that, for a physical interpretation in terms of (wrapped) D-branes in critical type IIB and type I string theory, we should restrict to values of $n$ such that
\begin{equation}
n+ k \leq 10\,
\end{equation}
where we defined $k={\rm dim}(X)$.

\subsection{Gauging and tadpoles}
\label{sec_gaugandtad}

As we briefly review in this section, at the mathematical level, cobordism and K-theory are deeply related. 
For Spin/Spin$^c$ cobordism on one side and real/complex K-theory on the other side, this fact is based on rigorous theorems \cite{ConnerFloyd,Hopkins1992}.
A physical interpretation of this inherent relation in terms of the gauging of global symmetries has been provided  in \cite{Blumenhagen:2021nmi}, which will also be recalled.
For more details, we refer the reader to that paper.

It is instructive to have a close look at the known cobordism and K-theory groups of the point.
In the table \ref{SpinSpincptgroups} we list the cobordism groups\footnote{In our conventions, the notation $n G$ means $G^n=\underbrace{G\oplus\ldots\oplus G}_{n\  {\rm times}}$.}
\begin{table}[h!]
\centering
\begin{tabular}{ c | c c c c c c c c c c c }
$n$ & 0 & 1 & 2 &  3 & 4 & 5 & 6 & 7 & 8 & 9 & 10\\
\midrule
 $\Omega^{\rm Spin}_n({\rm pt})$ &$\mathbb{Z}$ & $\mathbb{Z}_2$  & $\mathbb{Z}_2$  & 0 & $\mathbb{Z}$ & 0 & 0 &0 & 2$\mathbb{Z}$ & 2$\mathbb{Z}_2$ & 3$\mathbb{Z}_2$   \\
 $\Omega^{\rm Spin^c}_n ({\rm pt})$ &$\mathbb{Z}$ & 0  & $\mathbb{Z}$  & 0 & 2$\mathbb{Z}$ & 0 & 2$\mathbb{Z}$ &0 & 4$\mathbb{Z}$ &0 & 4$\mathbb{Z} \oplus \mathbb{Z}_2$\\
\end{tabular}
\caption{Spin and Spin$^c$ cobordism groups of the point up to $n=10$.}
\label{SpinSpincptgroups}
\end{table}

\vspace{0.2cm}
\noindent
and in table \ref{KKOptgroups}  the corresponding K-theory groups, where
Bott periodicity is manifest.
\begin{table}[h!]
\centering
\begin{tabular}{ c | c c c c c c c c c c c }
$n$ & 0 & 1 & 2 &  3 & 4 & 5 & 6 & 7 & 8 & 9 & 10\\
\midrule
 $KO^{-n} ({\rm pt})$ &$\mathbb{Z}$ & $\mathbb{Z}_2$  & $\mathbb{Z}_2$  & 0 & $\mathbb{Z}$ & 0 & 0 &0 & $\mathbb{Z}$ & $\mathbb{Z}_2$ & $\mathbb{Z}_2$   \\
   $K^{-n} ({\rm pt})$ &$\mathbb{Z}$ & 0  & $\mathbb{Z}$  & 0 & $\mathbb{Z}$ & 0 & $\mathbb{Z}$ &0 & $\mathbb{Z}$ &0 & $\mathbb{Z}$ \\ 
\end{tabular}
\caption{K- and KO-groups of the point up to $n=10$.}
\label{KKOptgroups}
\end{table}

\noindent One can see that they are isomorphic up to $n=8$, for Spin cobordism and real K-theory, and up to $n=4$, for Spin$^c$ cobordism and complex K-theory.
Indeed, one can introduce a map known as the Atiyah--Bott--Shapiro (ABS) orientation \cite{Atiyah:1964zz}
\eq{
\alpha:\quad &\Omega_n^{\rm Spin}({\rm pt}) \,\to KO^{-n}({\rm pt})\\[0.1cm]
\alpha^c:\quad &\Omega_n^{{\rm Spin}^c}({\rm pt}) \to K^{-n}({\rm pt})
}
and then show that there exist isomorphisms\cite{KreckStolz}
\begin{align}
\label{isom1}
    \Omega_n^{\rm Spin}({\rm pt}) / \ker \alpha &\cong KO^{-n}({\rm pt})\, ,\\[0.1cm]
    \label{isom2}
     \Omega_n^{{\rm Spin}^c}({\rm pt})  / \ker \alpha^c&\cong K^{-n}({\rm pt})\,.
\end{align}
Explicitly, the two ABS orientations at fixed degree $n$  are given by the Todd genus, i.e.~the index of the Spin$^c$ Dirac operator
\begin{equation}
\alpha^c_n([M])= {\rm Td}(M) \equiv \int_M {\rm td}_n(M)\,,
\end{equation}
and by the index of the Dirac operator on $M$, respectively \cite{HITCHIN19741}
\begin{equation}
\label{ABSorient}
\alpha_n([M]) = \left\{\begin{array}{ccl}
 \hat A(M) & &n=8m,\\
 \hat A(M)/2   & & n = 8m+4,\\
\text{dim} \, H \quad \,\,\,{\rm mod}\ 2 & & n = 8m+1,\\
\text{dim} \, H^+ \quad {\rm mod}\ 2 & & n = 8m+2,\\
0 & &\text{otherwise},
\end{array}\right.
\end{equation} 
where $\hat A(M)$ is the $\hat A$ genus and $H$ ($H^+$) the space of (positive) harmonic spinors.

In \cite{Blumenhagen:2021nmi}, it has been proposed that whenever the ABS orientation exists, as for example in type IIB string theory on a Spin$^c$ manifold, it is not just the K-theory charge that is gauged but actually a combination of the K-theory and the respective cobordism charge.
Hence, schematically the gauging can proceed  like
\begin{equation}
\label{schemabianchi}
dF_{n-1}=  J^{K}_n + a^{(n)}\, J_n^{\rm cobord}
\end{equation}
(no sum over $n$ in the second term on the right hand side), i.e.~it is a linear combination of K-theory and cobordism global charges that is gauged.
The constant $a^{(n)}$ is not a priori fixed 
and in certain cases can also be vanishing.
It would be interesting to uncover whether and how it can be fixed just from mathematical data, without any reference to a concrete
quantum gravity theory, like string theory.
Upon integration over a compact space, such a Bianchi identity implies a charge neutrality condition, which was shown in \cite{Blumenhagen:2021nmi} to match some of the tadpole cancellation conditions known from string theory.

More precisely, given non-vanishing groups $\Omega_n^{{\rm Spin}^c}({\rm pt})$ and  $K^{-n}(\rm pt)$, then the current $J_n^{\rm cobord}$ on the right hand side of \eqref{schemabianchi}  is given  by the sum over the cobordism invariants of $\Omega_n^{{\rm Spin}^c}({\rm pt})$, denoted as $\mu_n^j$.
For $0\le n\le 6$, a list of the independent cobordism invariants of $\Omega_n^{{\rm Spin}^c}({\rm pt})$ is\footnote{For the Chern classes, we use the shorthand notation $c_i(M) \equiv c_i(TM)$. Similarly for Stiefel-Whitney classes.}
\begin{equation}
\begin{aligned}
\label{cobinvspinc}
  \mu_0&={\rm td}_0(M)=1\,,\\
  \mu_2&={\rm td}_2(M)={\frac{1}{2}} c_1(M)\,,\\
  \mu^1_4&={\rm td}_4(M)={\frac{1}{12}}\left(c_2(M)+c_1^2(M)\right)\,,\qquad
  \mu^2_4=c^2_1(M)\,,\\
  \mu^1_6&={\rm td}_6(M)={\frac{1}{24}} c_2(M)\,
  c_1(M)\,,\qquad\qquad\ 
  \mu^2_6={\frac{1}{2}}c^3_1(M) \,.
  \end{aligned}
\end{equation}
Notice that the ABS orientation always provides one cobordism invariant, but in general there can be more and they all have to be taken into account since they all contribute as global charges in general. 
The K-theory theory current $ J^{K}_n$ is simply defined   by the delta functions for the localised branes classified by $K^{-n}(\rm pt)$.
Upon integration over a representative manifold $M\in[M]$, one thus obtains a tadpole constraint of the form
\begin{equation}
\label{tadbotup}
    0 = \int_{M} dF_{n-1} =  \int_{M}\sum_{i \in \text{def}} Q_i \,\delta^{(n)}(\Delta_{10-n,i})+\int_{M} \sum_{j\in \text{inv}} a_j^{(n)} \,\mu^j_n\,,
\end{equation}
where $\Delta_{10-n,i}$ is the submanifold wrapped by the $i$-th $Dp$-brane (with $p=9-n$ in type I/IIB) with charge $Q_i$.
Notice  that \eqref{tadbotup} is valid off-shell, for all compact manifolds cobordant with $M$.

Let us give an illustrative example taken from \cite{Blumenhagen:2021nmi}. Consider $\Omega_6^{{\rm Spin}^c}({\rm pt}) = \mathbb{Z}\oplus\mathbb{Z}$. 
This is a three-form global symmetry in ten dimensions and $K^{-6}(\rm pt)$ classifies D3-branes.
Specialising \eqref{tadbotup} to this case, we get
\begin{equation}
    0 = \int_{M} \sum_{i \in \text{def}} Q_i \,\delta^{(6)}(\Delta_{4,i}) + \int_{M}\left(a^{(6)}_1 \, \frac{c_1\,c_2}{24} + a^{(6)}_2 \,\frac{c_1^3}{2} \right)\,.
\end{equation} 
For the particular choice $a^{(6)}_1=-12$ and $a^{(6)}_2=-30$, this tadpole is realised by F-theory on a smooth Calabi-Yau fourfold which is elliptically fibered over a base $M$.
Many more examples were discussed in \cite{Blumenhagen:2021nmi}, supporting the above conjecture about the simultaneous gauging of K-theory and cobordism global symmetries.

For the type I string, the Spin cobordism and the $KO$ groups are relevant. 
For continuous global symmetries, corresponding to $\mathbb Z$ groups in tables \ref{SpinSpincptgroups} and \ref{KKOptgroups}, one similarly gets tadpole cancellation conditions known from the type I literature. 
For gauging torsion $\mathbb Z_2$ groups, even though there are no gauge fields that one can introduce, there is an in principle  mixed $\mathbb Z_2$-valued charge neutrality condition.
For instance, for the case $\Omega_2^{{\rm Spin}}({\rm pt})/KO_2({\rm pt})$ this reads 
\begin{equation}
\label{tadS1xS1}
0=\int_{M} \sum_i Q_i\, \delta^{(2)}(\Delta_{8,i}) - a^{(2)}\, \alpha_2(M)\qquad {\rm mod}\ 2\,,
\end{equation}
where  we have a contribution from non-BPS $\widehat{D7}$-branes and from the cobordism group.
However, for $a^{(2)}$ even, the latter contribution decouples in this equation and the global symmetry $\Omega_2^{{\rm Spin}}=\mathbb Z_2$ still needs to be broken by appropriate defects. 
Since the generator of $\Omega_2^{{\rm Spin}}$ is $M=S^1_p\times S^1_p$, a manifold that is a valid background of the type I string (without needing the presence of a $\widehat{D7}$-brane), in this case $a^{(2)}$  should indeed be even. 
Whether such a decoupling  holds generically for all $\mathbb Z_n$ torsion charges remains to be understood.

Since the Hopkins-Hovey isomorphism holds also for cobordism and K-theory groups on generic backgrounds $X$, the gauging should carry over to this more general case as well, if the assertion of \cite{Blumenhagen:2021nmi} is really correct.
To check this quantitatively, we need to understand how such groups are actually computed, at least for some treatable classes of background manifolds $X$, and what kind of physical information they contain.
The expectation is that, when specifying a background $X$, at least part of the cobordism and K-theory groups can be understood via dimensional reduction on $X$.

 \subsection{Dimensional reduction of symmetries}
\label{sec_dimredsymm}

Before we delve into the mathematically rather involved evaluation of cobordism
and K-theory groups, it is instructive to review how dimensional reduction is usually performed in (co)homology. This will help us  in appreciating  what one really gains from using the description in terms of cobordism and K-theory.

Let us consider an effective theory in $d$ dimensions, where of course we have $d=10$ in mind. Recall that to a continuous global $p$-form symmetry, there exists an associated   current $J_{n}$, with $n=d-p-1$, which is closed
\begin{equation}
dJ_n=0\,.
\end{equation}
To break the symmetry, one introduces defects such that the current ceases to be  closed anymore
\begin{equation}
d J_{n} = \delta^{(n+1)}(\Delta_p) \neq 0\,,
\end{equation}
where $\Delta_{p}$ is the cycle wrapped by the defect.
To gauge the symmetry, one introduces (and further integrates over) gauge fields coupling minimally to the current
\begin{equation}
S = \int\left( -\frac12 F_{p+2} \wedge * F_{p+2} + C_{p+1} \wedge J_{n} + \dots\right), \qquad F_{p+2} = d C_{p+1},
\end{equation}
such that from the equations of motion of $C_{p+1}$ it follows that the current is trivial in cohomology (i.e.~exact)
\begin{equation}
\label{JdsF}
J_{n} = (-1)^{p} \,d * F_{p+2}\,.
\end{equation}

When performing a dimensional reduction over a compact space $X$, one typically expands the various objects, such as currents and gauge fields, in a cohomological basis of $X$. The expansion coefficients are fields propagating along the external non-compact dimensions. Classically, the expansion can be performed in de Rham cohomology, $H^p(X; \mathbb{R})$. Quantum mechanics typically imposes  charges to be quantised, and thus one would rather consider singular cohomology with integer coefficients, $H^p(X;\mathbb{Z})$.

Let us now compactify the theory on a $k$-dimensional space $X$. We get a $D=d-k$ dimensional effective theory with broken and gauged symmetries inherited from the parent theory. 
In general, a given $p$-form symmetry in $D$ dimensions can receive contributions from different $(p+q)$-form symmetries of the $d$-dimensional theory. To these contributions, we associate currents $J_{n+m}$, now with $p=D-n-1$ and $q=k-m$, wrapping $m=0,1,\dots,k$ cycles in $X$ and extending along $n$ directions in the non-compact space. 
Let us consider a basis of cohomology, $\omega_{(m)a} \in H^{m}(X;\mathbb{Z})$, where $a=1,\dots,b_m$, with $b_m$ the Betti numbers. We can decompose the currents as
\begin{equation}
  J_{n+m} =\sum_{a=1}^{b_m} j_{n}^{(m)a} \wedge \omega_{(m)a}  \,.
\end{equation}
Thus, $p$-form symmetries in $D$ dimensions arise from the set of currents $j_{n}^{(m)a}$, for $a=1,\dots b_m$ and $m=0,\dots,k$. Since we are performing an expansion in cohomology, we see that if the $J_{n+m}$ are closed, $dJ_{n+m}=0$, they produce a lattice of global $p$-form symmetries in $D$ dimensions, namely
\begin{equation}
\label{djnma}
d j_{n}^{(m)a}=0, \qquad \forall \, a=1,\dots b_m, \quad \forall\, m=0,\dots k\,.
\end{equation}

Breaking or gauging global symmetries in $d$ dimensions does not jeopardise this structure and one generically expects a lattice of broken or gauged symmetry in the lower dimensional theory to be produced. They arise from different broken or gauged symmetries of the original theory.
As we will discuss next, delta functions for the defects breaking the symmetry and gauge fields can be similarly expanded in cohomology. 

To break the currents $J_{n+m}$ we need forms $\delta^{(n+m+1)}$ in the $d$-dimensional theory such that $dJ_{n+m} = \delta^{(n+m+1)}(\Delta_{p+q}) \neq 0$, with $p+q=d-n-m-1$. These forms represent defects wrapping submanifolds $\Delta_{p+q} = \Pi_p \times \Sigma_q$ of the $d$-dimensional space, where $\Pi_p$ is a $p$-dimensional submanifold of the non-compact space, while $\Sigma_q$ is a $q$-dimensional cycle of $X$. For the global symmetry to be broken in the $D$-dimensional theory, we take $p=D-n-1$ and $q=k-m$, in such a way that the defect in the dimensionally reduced theory has codimension $n+1$ (for $n=0$, namely a $(D-1)$-form global symmetry, we get a domain wall in $D$ dimensions).  We can then formally expand in cohomology
\begin{equation}
\delta^{(n+m+1)}(\Delta_{p+q}) = \sum_{a=1}^{b_m} \delta^{(n+1)}(\Pi_p)^{(m)a} \wedge \omega_{(m)a}.
\end{equation}
Thus, from any defect $\delta^{(n+m+1)}$ in $d$ dimensions we generate a lattice of codimension $n+1$ defects in $D$ dimensions, $ \delta^{(n+1)}(\Pi_p)^{(m)a}$.
They can be used to break the lattice of global currents \eqref{djnma}, i.e.
\begin{equation}
dj_n^{(m)a} = \delta^{(n+1)}(\Pi_{D-n-1})^{(m)a} \neq 0,
\end{equation}
where again this is really a set of equations for $a=1,\dots b_m$ and $m=0,\dots, k$.

To gauge the currents $J_{n+m}$ we need gauge field strengths $F_{n+m-1}$ in the $d$-dimensional theory such that $J_{n+m} = d F_{n+m-1}$ (here $F_{n+m-1}$ is the magnetic dual of the field strength in \eqref{JdsF}). The dimensional reduction of these Bianchi identities can be performed in analogy to what was previously done. In particular, one can replace $J_{n+m}\to F_{n+m-1}$ in the above analysis and repeat the same steps. One thus finds a lattice of $(n-1)$-form field strengths $f_{n-1}^{(m)a}$ in $D$ dimensions which are gauging the $n$-form currents $j_{n}^{(m)a}$, thus giving the Bianchi identities
\begin{equation}
j_n^{(m)a} = df_{n-1}^{(m)a}.
\end{equation}
In general, the currents $j_n^{(m)a}$ will also  contain  contributions from localised delta functions $\delta^{(n)}(\Pi_{D-n})$, arising from the reduction of D-branes in $d$ dimensions.

After computing cobordism and K-theory groups of $X$ in the next section, we will show that they exhibit exactly the same pattern explained here for the dimensional reduction of broken and gauged symmetries on $X$.
We will see that the description in terms of cobordism and K-theory provides by itself an organising principle for the various symmetries in the dimensionally reduced theory, something which is not transparent from the above analysis. 
Indeed, contributions to a given (broken or gauged) $p$-form symmetry in $D$-dimensions
and its corresponding charged objects will be encoded into $K^{-n}(X)$ and $\Omega^{\rm Spin^c}_{k+n}(X)$, for $p=D-1-n$ and $n \geq 0$. 
We will see that for $-k\leq n\leq 0$ the corresponding D-brane, respectively gravitational soliton, does not consistently fit into  the $D$-dimensional space so that  there does not exist any obvious physical interpretation of the cobordism and K-theory groups.
For type I, we have a similar story for  $\Omega^{\rm Spin}_{k+n}(X)$ and   $KO^{-n}(X)$.
This behavior under compactification gives further support to the interpretation of K-theory and cobordism groups as higher-form charges in an effective field theory.

As said, the above analysis was the classical dimensional reduction using (de Rham) singular (co)homology without torsion. 
Therefore, all objects in $D$-dimensions are the result of a naive dimensional reduction along homological cycles in $X$, nothing is lost and nothing new arises in $D$-dimensions. 
However, the appearance of torsion through the refinement to generalised (co)homology theories can open up new decay channels of non-BPS branes, and it is known that new stable torsion branes can appear on $X$, even if they were not present in $d$ dimensions.
Moreover, for wrapped D-branes there can be quantum effects that spoil these simple (classical) expectations.
For instance, some wrapped branes can develop a Freed-Witten anomaly so that they  should actually not be  present in the $D$-dimensional theory.
All these effects are taken into account by the description in terms of cobordism and K-theory rather than (co)homology.

\section{Computing cobordism and K-theory on $X$}
\label{sec_AHSS}

This section concerns the computation of cobordism and K-theory groups using a technique known as Atiyah-Hirzebruch spectral sequence. We first introduce spectral sequences for (generalised) homology and cohomology and then we use them to compute physically relevant cobordism and K-theory groups, respectively. This material is well-known to experts, but it might be less familiar to physicists not directly working on the subject. Thus, we believe that there is some pedagogical value in reviewing it. The reader interested in the physical interpretation might skip this section at first reading and go directly to section \ref{sec_interpretation}. We collect all results of this section in appendices \ref{sec_app_cobordism} and \ref{sec_app_Ktheory}.

\subsection{The Atiyah--Hirzebruch spectral sequence}

The Atiyah--Hirzebruch spectral sequence (AHSS) is a tool for calculating generalised (co)homology groups. 
We will use the homological AHSS to determine the cobordism groups $\Omega_n^{\xi}(X)$ and the cohomological AHSS for K-groups $K^{-n}(X)$, with $X$ a compact manifold of dimension up to ten. 
We will specialise to the choices of $X$ mentioned in the previous section, namely $X=\{S^k, T^k, K3, CY_3\}$, and to $\xi={\rm Spin}, {\rm Spin}^c$.
Standard references in the mathematical literature are for example \cite{Hatcher:478079,davis2012lecture,mccleary_2000}, for introductory material, and \cite{Husemoeller2008} for a physics-motivated treatment. 
A nice, recent review with applications to anomaly cancellation in physics can be found e.g.~in \cite{Garcia-Etxebarria:2018ajm}. 
With the goal of providing a self-contained exposition of the subject, in the present section we will briefly review the main steps of these techniques and some relevant mathematical results. 
Further background is presented in appendix \ref{sec_app_math}.

\subsubsection{Homological spectral sequence}

To get some intuition on the framework we will be working in, consider a fibration 
$F\to E\to B$, the three spaces $F$, $E$ and $B$ being fiber, total space and base respectively.\footnote{In the literature, the case with a trivial fibration is sometimes referred to as the AHSS exclusively. Here, following \cite{Hatcher:478079,davis2012lecture,mccleary_2000,Garcia-Etxebarria:2018ajm}, we discuss the more general case of a not necessarily trivial fibration and still refer to it as AHSS.} 
The goal is to compute the (generalised) homology of $E$, denoted $G_n(E)$, of which the cobordism groups $\Omega^\xi_n(E)$ are a particular case.

As a starting point, one typically needs some knowledge about the (generalised) homology of $F$ or $B$. 
Then, to compute $G_n(E)$, one  can run the Atiyah--Hirzebruch spectral sequence for homology, that is based on  a filtration of $G_n(E)$, i.e.~a sequence of subspaces $\ldots \subset F_p\subset F_{p+1} \subset \ldots$ whose union is $G_n(E)$.\footnote{In certain cases, it is also possible to use the spectral sequence ``backwards" and compute for example $G_n(F)$ from the $G_n(E)$.}
This can be interpreted as an approximate method becoming more and more accurate with each iterative step and stabilising after a finite number of steps.
However, this still does not give directly $G_n(E)$, rather it produces an associated graded module, ${\rm Gr}(G_n(E))$, which determines $G_n(E)$ up to an extension problem.  
In general this has to be solved on a case by case basis, as it requires additional information beyond the AHSS. 
Let us describe the whole method in more detail below.

By definition, a spectral sequence consists of a sequence of objects $E^r$, called pages, together with endomorphisms $d^r$, called differentials (since they square to zero), with $r$ non-negative integers. 
The pairs $(E^r,d^r)$ are such that the $(r+1)$-st page $E^{r+1}$ is given by the homology of the $r$-th page $E^r$,
\begin{equation}
    E^{r+1}\cong H(E^r) = \frac{\ker \,\, d^r\,:\, E^r\to E^r}{{\rm Im} \,\, d^r\,:\, E^r\to E^r}.
\end{equation}
The page $E^r$ together with the differential $d^r$ fully determine the next page $E^{r+1}$, but then additional input is necessary to determine the differentials $d^{r+1}$. 
Intuitively, the spectral sequence calculates a generalised (co)homology by first approximating it with ordinary (co)homology and then refining the approximation by acting with differentials.

In many circumstances, as it happens for the AHSS, there is more structure. 
Indeed, the pages can be bi-graded, i.e.~$E^r=\oplus_{p,q} E^r_{p,q}$ with $p,q\in \mathbb Z$, and the differential $d^r$ have a bi-degree $(-r,r+1)$, hence it maps between the bi-graded page elements as $d^r:E^r_{p,q}\to E^r_{p-r,q+r-1}$.

It is customary to have a pictorial representation of the pages and the relevant differentials, as in the figures \ref{table:AHSS_E2} and \ref{table:AHSS_E3} below, where we assume all entries outside the first quadrant to vanish.\footnote{This assumption is actually sufficient for the homological spectral sequence to terminate after a finite number of steps \cite{davis2012lecture}.} 
Conventionally, the horizontal axis refers to the $p$-value and the vertical to the $q$-value of an element of the $n^{th}$ page, $E^n_{p,q}$.  
As for the specific spectral sequences defined later on, the starting point is usually taken to be the second page, $E^2$.

\begin{table}[h]
  \centering
\scalebox{1}{
\begin{tabular}{ c | c c c c c c}

 $5$ & $\vdots$ &$\vdots$ &$\vdots$ &$\vdots$ &$\vdots$ & \\ 

  4 & $E^2_{0,4}$\tikzmark{4_0_r2} & \tikzmark{4_1_l2}$E^2_{1,4}$\tikzmark{4_1_r2} & \tikzmark{4_2_l2}$E^2_{2,4}$\tikzmark{4_2_r2}& \tikzmark{4_3_l2}$E^2_{3,4}$ &$E^2_{4,4}$& $\ldots$\\
 
 3 & $E^2_{0,3}$\tikzmark{3_0_r2} & \tikzmark{3_1_l2}$E^2_{1,3}$\tikzmark{3_1_r2} & \tikzmark{3_2_l2}$E^2_{2,3}$\tikzmark{3_2_r2} & \tikzmark{3_3_l2}$E^2_{3,3}$ &\tikzmark{3_4_l2}$E^2_{4,3}$& $\ldots$\\

 2 & $E^2_{0,2}$\tikzmark{2_0_r2} & \tikzmark{2_1_l2}$E^2_{1,2}$\tikzmark{2_1_r2} & \tikzmark{2_2_l2}$E^2_{2,2}$\tikzmark{2_2_r2} & \tikzmark{2_3_l2}$E^2_{3,2}$ &\tikzmark{2_4_l2}$E^2_{4,2}$& $\ldots$\\

 1 & $E^2_{0,1}$\tikzmark{1_0_r2} & \tikzmark{1_1_l2}$E^2_{1,1}$\tikzmark{1_1_r2} & \tikzmark{1_2_l2}$E^2_{2,1}$\tikzmark{1_2_r2}& \tikzmark{1_3_l2}$E^2_{3,1}$\tikzmark{1_3_r2}& \tikzmark{1_4_l2}$E^2_{4,1}$\tikzmark{1_4_r2}& $\ldots$\\

 0 & $E^2_{0,0}$\tikzmark{0_0_r2}& \tikzmark{0_1_l2}$E^2_{1,0}$ & \tikzmark{0_2_l2}$E^2_{2,0}$\tikzmark{0_2_r2}& \tikzmark{0_3_l2}$E^2_{3,0}$ & \tikzmark{0_4_l2}$E^2_{4,0}$& $\ldots$ \\ 
 \hline
& 0 & 1 &2 &3&4 &5
\end{tabular}
}
\scalebox{1}{
\begin{tikzpicture}[ overlay , remember picture, shorten >=2pt,shorten <=.5pt]
      \draw [thick,densely dashed,purple,<-] ({pic cs:1_0_r2})  to ({pic cs:0_2_l2});
      \draw [densely dashed,<-] ({pic cs:2_0_r2})  to ({pic cs:1_2_l2});
      \draw [thick,densely dashed, blue,<-] ({pic cs:3_0_r2})  to ({pic cs:2_2_l2});
      \draw [densely dashed,<-] ({pic cs:4_0_r2})  to ({pic cs:3_2_l2});
      \draw [thick,densely dashed,blue,<-] ({pic cs:1_1_r2})  to ({pic cs:0_3_l2});
      \draw [thick,densely dashed,blue,<-] ({pic cs:2_1_r2})  to ({pic cs:1_3_l2});
      \draw [densely dashed,<-] ({pic cs:3_1_r2})  to ({pic cs:2_3_l2});
      \draw [densely dashed,<-] ({pic cs:4_1_r2})  to ({pic cs:3_3_l2});
      \draw [thick,densely dashed,purple,<-] ({pic cs:1_2_r2})  to ({pic cs:0_4_l2});
      \draw [densely dashed,<-] ({pic cs:2_2_r2})  to ({pic cs:1_4_l2});
      \draw [densely dashed,<-] ({pic cs:3_2_r2})  to ({pic cs:2_4_l2});
      \draw [thick,densely dashed,blue,<-] ({pic cs:4_2_r2})  to ({pic cs:3_4_l2});
  \end{tikzpicture}
}
\captionof{figure}{Example of a second page $E^2$ of a first quadrant homological spectral sequence and all possible $d^2$ differentials. The non-vanishing $d^2$ are shown by purple and blue arrows.} \label{table:AHSS_E2}

\end{table}

\begin{table}[h]
  \centering
\scalebox{1}{
\begin{tabular}{ c | c c c c c c}

 5 & $\vdots$ &$\vdots$ &$\vdots$ &$\vdots$ &$\vdots$ & \\ 
 
 4 & $E^3_{0,4}$\tikzmark{4_0_r} & \tikzmark{4_1_l}$E^3_{1,4}$\tikzmark{4_1_r} & \tikzmark{4_2_l}{\color{blue} 0}\tikzmark{4_2_r}& \tikzmark{4_3_l}$E^3_{3,4}$ &$E^3_{4,4}$& $\ldots$\\ 
 
 3 & {\color{blue} 0}\tikzmark{3_0_r} & \tikzmark{3_1_l}$E^3_{1,3}$\tikzmark{3_1_r} & \tikzmark{3_2_l}$E^3_{2,3}$\tikzmark{3_2_r} & \tikzmark{3_3_l}$E^3_{3,3}$ &\tikzmark{3_4_l}{\color{blue} 0}& $\ldots$\\

 2 & $E^3_{0,2}$\tikzmark{2_0_r} & \tikzmark{2_1_l}{\color{blue} 0}\tikzmark{2_1_r} & \tikzmark{2_2_l}{\color{blue} 0}\tikzmark{2_2_r} & \tikzmark{2_3_l}$E^3_{3,2}$ &\tikzmark{2_4_l}$E^3_{4,2}$& $\ldots$\\

 1 & {\color{purple} $E^3_{0,1}$}\tikzmark{1_0_r} & \tikzmark{1_1_l}{\color{blue} 0}\tikzmark{1_1_r} & \tikzmark{1_2_l}{\color{purple}$E^3_{2,1}$}\tikzmark{1_2_r}& \tikzmark{1_3_l}{\color{blue} 0}\tikzmark{1_3_r}& \tikzmark{1_4_l}$E^3_{4,1}$\tikzmark{1_4_r}& $\ldots$\\

 0 & $E^3_{0,0}$\tikzmark{0_0_r}& \tikzmark{0_1_l}$E^3_{1,0}$ & \tikzmark{0_2_l}{\color{purple}$E^3_{2,0}$}\tikzmark{0_2_r}& \tikzmark{0_3_l}{\color{blue} 0} & \tikzmark{0_4_l}{\color{purple}$E^3_{4,0}$}& $\ldots$ \\ 
 \hline
& 0 & 1 &2 &3&4 &5
\end{tabular}
}
\scalebox{1}{
\begin{tikzpicture}[ overlay , remember picture, shorten >=2pt,shorten <=.5pt
]
   \draw [densely dashed,<-] ({pic cs:4_0_r})  to ({pic cs:2_3_l});
   \draw [densely dashed,<-] ({pic cs:4_1_r})  to ({pic cs:2_4_l});
  \end{tikzpicture}
  }
\captionof{figure}{Third page $E^3$ of the same spectral sequence and all possible $d^3$ differentials. The blue differentials have (co-)killed the page elements they were acting on, while the purple ones let them partially survive. The black elements, on which no differential acted, carried over intact to the next page, i.e.~$E_{p,q}^3\cong E_{p,q}^2$.} \label{table:AHSS_E3}
\end{table}

The procedure of acting on a page with the differential leading to the next page is often referred to as ``turning the page". 
The only elements that might be different\footnote{Different in this case means either vanishing, or possibly a subgroup of the original page element.} once we turn the page are those that non-vanishing differentials act on, while the rest carries over intact to the next page.
For a sequence confined to the first quadrant, such as a homological spectral sequence, after a finite number of iterations no non-trivial differential can act anymore. 
There, the sequence stabilises and we reach the so-called $E^\infty$-page.

$E^{\infty}$ is related to the desired (generalised) homology theory. 
In particular, one computes the generalised homology groups $G_{n}(E)$ using all diagonal elements of $E^{\infty}_{p,q}$, with $p+q=n$.
One says that the spectral sequence converges to $G_n(E)$ and writes $E^2_{p,q}\Rightarrow G_{p+q}$.
In the simplest case there is just one element on the diagonal of $E^\infty$, so a direct identification is possible, but usually one has to deal with a non-trivial extension problem, especially when torsion is present. 
In the general case, one has  ${\rm Gr}(G_n(E)) \cong \bigoplus_{p=0}^n E^\infty_{p,n-p}$ and to obtain $G_n(E)$ one needs extra information.
We will discuss specific examples and possible ways around the extension problem in the following sections \ref{sec_AHSS_cob} and \ref{sec_AHSS_K}.

Having explained the general idea of a spectral sequence, let us now apply it to our initial  problem of computing (generalised) homologies of the total space  $E$ for a fibration $F\to E \to B$.
Depending on what kind of structure one has, one can distinguish three types of spectral sequences.
First, let $M$ be an abelian group 
and  $B$ path-connected. 
The \emph{homological Serre spectral sequence} is  a first quadrant spectral sequence defined as
\begin{equation}
E^2_{p,q}\cong H_p(B; H_q(F;M))\Rightarrow H_{p+q}(E;M)\,.
\label{2nd_page_hom_Serre}
\end{equation}
Second, for $R$ a ring and  $B$ simply connected, we have the \emph{Leray-Serre spectral sequence}
\begin{equation}
E^2_{p,q}\cong H_p(B;R\otimes  H_q(F;R))\Rightarrow H_{p+q}(E;R)\,.
\label{2nd_page_hom_LS}
\end{equation}
Finally, the \emph{Leray--Serre--Atiyah--Hirzebruch spectral sequence}, or simply \emph{Atiyah--Hirzebruch Spectral Sequence} is defined
for  an  additive homology theory $G_*$  and a path-connected $B$
\begin{equation}
E^2_{p,q}\cong H_p(B;G_q(F))\Rightarrow G_{p+q}(E)\,.
\label{2nd_page_hom_AHSS}
\end{equation}
Note that $H_p(B;G_q(F))=0$ for $p<0$. This spectral sequence can be used for the computation of cobordism groups.

\subsubsection{Cohomological Spectral Sequence}

As mentioned, for the computation of the  K-theory groups $K^{-n}(X)$ we will employ the cohomological version of the AHSS.
Indeed, analogously to the discussion in the previous section one   constructs spectral sequences to compute generalised cohomology groups.
One starts with a fibration fulfilling certain requirements and uses knowledge over cohomological groups of some space (such as fiber or base) to deduce what the generalised cohomology of the desired space is (such as the total space). Once again, we have a collection of objects $(E_r,d_r)$, where now the bi-grading of the differential is $(r, -r +1 )$, i.e. $d_r: E_r^{p,q}\to E_r^{p+r,q-r+1}$ and the $(r+1)$-st page $E_{r+1}$ given by the cohomology of the $E_r$ page. In the pictorial representation, 
the pages of a cohomological spectral sequence look very similar to those of a homological one, with the exception that the differential arrows now point in the opposite direction. Another difference is that now the sequence possesses a cup product structure which may allow for a formal computation of the differentials.

The \emph{cohomological Serre spectral sequence} is defined  
similarly to the homological one. 
For the usual fibration $F\to E \to B$, with $B$ path-connected and $R$ a ring, there is a first quadrant cohomological spectral sequence of algebras,  converging (as a graded algebra) as
\begin{equation}
E_2^{p,q}=H^p(B; H^q(F;R))\Rightarrow H^{p+q}(E;R)\,.
\label{2nd_page_cohom}
\end{equation}
If $\pi_1(B)=0$ and $R$ a field, the previous equation simplifies to
\begin{equation}
E_2^{p,q}=H^p(B)\otimes H^q(F;R)\Rightarrow H^{p+q}(E;R)\,.
\label{2nd_page_cohom_2}
\end{equation}
Since K-theory is a generalised cohomology theory, a generalisation of the Serre spectral sequence is necessary. This is the \emph{Atiyah--Hirzebruch spectral sequence}, defined now for $G^*$ a generalised cohomology theory and the fibration as in \ref{2nd_page_cohom}. 
Namely, there is a half-plane cohomological spectral sequence 
\begin{equation}
E_2^{p,q}=H^p(B;G^q(F))\Rightarrow G^{p+q}(E)\,.
\label{2nd_page_cohom_AHSS}
\end{equation}

\subsubsection{Trivial fibration and vanishing differentials on the edge}
\label{sec:edgehom}

Besides the extension problem, computing the differentials in a spectral sequence can also be a tedious task.  However, there are instances where one can generally show that they vanish. This is the case for differentials from/to the edge of a given page, when the AHSS involves particularly simple fibrations.

Consider the trivial fibration 
\begin{equation}
\label{eq_ses1}
{\rm pt} \xhookrightarrow{} X \overset{\rm id}{\to} X. 
\end{equation}
The inclusion ${\rm pt} \xhookrightarrow{} X$ is split by the constant map $X \to {\rm pt}$, implying that
\begin{equation}
\label{edgehom}
G_n ({\rm pt}) \to G_n(X)
\end{equation}
is a split injection ($G_*$ being a generalised homology theory).
On the other hand, this is also a special case of a map known as the edge homomorphism.
Indeed, consider the fibration $F \to E \to B$, which generalises \eqref{eq_ses1}. An edge homomorphism is defined as
\begin{equation}
\label{edgehomgen}
G_n(F)\to H_0 (B;G_n(F)) = E^2_{0,n} \to E^\infty_{0,n} \to G_n(E),
\end{equation}
where the last arrow is an injection while the others are surjections. 
As stated e.g.~in Theorem 9.10 of  \cite{davis2012lecture}, this is equal to the map
\begin{equation}
\label{edgehom2}
G_n(F) \to G_n(E) ,
\end{equation}
induced by the inclusion $F \xhookrightarrow{} E$. For $F={\rm pt}$ and $B=E=X$, one should recover the split injection \eqref{edgehom} and thus
\begin{equation}
E^2_{0,n} \cong E^\infty_{0,n}.
\end{equation}
In other words, in this case the entries survive to the final page and any differential acting on them,
\begin{equation}
d^r: E^{r}_{r,q} \to E^{r}_{0,q+r-1},
\end{equation}
has to be zero. This observation greatly simplifies the calculation of the related spectral sequences and will have a direct application in the upcoming  computation of cobordism groups.

\subsection{Application to cobordism}
\label{sec_AHSS_cob}

In this section, we employ the homological version of the AHSS to compute cobordism groups $\Omega^\xi_n(X)$ for non-trivial $k$-dimensional spaces $X$. 
Considering the trivial fibration\footnote{The choice of the trivial fibration allows us to avoid a complication we have not discussed yet. Notice that we assumed $B$ to be path connected, but in general not simply connected. When $\pi_1(B)\neq 0$, one deals with a system of local coefficients over $B$ with fiber $G_q(F)$ \cite{davis2012lecture}. As a consequence, in \eqref{2nd_page_hom_AHSS} one has to consider ordinary homology with local coefficients. However, if the fibration is trivial this complication can be ignored \cite{davis2012lecture}.} $\,{\rm pt} \to X \to X$, the AHSS allows us to determine $\Omega^\xi_n(X)$ from the known cobordism groups of the point given in table \ref{SpinSpincptgroups}.
Then, the second page of the AHSS is given by
\begin{equation}
    E^2_{p,q}=H_p(X; \Omega_q^\xi )\,.
    \label{2nd_page_cob}
\end{equation}
To avoid cluttering the expressions, in the remainder of this section we use the shorthand notation $\Omega^\xi_n({\rm pt})\equiv \Omega^\xi_n$.
Note that  we will only show the parts of the pages with $p,q \leq 10$, as  this is sufficient to study the manifolds of interest for physical applications.

\subsubsection{Computing  $\Omega^\xi_n(S^k)$}

Before passing to higher-dimensional spheres,
we start with the straightforward, yet illustrative, computation of $\Omega^\xi_n(S^2)$.  We present here the case where $\xi =\rm{Spin}$, while the similarly computed results for $\xi=\rm{Spin}^c$ are relegated to the appendix \ref{sec_app_cobordism}.

While a direct computation of $H_p(S^2, \Omega_q^{\rm Spin})$ is straightforward for low $q$, in general one turns to the universal coefficient theorem (see appendix \ref{sec_app_UCT}), according to which there is a short exact sequence
\begin{equation}
0 \to H_n(S^2; \mathbb{Z}) \otimes \Omega_q^{\rm Spin} \to H_n(S^2; \Omega_q^{\rm Spin} )\to {\rm Tor}_1 (H_{n-1}(S^2;\mathbb{Z}),\Omega^{\rm Spin}_q) \to 0.
\end{equation}
Recalling the well known homology groups
\begin{equation}
H_n(S^2; \mathbb{Z}) =\left\{ \begin{array}{cl} \mathbb{Z} & \text{for $n=0,2$},\\
0 & \text{otherwise} \end{array}\right.
\end{equation}
and the fact that  $\mathbb{Z}$ is torsion-free, 
\eqref{2nd_page_cob} can be directly evaluated as
\begin{equation}
E^2_{p,q} = H_p(S^2; \Omega_q^{\rm Spin} )\cong H_p(S^2; \mathbb{Z}) \otimes \Omega_q^{\rm Spin} =\left\{ \begin{array}{cl} \Omega_q^{\rm Spin} & \text{for $p=0,2$},\\
0 & \text{otherwise} .\end{array}\right. 
\end{equation}
Hence, the second page of the AHSS takes the following form.
\begin{table}[h!]
\begin{center}
\begin{tabular}{ c| c c c c c c}
 10 & $\Omega^{\rm Spin}_{10}$&0   & $\Omega^{\rm Spin}_{10}$  & 0 & 0 & 0 \\
 9 & $\Omega^{\rm Spin}_9$&0   & $\Omega^{\rm Spin}_9$  & 0 & 0 & 0 \\
 8 & $\Omega^{\rm Spin}_8$&0   & $\Omega^{\rm Spin}_8$  & 0 & 0 & 0 \\
 7 & $\Omega^{\rm Spin}_7$ &0& $\Omega^{\rm Spin}_7$   & 0 & 0 & 0 \\
 6 & $\Omega^{\rm Spin}_6$&0 & $\Omega^{\rm Spin}_6$   & 0 & 0 & 0 \\
 5 &$\Omega^{\rm Spin}_5$ &0& $\Omega^{\rm Spin}_5$   & 0 & 0 & 0 \\
 4 &$\Omega^{\rm Spin}_4$&0 &     $ \Omega^{\rm Spin}_4$  & 0 & 0 & 0  \\
 3 & $\Omega^{\rm Spin}_3$ &0& $\Omega^{\rm Spin}_3$  & 0 & 0 & 0 \\
 2 &  $\Omega^{\rm Spin}_2$ &0  &    $\Omega^{\rm Spin}_2$  & 0 & 0 & 0\\
 1 &  $\Omega^{\rm Spin}_1$ &0  &    $\Omega^{\rm Spin}_1$  & 0 & 0 & 0 \\
0 &  $\Omega^{\rm Spin}_0$&0 &     $\Omega^{\rm Spin}_0$ & 0 & 0 & 0 \\
\hline
&0 &1 &2 &3 &4 &5
\end{tabular}
\quad
=
\begin{tabular}{ c | c c c c c c}
 10 & 3$ \mathbb{Z}_2$\tikzmark{b1}&0  & 3$ \mathbb{Z}_2$ & 0 & 0 &0\\
 9 & 2$ \mathbb{Z}_2$\tikzmark{a1}&0  &\tikzmark{b2}2$ \mathbb{Z}_2$ & 0 & 0 &0\\
 8 & 2$ \mathbb{Z}$  &0 & \tikzmark{a2}2$ \mathbb{Z}$ & 0 & 0&0 \\
 7 & 0 & 0  &    0 & 0 & 0 & 0 \\
 6 & 0 & 0  &    0 & 0 & 0 & 0 \\
 5 &0 & 0  &    0 & 0 & 0 & 0 \\
 4 &$\mathbb{Z}$ & 0  &    $\mathbb{Z}$ & 0 & 0 & 0 \\
 3 & 0 & 0  &    0 & 0 & 0 & 0 \\
 2 &  $\mathbb{Z}_2$\tikzmark{c}& 0  &    $\mathbb{Z}_2$ & 0 & 0 & 0 \\
 1 &  $\mathbb{Z}_2$\tikzmark{a}& 0  &    \tikzmark{d}$\mathbb{Z}_2$ & 0 & 0 & 0 \\
0 &  $\mathbb{Z}$ & 0  &  \tikzmark{b}$\mathbb{Z}$  & 0 & 0 & 0 \\
\hline
& 0 & 1 &2 &3&4&5
\begin{tikzpicture}[ overlay , remember picture, shorten >=2pt,shorten <=.5pt]
    \draw [<-] ({pic cs:a})  to ({pic cs:b});
      \draw [<-] ({pic cs:c})  to ({pic cs:d});
      \draw [<-] ({pic cs:a1})  to ({pic cs:a2});
         \draw [<-] ({pic cs:b1})  to ({pic cs:b2});
\end{tikzpicture}
  \end{tabular}
\end{center}
\captionof{figure}{Second (and final) page of AHSS for $\Omega_n^{\rm Spin}(S^2)$.} 
\end{table}

We see that there exist four differentials that could kill some of the page entries.
However, they all end on the first column of the page and thus they vanish according to the edge homomorphism reviewed in section  \ref{sec:edgehom}. 
Thus, one can immediately conclude that $E^2_{p,q}\cong E^3_{p,q}$. 
From the third page, no differentials can act on the page elements, as its degree would be larger than any possible difference of degree between non-zero elements of the page. 
Therefore, $E_{p,q}^2 \cong E_{p,q}^\infty$ and we arrive at the results in table \ref{tableOmegaSpinS2}.\footnote{As explained in appendix \ref{sec_app_ses}, we denote by $e(A,B)$ the extension of $A$ by $B$. 
The opposite convention is also used in the literature, e.g.~in\cite{Garcia-Etxebarria:2018ajm}.}
\begin{table}[h!]
\begin{center}
\resizebox{\textwidth}{!}{
\begin{tabular}{ c | c c c c c c c c c c c }
n & 0 & 1 & 2 &  3 & 4 & 5 & 6 & 7 & 8 & 9 & 10\\
\midrule
 $\Omega^{\rm Spin}_n (S^2)$ &$\mathbb{Z}$ & $\mathbb{Z}_2$  & $e(\mathbb{Z},\mathbb{Z}_2)$  & $\mathbb{Z}_2$ & $e(\mathbb{Z}_2,\mathbb{Z})$ & 0& $\mathbb{Z}$  &0 & 2$\mathbb{Z}$ & $2\mathbb{Z}_2$ & $e($2$\mathbb{Z}, $3$\mathbb{Z}_2)$\\[-0.3cm]
\end{tabular}
}
\end{center}
\captionof{table}{Cobordism groups $\Omega^{\rm Spin}_n (S^2)$, $n=0,\ldots,10$, up to extensions. }
\label{tableOmegaSpinS2}
\end{table}

Let us now tackle the extension problems one by one. Our main tools are briefly reviewed in appendix \ref{sec_app_ses}.

\begin{itemize}
\item $e(\mathbb{Z},\mathbb{Z}_2)$: We have ${\rm Ext}^1(\mathbb{Z},\mathbb{Z}_2 ) = 0$ and thus there is only the trivial extension, $e(\mathbb{Z},\mathbb{Z}_2)= \mathbb{Z} \oplus \mathbb{Z}_2$.
\item $e(\mathbb{Z}_2,\mathbb{Z})$: We have from \eqref{Extprop3} that ${\rm Ext}^1(\mathbb{Z}_2,\mathbb{Z}) =\mathbb{Z}_2 $. The two possible extensions are $\mathbb{Z}$ and $\mathbb{Z}_2\oplus \mathbb{Z}$, so we need some additional input to select the appropriate one. One simple strategy would be to use the splitting lemma \eqref{SplittLemmaCob}, which tells us that $\Omega_4^{\rm Spin}(S^2)$ should contain a factor $\Omega_4^{\rm Spin}=\mathbb{Z}$. Unfortunately such factor is present in both extension options, so we cannot draw any conclusion. In appendix \ref{sec:altstrat}, we show (indirectly) that for $\Omega_n^\xi(S^k)$ the extension is always trivial, therefore even in this case $e(\mathbb{Z}_2,\mathbb{Z})=\mathbb{Z}\oplus\mathbb{Z}_2$.
\item $e(2$$\mathbb{Z},3$$\mathbb{Z}_2)$: We have ${\rm Ext}^1(2$$\mathbb{Z},3$$\mathbb{Z}_2)=2{\rm Ext}^1(\mathbb{Z},3$$\mathbb{Z}_2)=5{\rm Ext}^1(\mathbb{Z},\mathbb{Z}_2)=0$, so the trivial extension must be chosen, in accordance with the general proof of appendix \ref{sec:altstrat}.
\end{itemize}

\noindent
We summarize our findings in the following table.
\begin{table}[h!]
\begin{center}
\begin{tabular}{ c | c c c c c c c c c c c }
n & 0 & 1 & 2 &  3 & 4 & 5 & 6 & 7 & 8 & 9 & 10\\
\midrule
 $\Omega^{\rm Spin}_n (S^2)$ &$\mathbb{Z}$ & $\mathbb{Z}_2$  & $\mathbb{Z}\oplus \mathbb{Z}_2$  & $\mathbb{Z}_2$ & $\mathbb{Z}_2\oplus\mathbb{Z}$ & 0& $\mathbb{Z}$  &0 & 2$\mathbb{Z}$ & 2$\mathbb{Z}_2$ & 2$\mathbb{Z}\oplus 3\mathbb{Z}_2$\\[-0.3cm]
\end{tabular}
\end{center}
\captionof{table}{Cobordism groups $\Omega_n^{\rm Spin}(S^2)$.} 
\end{table}

The calculation of $\Omega_n^{\rm Spin}(S^k)$ for higher $k$ proceeds similarly. 
Since the only non-vanishing homology classes are $H_0(S^k;\mathbb{Z})=H_k(S^k;\mathbb{Z})=\mathbb{Z}$ and the universal coefficient theorem applies, the second page for the trivial fibration ${\rm pt}\to S^k \to S^k$ looks very similar to the one for $S^2$, with the non-vanishing entries along the $p=0,k$ columns. 
The only possibly non-vanishing differentials are $d_k$, but since they end on the first column they vanish due to the edge homomorphism. 
Hence, the computation proceeds exactly as before.
For $S^1$ the computation is even simpler, since for degree reasons no differential can act. 
As explained at the beginning of the present section, the fact that $\pi_1(S^1)\neq 0$ does not concern us since we are using a trivial fibration.

For the  computation of the Spin$^c$ cobordism groups  $\Omega_n^{{\rm Spin}^c}(S^k)$ one follows similar steps. 
Now the second page is
\begin{equation}
E^2_{p,q} = H_p(S^k; \Omega_q^{{\rm Spin}^c} )\cong H_p(S^k; \mathbb{Z}) \otimes \Omega_q^{{\rm Spin}^c} =\left\{ \begin{array}{cl} \Omega_q^{{\rm Spin}^c} & \text{for $p=0,k$},\\
0 & \text{otherwise},\end{array}\right. 
\end{equation}
and the same arguments as for the $\Omega^{\rm Spin}_n(S^k)$ computation still go through. 
As proven in appendix \ref{sec:altstrat}, for both structures $\xi={\rm Spin},{\rm Spin}^c$ the final result
can be compactly written as
\begin{equation}
\label{cob_final_sphere}
\Omega_n^{\xi}(S^k) =\Omega_n^{\xi}({\rm pt}) \oplus \Omega^{\xi}_{n-k}({\rm pt})\,.
\end{equation}
Explicitly, the groups for $n,k \geq 10$ are given in the appendix \ref{sec_app_cobordism}.

\subsubsection{Computing $\Omega^{\xi}_n(T^2)$}

For the two-torus, $T^2=S^1\times S^1$, we present the computation for both $\xi = {\rm Spin}$ and  ${\xi =\rm Spin}^c$ in parallel. 
Starting from the known homology groups (recall the Betti numbers of the torus $b_0=b_2=1$, $b_1=2$)
\begin{equation}
H_n(T^2; \mathbb{Z}) =\left\{\begin{array}{cl} \mathbb{Z} & \text{for}\,\, n=0,2,\\
2\mathbb{Z} & \text{for}\,\, n=1,\\
0& \text{otherwise},\end{array}\right.
\end{equation}
and using the universal coefficient theorem again (with vanishing Tor$_1$ group),
one can compute the second page 
\begin{equation}
E_{p,q}^2 = H_p(T^2; \Omega_q^{\xi}) \cong H_p(T^2; \mathbb{Z}) \otimes 
\Omega_q^\xi =\left\{\begin{array}{cl}\Omega_q^{\xi} & \text{for}\,\, p=0,2,\\
2\Omega_q^{\xi} & \text{for}\,\, p=1,\\
0& \text{otherwise}\,.\end{array}\right.
\end{equation}

The second pages for the two structures $\xi= {\rm Spin}, {\rm Spin}^c$ are shown in figure \ref{table:AHSS_E2T2}. 
\begin{table}[h!]
\begin{center}
\begin{tabular}{ c | c c c c c c  }
 10 & $3 \mathbb{Z}_2$\tikzmark{B1}   & $6 \mathbb{Z}_2$ & $3 \mathbb{Z}_2$ & 0 & 0 & 0 \\
 9 & $2 \mathbb{Z}_2$\tikzmark{A1}  & $4 \mathbb{Z}_2$ & \tikzmark{B2}$2 \mathbb{Z}_2$ & 0 & 0 & 0\\
 8 & $2 \mathbb{Z}$   & $4 \mathbb{Z}$ &  \tikzmark{A2}$2 \mathbb{Z}$ & 0 & 0 & 0 \\
 7 & 0 & 0  &    0 & 0 & 0 & 0 \\
 6 & 0 & 0  &    0 & 0 & 0 & 0 \\
 5 &0 & 0  &    0 & 0 & 0 & 0 \\
 4 &$\mathbb{Z}$ &   2$\mathbb{Z}$ &  $\mathbb{Z}$  & 0 & 0 & 0 \\
 3 & 0 & 0  &    0 & 0 & 0 & 0 \\
 2 &  $\mathbb{Z}_2 \tikzmark{C}$   &    2$\mathbb{Z}_2$ & $\mathbb{Z}_2$ & 0 & 0 & 0 \\
 1 &  $\mathbb{Z}_2  \tikzmark{A}$   &    2$\mathbb{Z}_2$ & \tikzmark{D}$\mathbb{Z}_2$ & 0 & 0 & 0 \\
0 &  $\mathbb{Z}$ &    2$\mathbb{Z}$&  \tikzmark{B}$\mathbb{Z}$ & 0 & 0 & 0  \\
\hline
& 0 & 1 &2 &3&4&5
\end{tabular}
\quad 
\begin{tabular}{ c | c c c c c   }
 10 & $4 \mathbb{Z} \oplus \mathbb{Z}_2$   &$8 \mathbb{Z} \oplus 2\mathbb{Z}_2$& $4 \mathbb{Z} \oplus \mathbb{Z}_2$ &0& 0 \\
 9 & 0 & 0 & 0 & 0 & 0 \\
 8 & $4 \mathbb{Z}$   & $8 \mathbb{Z}$ &  $4 \mathbb{Z}$ & 0  & 0 \\
 7 & 0 & 0  &    0 & 0 & 0  \\
 6 & $2\mathbb{Z}$  &  $4 \mathbb{Z}$  &     $2 \mathbb{Z}$  & 0 & 0 \\
 5 &0 & 0  &    0 & 0 & 0  \\
 4 &$2\mathbb{Z}$ &   4$\mathbb{Z}$ &  $2\mathbb{Z}$  & 0  & 0 \\
 3 & 0 & 0  &  0& 0& 0 \\
 2 &  $\mathbb{Z}$   &    2$\mathbb{Z}$ & $\mathbb{Z}$ & 0  & 0\\
 1 & 0  &    0 & 0 & 0 & 0 \\
0 &  $\mathbb{Z}$ &    2$\mathbb{Z}$&   $\mathbb{Z}$ & 0  & 0 \\
\hline
& 0 & 1 &2 &3&4
\end{tabular}
\begin{tikzpicture}[ overlay , remember picture, shorten >=2pt,shorten <=.5pt
]
    \draw [<-] ({pic cs:A})  to ({pic cs:B});
      \draw [<-] ({pic cs:C})  to ({pic cs:D});
      \draw [<-] ({pic cs:A1})  to ({pic cs:A2});
         \draw [<-] ({pic cs:B1})  to ({pic cs:B2});
  \end{tikzpicture}
\end{center}
\captionof{figure}{Second (and final) pages of AHSS for $\Omega_n^{\rm Spin}(T^2)$ (left) and $\Omega_n^{{\rm Spin}^c}(T^2)$ (right).}
\vspace{0.5cm}
\label{table:AHSS_E2T2}
\end{table}
For the Spin case we have four differentials which could be non-trivial, but they vanish due to the edge homomorphism for the trivial fibration. For the Spin$^c$ case, no differential can act for degree reasons. Hence, the second pages above are in fact the final pages and we have the results displayed in table \ref{OmegaspinT2ext}, where we used the notation $e(A,B,C)=e(A,e(B,C))$.

\begin{table}[h!]
\begin{center}
\hspace{-1.8cm}\begin{tabular}{ c | c c c c c }
n & 0 & 1 & 2 &  3 & 4  \\
\midrule
 $\Omega^{\rm Spin}_n (T^2)$ &$\mathbb{Z}$ & $e(2\mathbb{Z},\mathbb{Z}_2)$  & $e(\mathbb{Z},2\mathbb{Z}_2,\mathbb{Z}_2)$  & $e(\mathbb{Z}_2,2\mathbb{Z}_2)$ & $e(\mathbb{Z}_2,\mathbb{Z})$ \\
 $\Omega^{\rm Spin^c}_n (T^2)$ &$\mathbb{Z}$ & $2\mathbb{Z}$  & $e(\mathbb{Z},\mathbb{Z})$ & $2\mathbb{Z}$ & $e(\mathbb{Z},2\mathbb{Z})$ 
\end{tabular}

\vspace{0.5cm}
\begin{tabular}{ c| c c c c c c }
n & 5 & 6 & 7 & 8 & 9 & 10\\
\midrule
 $\Omega^{\rm Spin}_n (T^2)$ & $2\mathbb{Z}$ & $\mathbb{Z}$ &0 & 2$\mathbb{Z}$ & $e(4\mathbb{Z},2\mathbb{Z}_2)$ & $ e(2\mathbb{Z},4\mathbb{Z}_2,3\mathbb{Z}_2)$\\
 $\Omega^{\rm Spin^c}_n (T^2)$  & $4\mathbb{Z}$ & $e(2\mathbb{Z},2\mathbb{Z})$ &$4\mathbb{Z}$ & $e(2\mathbb{Z},4\mathbb{Z})$ &$8\mathbb{Z}$ &$e(4\mathbb{Z},4\mathbb{Z} \oplus \mathbb{Z}_2)$\\[-0.0cm]
\end{tabular}
\end{center}
\caption{Cobordism groups $\Omega^{\rm Spin}_n (T^2)$ and  $\Omega^{\rm Spin^c}_n (T^2)$, $n=0,\ldots,10$, up to extensions.}
\label{OmegaspinT2ext}
\end{table}

Two facts are crucial to solve the extension problem for these cobordism groups. First, the extensions of all free abelian groups are trivial. Second, $e(m\mathbb{Z}, n \mathbb{Z}_k) = m\mathbb{Z}\oplus n\mathbb{Z}_k$ since ${\rm Ext}^1(m\mathbb{Z}, n \mathbb{Z}_k) = 0$. However, since ${\rm Ext}^1(\mathbb{Z}_2,\mathbb{Z}_2)=\mathbb{Z}_2$, we cannot conclude anything about $e(\mathbb{Z}_2,\mathbb{Z}_2)$, which is either $2\mathbb{Z}_2$ or $\mathbb{Z}_4$.
A similar story applies for  $e(\mathbb{Z}_2,\mathbb{Z})$. 
Up to this point, our results are  shown in table \ref{OmegaspinT2}.
\begin{table}[h!]
\begin{center}
\begin{tabular}{ c | c c c c c }
n & 0 & 1 & 2 &  3 & 4  \\
\midrule
 $\Omega^{\rm Spin}_n (T^2)$ &$\mathbb{Z}$ & $2\mathbb{Z}\oplus\mathbb{Z}_2$  & $e(\mathbb{Z},2\mathbb{Z}_2,\mathbb{Z}_2)$  & $e(\mathbb{Z}_2,2\mathbb{Z}_2)$ & $e(\mathbb{Z}_2,\mathbb{Z})$ \\
 $\Omega^{\rm Spin^c}_n (T^2)$ &$\mathbb{Z}$ & $2\mathbb{Z}$  & $2\mathbb{Z}$ & $2\mathbb{Z}$ & $3\mathbb{Z}$\\
\end{tabular}

\vspace{0.5cm}
\hspace{-0.5cm}\begin{tabular}{ c | c c c c c c }
n & 5 & 6 & 7 & 8 & 9 & 10\\
\midrule
 $\Omega^{\rm Spin}_n (T^2)$ & $2\mathbb{Z}$ & $\mathbb{Z}$ &0 & 2$\mathbb{Z}$ & $4\mathbb{Z}\oplus 2\mathbb{Z}_2$ & $ e(2\mathbb{Z},4\mathbb{Z}_2,3\mathbb{Z}_2)$\\
 $\Omega^{\rm Spin^c}_n (T^2)$  & $4\mathbb{Z}$ & $4\mathbb{Z}$ &$4\mathbb{Z}$ & $6\mathbb{Z}$ &$8\mathbb{Z}$ &$8\mathbb{Z} \oplus \mathbb{Z}_2$\\[-0.5cm]
\end{tabular}
\end{center}
\caption{Cobordism groups $\Omega^{\rm Spin}_n (T^2)$ and  $\Omega^{\rm Spin^c}_n (T^2)$, $n=0,\ldots,10$.}
\label{OmegaspinT2}
\end{table}
According to the general proof given in appendix \ref{sec:altstrat}, the remaining extension problems should be trivial.
Indeed, there we generically show that the cobordism groups of $k$-dimensional tori have a simple decomposition,
\begin{equation}
  \Omega_n^{\xi}(T^k) = \bigoplus_{m = 0}^{k} \binom{k}{m} \,\Omega_{n-m}^{\xi}({\rm pt}),
\end{equation}
for a generic structure $\xi$, which can be taken to be Spin or Spin$^c$.
The binomial coefficient can be interpreted as the number of $m$-cycles on $T^k$.
Explicit  results with all extensions solved are reported in appendix \ref{sec_app_cobordism}.

\subsubsection{Computing $\Omega_n^{\rm Spin^c}(K3)$}

For the determination of the cobordism groups of $K3$ we again start with the known result for $H_n(K3; \mathbb{Z})$.
\begin{equation}
H_n(K3; \mathbb{Z}) =\left\{\begin{array}{cl} \mathbb{Z} & \text{for}\,\, n=0,4,\\
22\mathbb{Z} & \text{for}\,\, n=2,\\
0& \text{otherwise},\end{array}\right.
\end{equation}
where the non-vanishing Betti numbers of $K3$ are $b_0=b_4=1$, $b_2=22$.
Once again using the trivial fibration and the universal coefficient theorem we compute the second page entries shown in figure \ref{fig:cobordk3}.
For ${\rm Spin}^c$ all differentials are trivial for degree reason, so
that we can conclude $E_{p,q}^{2} = E_{p,q}^{\infty}$ with
\begin{equation}
E^2_{p,q}= H_p(K3; \Omega_q^{\rm Spin^c} )\cong H_p(K3; \mathbb{Z}) \otimes \Omega_q^{\rm Spin^c} =\left\{ \begin{array}{cl} \Omega_q^{\rm Spin^c} & \text{for $p=0,4$},\\
 22 \, \Omega_q^{\rm Spin^c} & \text{for $p=2$},\\
0 & \text{otherwise}\,.\end{array}\right. 
\end{equation}
\begin{table}
\begin{center}
\begin{tabular}{ c | c c c c c c c c }
 10 & $4 \mathbb{Z} \oplus  \mathbb{Z}_2$ & 0  & $88 \mathbb{Z} \oplus 22 \mathbb{Z}_2$ & 0 & $4 \mathbb{Z} \oplus  \mathbb{Z}_2$ & 0 & 0 & 0\\
 9 & 0 & 0  & 0 & 0 & 0 & 0 & 0 & 0\\
 8 & $4 \mathbb{Z}$ & 0  & $88\mathbb{Z}$ & 0 & $4 \mathbb{Z}$ & 0 & 0 & 0\\
 7 & 0 & 0  &    0 & 0 & 0 & 0 & 0 & 0\\
 6 & $2 \mathbb{Z}$ & 0  & $44\mathbb{Z}$ & 0 & $2 \mathbb{Z}$ & 0 & 0 & 0\\
 5 &0 & 0 & 0 & 0 & 0 & 0 & 0 & 0\\
 4 & $2 \mathbb{Z}$ & 0  & $44\mathbb{Z}$ & 0 & $2\mathbb{Z}$ & 0 & 0 & 0\\
 3 & 0 & 0  &    0 & 0 & 0 & 0 & 0 & 0\\
 2 &  $\mathbb{Z}$ & 0  & $22\mathbb{Z}$ & 0 & $\mathbb{Z}$ & 0 & 0 & 0\\
 1 &  0 & 0  & 0 & 0 & 0 & 0 & 0 & 0\\
0 &  $\mathbb{Z}$ & 0  & $22\mathbb{Z}$& 0 & $\mathbb{Z}$ & 0 & 0 & 0\\
\hline
& 0 & 1 &2 &3&4&5&6&7
\end{tabular}
\end{center}
\captionof{figure}{Second (and final) page of the AHSS for the computation of $\Omega_n^{\rm Spin^c}(K3)$.}
\label{fig:cobordk3}
\end{table}

\noindent
Up to $n = 10$ all extension problems are trivial, so that we can express the final result as
\begin{equation}
\begin{aligned}
\Omega_n^{{\rm Spin}^c}(K3) &= \Omega_n^{{\rm Spin}^c}({\rm pt}) \, \oplus \, \tilde{\Omega}_n^{{\rm Spin}^c}(K3) \\
&= \Omega_n^{{\rm Spin}^c}({\rm pt}) \, \oplus \, 22 \, \Omega_{n-2}^{{\rm Spin}^c}({\rm pt}) \, \oplus \, \Omega_{n-4}^{{\rm Spin}^c}({\rm pt})\,. 
\end{aligned}
\end{equation}

\noindent
In this formula, it is understood that cobordism groups with negative index are set to zero.
The explicit groups resulting from the formula above are reported in table \ref{OmegaspincK3}.
\begin{table}[h!]
\vspace{0.3cm}  
\begin{center}
\resizebox{\textwidth}{!}{
\begin{tabular}{ c | c c c c c c c c c c c c}
n & 0 & 1 & 2 &  3 & 4 & 5 & 6 & 7 & 8 & 9 & 10\\
\midrule
 $\Omega^{{\rm Spin}^c}_n (K3)$ &$\mathbb{Z}$ & 0 & $23\mathbb{Z}$  & 0 & $25\mathbb{Z}$  &0 & $47\mathbb{Z}$ & 0 &  $50\mathbb{Z}$ & 0 & $94\mathbb{Z}\oplus \mathbb{Z}_2$ \\[-0.3cm]
\end{tabular}
}
\end{center}
\caption{Cobordism groups $\Omega^{{\rm Spin}^c}_n (K3)$, $n=0,\ldots,10$.}
\label{OmegaspincK3}
\end{table}

\subsubsection{Computing $\Omega^{{\rm Spin}^c}_n (CY_3)$}

The computation for the cobordism groups of a Calabi-Yau threefold are obtained similarly to those of $K3$. We start from the known result\footnote{By assumption, the Calabi-Yau threefolds we consider in this work are such that $\pi_1(CY_3) =0$. In general, there exist Calabi-Yau manifolds with $\pi_1 (CY_3) = \mathbb{Z}_n$, for some integer $n$, i.e.~with torsion in $H^1(CY_3;\mathbb{Z})$. Typical examples are free quotient of Calabi-Yaus without torsion, such as the free quotient of the quintic $\mathbb{P}_4[5]/\mathbb{Z}_5$. They have been investigated, especially in a K-theory context, for instance in \cite{Brunner:2001eg, Brunner:2001sk}. For Calabi-Yau twofolds, one has $\pi_1(K3)=0$. Taking a free quotient by $\mathbb{Z}_n$ reduces the Euler number to $\chi/24 n$, so that the quotient manifold is not $K3$ anymore.}

\begin{equation}
H_n(CY_3; \mathbb{Z}) =\left\{\begin{array}{cl} \mathbb{Z} & \text{for}\,\, n=0,6,\\
b_2\,\mathbb{Z} & \text{for}\,\, n=2,4,\\
b_3\,\mathbb{Z} & \text{for}\,\, n=3,\\
0& \text{otherwise},\end{array}\right.
\end{equation}
where $b_p$ are the $CY_3$ Betti numbers (recall that $b_{p}=b_{6-p}$).
The second page is then given by
\begin{equation}
\begin{aligned}
E^2_{p,q}&= H_p(CY_3; \Omega_q^{\rm Spin^c} )\\
&\cong H_p(CY_3; \mathbb{Z}) \otimes \Omega_q^{\rm Spin^c} =\left\{\begin{array}{cl}  \Omega_q^{\rm Spin^c} & \text{for}\,\, p=0,6,\\
b_2\, \Omega_q^{\rm Spin^c} & \text{for}\,\, p=2,4,\\
b_3\, \Omega_q^{\rm Spin^c} & \text{for}\,\, p=3,\\
0& \text{otherwise}\end{array}\right.
\end{aligned}
\end{equation}
and shown explicitly in figure \ref{fig:cobordcy}.
One realises  that this time five non-vanishing columns $E^2_{p,q}$ exist in the second page, the elements of which are given by $b_p \Omega_q^{{\rm Spin}^c}$.

\begin{table}[h!]
\begin{center}
\begin{tabular}{ c | c c c c c c c c }
 10 & $4 \mathbb{Z} \oplus  \mathbb{Z}_2$ & 0  & $b_2(4 \mathbb{Z} \oplus \mathbb{Z}_2)$ & $b_3(4 \mathbb{Z} \oplus \mathbb{Z}_2)$ & $b_2(4 \mathbb{Z} \oplus  \mathbb{Z}_2)$ & 0 & $4 \mathbb{Z} \oplus  \mathbb{Z}_2$ & 0\\
 9 & 0 & 0  & 0 & 0 & 0 & 0 & 0 & 0\\
 8 & $4 \mathbb{Z}$ & 0  & $4b_2\mathbb{Z}$ & $4b_3\mathbb{Z}$ & $4b_2 \mathbb{Z}$ & 0 & $4\mathbb{Z}$  & 0\\
 7 & 0 & 0  &    0 & 0 & 0 & 0 & 0 & 0\\
 6 & $2 \mathbb{Z}$ & 0  & $2b_2\mathbb{Z}$ & $2b_3\mathbb{Z}$ & $2b_2 \mathbb{Z}$ & 0 & $2\mathbb{Z}$  & 0\\
 5 &0 & 0 & 0 & 0 & 0 & 0 & 0 & 0\\
 4 & $2 \mathbb{Z}$ & 0  & $2b_2\mathbb{Z}$ & $2b_3\mathbb{Z}$ & $2b_2\mathbb{Z}$ & 0 & $2\mathbb{Z}$  & 0\\
 3 & 0 & 0  &    0 & 0 & 0 & 0 & 0 & 0\\
 2 &  $\mathbb{Z}$\tikzmark{AAA} & 0  & $b_2\mathbb{Z}$ & $b_3\mathbb{Z}$\tikzmark{CCC} & $b_2\mathbb{Z}$ & 0 & $\mathbb{Z}$  & 0\\
 1 &  0 & 0  & 0 & 0 & 0 & 0 & 0 & 0\\
0 &  $\mathbb{Z}$ & 0  & $b_2\mathbb{Z}$& \tikzmark{BBB}$b_3\mathbb{Z}$ & $b_2\mathbb{Z}$ & 0 & \tikzmark{DDD}$\mathbb{Z}$  & 0\\
\hline
& 0 & 1 &2 &3&4&5&6&7
\end{tabular}
\begin{tikzpicture}[ overlay , remember picture, shorten >=2pt,shorten <=.5pt]
\draw [<-] ({pic cs:CCC})  to ({pic cs:DDD});
\end{tikzpicture}
\end{center}
\captionof{figure}{Second (and final) page of the AHSS for the computation of $\Omega_n^{\rm Spin^c}(CY_3)$. One of the possibly non-vanishing differentials $d^3:E_{6,q}^3\to E_{3,q+2}^3$ is displayed (for $q=0$). They eventually vanish for $q\leq 6$.}
\label{fig:cobordcy}
\end{table}

\noindent
None of the  differentials $d_r$ with even $r$ can act for degree reasons. 
However, there are two kinds of third differentials that can be non-trivial. 
The first class is
\begin{equation}
    d^3: E^3_{3,q} \to E^3_{0,q+2}\,,
\end{equation}
which vanish due to the edge homomorphism (see section \ref{sec:edgehom}).
The second class acts as
\begin{equation}
d^3:E_{6,q}^3\to E_{3,q+2}^3\,,
\end{equation}
which is in principle non-vanishing.\footnote{This differential is given by the homological dual of the cohomology operation $Sq_{\mathbb{Z}}^3$, the (integral) third Steenrod  square (the operations $Sq^i$ are introduced briefly later on; see also the appendix \ref{app_steenrodsq}). Interestingly, its triviality is the homological dual statement of the Freed-Witten anomaly cancellation \cite{Diaconescu:2000wy,Maldacena:2001xj}, which we are going to discuss later on in the K-theory calculations.}. 
That this differential is trivial up to $q=6$, too, follows from Lemma 3.1 of \cite{Arlettaz:1992}. 
We thus get the results in table \ref{OmegaspincCY3}.

\begin{table}[h!]
\begin{center}
\hspace{-2.5cm}\begin{tabular}{ c | c c c c c c  }
n & 0 & 1 & 2 &  3 & 4 & 5 \\
\midrule
 $\Omega^{\rm Spin^c}_n (CY_3)$ &$\mathbb{Z}$ & 0 & $(b_2+1)\mathbb{Z}$  & $b_3\mathbb{Z}$ & $(2+2b_2)\mathbb{Z}$ & $b_3\mathbb{Z}$ \\
\end{tabular}

\vspace{0.5cm}
\begin{tabular}{ c | c c c c c }
n &6 & 7 & 8 & 9 &10\\
\midrule
 $\Omega^{\rm Spin^c}_n (CY_3)$ & $(3+3b_2)\mathbb{Z}$ & $2b_3\mathbb{Z}$ & $(5+4b_2)\mathbb{Z}$ & $2b_3\mathbb{Z}$  & $(6+6b_2)\mathbb{Z}\oplus \mathbb{Z}_2$\\[-0.2cm]
\end{tabular}
\end{center}
\caption{Cobordism groups $\Omega^{\rm Spin^c}_n (CY_3)$, $n=0,\ldots,10$.}
\label{OmegaspincCY3}
\end{table}

\subsection{Application to K-theory}
\label{sec_AHSS_K}

Next, we perform similar computations for the K- and KO-theory groups on spheres, tori and Calabi-Yau manifolds. 
For this purpose we employ the cohomological version of the AHSS. 
Real K-theory turns out to be more involved, but we report some results in sections \ref{KOresults} and \ref{sec_KOK3}.

\subsubsection{Computing $K^{-n}(S^k)$}

The K-theory groups of spheres $S^k$ are known to be \cite{Hatcher:478079}

\begin{equation}
K^{-n}(S^k)= \left\{\begin{array}{cl} 
\mathbb{Z} & \text{for}\,\, k \ \text{odd} , \\2\mathbb{Z} & \text{for}\,\, n,k \ \text{even},\\
0 & \text{otherwise},\end{array}\right.
\end{equation}

\noindent
but it is instructive to reproduce these results using the cohomological AHSS \eqref{2nd_page_cohom_AHSS}. 
As usual, we use the trivial fibration ${\rm pt} \to S^k \to S^k$ and we do not have to worry about local coefficients. Recalling that
\begin{equation}
    K^{-n}({\rm pt}) = \left\{\begin{array}{cc} \mathbb{Z} & \text{for $n$ even},\\
    0 & \text{otherwise},\\
    \end{array}
    \right.
\end{equation}
we have the second page
\begin{equation}
    E^{p,q}_2= H^p(S^k;K^q({\rm pt})) = \left\{\begin{array}{cl} \mathbb{Z}, & \text{for $q$ even, $p=0,k$,}\\
    0, & \text{otherwise\,.}\\
    \end{array}
    \right.
\end{equation}
Note that it is essential to include the bottom quadrant (with $q<0$) to arrive at  reasonable results.  
Limiting our spectral sequence to the first quadrant only, as in the homological case, is not consistent as it would violate Bott periodicity.

For concreteness, let us  consider $X=S^3$. We are interested in the groups $K^{-n}(X)$, with $n>0$, so the relevant page elements lie on the $p+q=-n$ bands of the final page, which now intersect the axes only once.
\begin{table}[h!]
\begin{center}
\begin{tabular}{ c | c c c c c }
6&$\mathbb{Z}$ & 0 &0 & $\mathbb{Z}$ &0\\
5&0&0&0&0&0\\
4&$\mathbb{Z}$ & 0 &0 & $\mathbb{Z}$ &0\\
3&0&0&0&0&0\\
2&$\mathbb{Z}$ \tikzmark{ddd31}& 0  &0& $\mathbb{Z}$ &0\\
1&0&0&0&0&0\\
0&$\mathbb{Z}$ & 0 &0 & \tikzmark{ddd32}$\mathbb{Z}$ &0\\
\hline
-1&0&0&0&0&0\\
-2&$\mathbb{Z}$ & 0 &0& $\mathbb{Z}$ &0\\
-3&0&0&0&0&0\\
-4&$\mathbb{Z}$  & 0&0 & $\mathbb{Z}$ &0\\
-5&0&0&0&0&0\\
-6&$\mathbb{Z}$ & 0&0 & $\mathbb{Z}$ &0\\
\end{tabular}
\begin{tikzpicture}[ overlay , remember picture, shorten >=2pt,shorten <=.5pt]
\draw [->] ({pic cs:ddd31})  to ({pic cs:ddd32});
  \end{tikzpicture}
\end{center}
\captionof{figure}{Second (and final) page of the AHSS for the computation of $K^{-n}(S^3)$. One of the $d_3$ differentials is shown explicitly. They all eventually vanish.} 
\end{table}
The $d_2$ differential vanish so that $E^{p,q}_3=E^{p,q}_2$, but $d_3$ may act non-trivially
\begin{equation}
d_3: E_3^{0,q} \to E_3^{3,q-2}, \qquad \text{$q$ even}\,.
\end{equation}
This differential was found by Atiyah and Hirzebruch \cite{Atiyah1961} to be an instance of a cohomological operation known as (integral) Steenrod square ($Sq_{\mathbb{Z}}^i$)
\begin{equation}
Sq^3_{\mathbb Z}: H^{n}(X;\mathbb{Z}) \to H^{n+3}(X;\mathbb{Z}).
\end{equation}
Explicitly, it is given by the composition
\begin{equation}
\label{Sq3comp}
d_3 = Sq^3_{\mathbb Z} = \beta \circ Sq^2 \circ \rho,
\end{equation}
where $\rho$ is the reduction modulo 2 and $\beta$ the Bockstein homomorphism, namely
\begin{equation}
\label{Sq3comp2}
Sq^3_{\mathbb Z}:  H^{n}(X;\mathbb{Z}) \overset{\rho}{\longrightarrow} H^{n}(X;\mathbb{Z}_2) \overset{Sq^2}{\longrightarrow} H^{n+2}(X;\mathbb{Z}_2) \overset{\beta}{\longrightarrow}H^{n+3}(X;\mathbb{Z}).
\end{equation}
We refer the reader to the appendix \ref{app_steenrodsq} for a more precise definition of Steenrod squares and of the Bockstein homomorphism, together with a short summary of their main properties.

Fortunately, since no torsion is involved, according to Theorem 4.8 of \cite{Husemoeller2008} all differentials (including $d_3$) vanish. 
This fact will be used systematically in the other computations of $K^{-n}(X)$ groups below.\footnote{This is a consequence of the Chern isomorphism
\begin{align}
K^{0}(X) \otimes_{\mathbb{Z}}\mathbb{R} \cong \bigoplus_n H^{2n}(X; \mathbb{R}),\qquad
K^{-1}(X) \otimes_{\mathbb{Z}}\mathbb{R} \cong \bigoplus_n H^{2n+1}(X; \mathbb{R}),
\end{align}
which implies that if there is no torsion in cohomology, the AHSS for K-theory terminates already at the second page.
}
Moreover, the extension problem is always trivial, since only free abelian groups are present.
Thus, for  every odd value of $k$  we recover $K^{-n}(S^{2k+1})=\mathbb{Z}$.
The situation for even $k$ is simpler as for degree reasons no differentials can act, so that $E^{p,q}_2=E^{p,q}_{\infty}$. 
We recover then $K^{-2n-1}(S^{2k}) = 0$ and $K^{-2n}(S^{2k})=2\mathbb{Z}$. Notice that  the final result can be expressed as
\begin{equation}
K^{-n}(S^k) = K^{-n}({\rm pt}) \oplus K^{-k-n}({\rm pt})\,.
\label{K_final_spheres}
\end{equation}

\subsubsection{Comment of Freed-Witten anomalies}
\label{sec_FWanomalycanc}

Let us comment more on the role of $d_3 =Sq^3_{\mathbb Z}$ and on its physical consequences, beyond the computation of $K^{-n}(S^k)$. From \cite{Freed:1999vc}, it is known that type II D-branes (in absence of $B$ field) must wrap a Spin$^c$ manifold $Y$, otherwise there is a global Freed--Witten anomaly. Given an element $y \in H^{n}(X;\mathbb{Z})$, one has (see appendix \ref{app_steenrodsq})
\begin{equation}
Sq^3_{\mathbb Z}(y) = W_3(N) \cup y,
\end{equation}
where $N$ is the normal bundle of the codimension $n$ submanifold Poincar\'e dual to $y$, which we call $Y$ below, while $\cup$ is the cup product. Since $W_3(N)=0$, iff $Y$ is Spin$^c$\footnote{The obstruction to Spin$^c$ structure on $Y$ is really $W_3(Y) = \beta(w_2(Y))$. However, since in our case $X$ is Spin and $Y$ is oriented by assumption (in type II), one can show that $w_2(N) = w_2(Y)$, implying $W_3(N) = W_3(Y)$ \cite{Witten:1998cd,Freed:1999vc}.}, one can relate a trivial action of $d_3$ in the AHSS to the absence of Freed--Witten anomalies for a D-brane wrapping $Y$ \cite{Diaconescu:2000wy,Maldacena:2001xj}. Indeed, if  $E^4=\ker d_3 / {\rm Im} \, d_3$ is given in terms of the groups $H^n(X;\mathbb{Z})$ without further restrictions, all cohomology classes (and their dual cycles) survive. Otherwise, either some are removed when passing from cohomology to K-theory or they change to a torsion group \cite{Maldacena:2001xj}. Physically, they would correspond to D-branes which are anomalous or unstable.

\subsubsection{Computing $K^{-n}(T^k)$}

Next we consider the $k$-dimensional torus  $T^k=(S^1)^k$.
To proceed, one can either compute the groups by using the AHSS in a similar manner as done for the sphere  (extending also the page to include the fourth quadrant) or use the known results for the reduced K-theory groups $\widetilde{K}^{-n}(T^k)$ and the decomposition \eqref{Ksplitting}.

Starting with the second approach, we observe that according to \cite{Hatcher:478079} we have 
\begin{equation}
 \widetilde{K}^{-n}(T^k)=\left\{ \begin{array}{cl} 2^{k-1}\mathbb{Z} & \text{for $n$ odd},\\
 (2^{k-1}-1)\mathbb{Z} & \text{for $n$ even}.
 \end{array}\right. 
\end{equation}
Since $K^{-2n}({\rm pt})=\mathbb{Z}$ and $K^{-2n-1}({\rm pt})=0$, it follows that 
\begin{equation}
\label{Ktk}
K^{-n}(T^k)=2^{k-1}\mathbb{Z},
\end{equation}
for $n$ any integer. 
For the trivial case $k=1$, where the torus is just a circle, the above result coincides with the expected one from the sphere computation, i.e.~$K^{-n}(T^1)=\mathbb{Z}$.

Let us also comment on the calculation of $K^{-n}(T^k)$ using the spectral sequence. 
One has the second page
\begin{equation}
E^{p,q}_2=H^p(T^k; K^q({\rm pt}))   \, .
\end{equation}
The computation using the AHSS for the trivial fibration gives the same result \eqref{Ktk}, upon realising that once again all differentials vanish since there is no torsion, so $E^{p,q}_2=E^{p,q}_\infty$, and the extension problem is trivial.
We note that  the final result can be elegantly written  as
\begin{equation}
K^{-n}(T^k) = \bigoplus_{m = 0}^k {k\choose m}\, K^{-m-n}({\rm pt})\,,
\label{K_final_tori}
\end{equation}
where the binomial coefficient can be interpreted as the number of $m$-cycles on $T^k$.

\subsubsection{Computing $K^{-n}(K3)$}

The AHSS also allows to straightforwardly compute the K-theory groups on $K3$.
The second page of the sequence is given by
\begin{equation}
E^{p,q}_2=H^{p}(K3; K^{q}({\rm pt}))=    
\left\{ \begin{array}{cl} \mathbb{Z} & \text{for $p=0,4$, $q$ even},\\
 22\,\mathbb{Z} & \text{for $p=2$, $q$ even},\\
 0 & \text{otherwise}.
 \end{array}\right. 
\end{equation}
This is explicitly shown in figure \ref{table:E2KK3}.
\begin{table}[h!]
\begin{center}
\begin{tabular}{ c | c c c c c }
6&$\mathbb{Z}$ & 0 & $22\mathbb{Z}$ &0&$\mathbb{Z}$\\
5&0&0&0&0&0\\
4&$\mathbb{Z}$ & 0 & $22\mathbb{Z}$ &0&$\mathbb{Z}$\\
3&0&0&0&0&0\\
2&$\mathbb{Z}$ & 0 & $22\mathbb{Z}$ &0&$\mathbb{Z}$\\
1&0&0&0&0&0\\
0&$\mathbb{Z}$ & 0 & $22\mathbb{Z}$ &0&$\mathbb{Z}$\\
\hline
-1&0&0&0&0&0\\
-2&$\mathbb{Z}$ & 0 & $22\mathbb{Z}$ &0&$\mathbb{Z}$\\
-3&0&0&0&0&0\\
-4&$\mathbb{Z}$ & 0 & $22\mathbb{Z}$ &0&$\mathbb{Z}$\\
-5&0&0&0&0&0\\
-6&$\mathbb{Z}$ & 0 & $22\mathbb{Z}$ &0&0$\mathbb{Z}$\
\end{tabular}
\end{center}
\captionof{figure}{Second (and final) page of the AHSS for the computation of $K^{-n}(K3)$.} \label{table:E2KK3}
\end{table}
It is evident that no differentials can act non-trivially on the second page for degree reasons so that  the sequence promptly terminates. Thus, the final result reads
\begin{equation}
{K}^{-n}(K3)=\left\{ \begin{array}{cl} 0 & \text{for $n$ odd},\\
 24\mathbb{Z} & \text{for $n$ even}.
 \end{array}\right. 
\end{equation}
Note that the factor 24 arises as $b_0+b_2+b_4=1+22+1$ with $b_m$ being the Betti numbers of $K3$. Therefore, we can also express the K-theory groups on K3 as
\begin{equation}
{K}^{-n}(K3)=\bigoplus_{m = 0}^4 b_{4-m}(K3) \, K^{-m-n}({\rm pt})\,.
\end{equation}

\subsubsection{Computing $K^{-n}(CY_3)$}

The computation of $K^{-n}(CY_3)$ is similar to that of $K3$. Omitting unnecessary details, we present directly the second page in figure \ref{table:page_K(CY3)}.
\begin{table}[h!]
\begin{center}
\begin{tabular}{ c | c c c c c c c}
6&$\mathbb{Z}$&0 & $b_2\mathbb{Z}$ & $b_3\mathbb{{Z}}$ & $b_2\mathbb{Z}$ &0 &$\mathbb{Z}$ \\
5&0&0&0&0&0&0&0\\
4&$\mathbb{Z}$&0 & $b_2\mathbb{Z}$ & $b_3\mathbb{{Z}}$ & $b_2\mathbb{Z}$ &0&$\mathbb{Z}$\\
3&0&0&0&0&0&0&0\\
2&$\mathbb{Z}$&0 & $b_2\mathbb{Z}$  &$b_3\mathbb{Z}$& $b_2\mathbb{Z}$ & 0&$\mathbb{Z}$\\
1&0&0&0&0&0&0&0\\
0&$\mathbb{Z}$&0& $b_2\mathbb{Z}$ &$b_3\mathbb{Z}$ & $b_2\mathbb{Z}$ &0&$\mathbb{Z}$\\
\hline
-1&0&0&0&0&0&0&0\\
-2&$\mathbb{Z}$&0 & $b_2\mathbb{Z}$ & $b_3\mathbb{{Z}}$ & $b_2\mathbb{Z}$ &0 &$\mathbb{Z}$\\
-3&0&0&0&0&0&0&0\\
-4&$\mathbb{Z}$&0 & $b_2\mathbb{Z}$ & $b_3\mathbb{{Z}}$ & $b_2\mathbb{Z}$&0 &$\mathbb{Z}$\\
-5&0&0&0&0&0&0&0\\
-6&$\mathbb{Z}$&0 & $b_2\mathbb{Z}$ & $b_3\mathbb{Z}$ & $b_2\mathbb{Z}$&0 &$\mathbb{Z}$\\
\end{tabular}
\end{center}
\captionof{figure}{Second (and final) page of AHSS for computation of $K^{-n}(CY_3)$.} \label{table:page_K(CY3)}
\end{table}
The only possibly non-vanishing differential is $d_3:E^{1,q}_3\to E^{4,q-2}_3$. However, due to lack of torsion it is in fact vanishing and, given also that the extension problem is trivial, we conclude that
\begin{equation}
{K}^{-n}(CY_3)=\left\{ \begin{array}{cc} b_3\, \mathbb{Z} & \text{if n odd},\\
 (2+2b_2)\,\mathbb{Z} & \text{ if n even}.
 \end{array}\right. 
  \label{CY3}
\end{equation}
Notice the factor $(2+2b_2)$ arises as $b_0+b_2+b_4+b_6$, with $b_0=b_6=1$ and $b_2=b_4$, $b_p$ being the Betti numbers of the $CY_3$. 
The result can also be found in Corollary 1.9 of \cite{Doran_2007}.
Again, we can  elegantly express the K-theory groups on (simply connected) Calabi-Yau threefolds as
\begin{equation}
{K}^{-n}(CY_3)=\bigoplus_{m = 0}^6 b_{6-m}(CY_3) \, K^{-m-n}({\rm pt})\,.
\end{equation}

\subsubsection{KO-groups of spheres and tori}
\label{KOresults}

The KO groups can similarly be computed using the AHSS. However, in this case there is torsion, so the differentials can be non-vanishing. 
For spheres $S^k$, one can use the splitting lemma and determine  the relevant groups as
\begin{equation}
KO^{-n}(S^k)= \widetilde{KO}(S^{n+k})\oplus \widetilde{KO}(S^n)=KO^{-n-k}({\rm pt}) \oplus KO^{-n}({\rm pt})\,.
\end{equation}
The full results for $KO^{-n}(S^k)$ for $n,k \leq 10$ are provided in appendix \ref{sec_app_Ktheory}.
For tori, it was shown in \cite{Gukov:1999qb} that 
\begin{equation}
KO^{-n}(T^k) = \bigoplus_{m = 0}^k {k\choose m}KO^{-m-n}({\rm pt})\,.
\label{KO_tori}
\end{equation}

\subsubsection{Computing $KO^{-n}(K3)$}
\label{sec_KOK3}

For real K-theory, computations involving higher dimensional manifolds with a richer topology than the torus or the sphere can become potentially more complicated, due to more involved differentials and extension problems. 
Indeed, for Calabi-Yau threefolds the computation turned out to be fairly subtle, so that we postpone it to future work.
However, in the case of $K3$, as we will show now, all differentials are vanishing and the computations can be performed, up to extensions.
The second page of the spectral sequence is the following:
\begin{equation}
E^{p,q}_2=H^{p}(K3; KO^{q}({\rm pt}))
=\left\{ \begin{array}{cl} KO^{q}({\rm pt}) & \text{for $p=0,4$},\\
 22 \, KO^q ({\rm pt})& \text{for $p=2$},\\
0 & \text{otherwise}.\end{array}\right. 
\end{equation}
\begin{table}[h!]
\begin{center}
\begin{tabular}{ c | c c c c c }
7&$\mathbb{Z}_2$ & 0 & $22\mathbb{Z}_2$ &0&$\mathbb{Z}_2$\\
6&$\mathbb{Z}_2$ & 0 & $22\mathbb{Z}_2$ &0&$\mathbb{Z}_2$\\
5&0&0&0&0&0\\
4&$\mathbb{Z}$ & 0 & $22\mathbb{Z}$ &0&$\mathbb{Z}$\\
3&0&0&0&0&0\\
2&0&0&0&0&0\\
1&0&0&0&0&0\\
0&$\mathbb{Z}$ & 0 & $22\mathbb{Z}$ &0&$\mathbb{Z}$\\
\hline
-1&$\mathbb{Z}_2$ & 0 & $22\mathbb{Z}_2$ &0&$\mathbb{Z}_2$\\
-2&$\mathbb{Z}_2$ & 0 & $22\mathbb{Z}_2$ &0&$\mathbb{Z}_2$\\
-3&0&0&0&0&0\\
-4&$\mathbb{Z}$ & 0 & $22\mathbb{Z}$ &0&$\mathbb{Z}$\\
-5&0&0&0&0&0\\
-6&0&0&0&0&0\\
-7&0&0&0&0&0\\
\end{tabular}
\end{center}
\captionof{figure}{Second (and final) page of the AHSS for the computation of $KO^{-n}(K3)$.} \label{table:E2KOK3}
\end{table} 
We realise that the differentials $d_2$ and $d_4$ (depending on how $d_2$ acts) can possibly be non-vanishing. At second degree, we have
\begin{equation}
    d_2:  E^{p,q}_2 \to E^{p+2,q-1}_2,
\end{equation}
for $p=0,2$ and $q=0,-1$, together with all of its periodic copies. 
The explicit form of this differential is known to be \cite{ABP:1967,GradySati} \begin{equation}
\label{diff2_KO}
d_2 = \left\{ \begin{array}{cl} Sq^2 \rho: \, &H^{p}(K3; KO^{0}({\rm pt})) \to H^{p+2}(K3; KO^{-1}({\rm pt}))\,,\\
 Sq^2: \, &H^{p}(K3; KO^{-1}({\rm pt})) \to H^{p+2}(K3; KO^{-2}({\rm pt})) \end{array}\right. 
\end{equation}
corresponding to $q=0,-1$ respectively. Here, $Sq^2: H^p(X;\mathbb{Z}_2) \to H^{p+2}(X; \mathbb{Z}_2)$ is the second Steenrod square and $\rho$ is the reduction modulo 2. 
We argue now that $d_2$ is vanishing for $X=K3$. We discuss the case $q=-1$, but the analysis can be extended to $q=0$ in a similar way.
For any element $y \in H^{p}(X;\mathbb{Z}_2)$, we can represent $Sq^2 (y) = \iota_{*}(w_2(N)) \cup y$ \cite{MilnorStasheff}.
Here, $N$ is the normal bundle of the submanifold $Y \subset X$ Poincar\'e dual to $y$ and $\iota_{*}: H^p(Y) \to H^p(X)$ the cohomological push-forward. 
For $p=0$, the differential $d_2$ vanishes since $y$ is dual to the whole four dimensional manifold $X=K3$ which is Spin, thus $w_1(N) = w_2(N)=0$. Alternatively, it vanishes since $Sq^2(y) = 0$ for $y \in H^0(X;\mathbb{Z}_2)$ (see the properties of $Sq^i$ listed in appendix \ref{app_steenrodsq}).
For $p=2$, the differential vanishes as well, since from the condition $w_2(X) = w_1(X)=0$ (i.e.~$X$ is Spin), one can then prove $w_2(N)=0$ for a two dimensional manifold $Y$ not necessarily orientable \cite{Diaconescu:2000wy}. 
Alternatively, for $p=2$ we can also write $Sq^2(y) = \nu_2 \cup y$ (see equation \eqref{sqWuclass}) and then the second Wu class, $\nu_2 = w_2 (X) + w_1(X)^2$, vanishes since $X=K3$ is Spin. Thus, $d_2$ is trivial.

At degree four, we have the differential
\begin{equation}
d_4: E_4^{0,-1} \to E_4^{4,-4}.
\end{equation}
However, since there cannot be non-trivial homomorphisms\footnote{This can be seen directly as follows. Consider the case $\phi: \mathbb{Z}_2 \to \mathbb{Z}$, the generalisation to $k >2$ being straightforward. $\phi$ cannot be a non-trivial homomorphism since, choosing $\phi(0)=0$ and $\phi(1)=1$, one is lead to the contradiction $0=\phi(0)=\phi(2) =\phi(1) + \phi(1)= 2$. Thus, the only option is to set also $\phi(1)=0$ and $\phi$ is trivial.} $\mathbb{Z}_k \to \mathbb{Z}$ for $k \geq 2$ also this differential must vanish and $E_2^{p,q} \cong E_\infty^{p,q}$.

Thus, one can read off the $KO^{-n}(K3)$ groups, which we present in table \ref{tableKOK3} up to extensions. Note that we have already made use of the splitting lemma \eqref{Ksplitting} to simplify the results.

\begin{table}[h!]
\begin{center}
\hspace{1.5cm}\begin{tabular}{ c | c c c c }
n & 0 & 1 & 2 &  3 \\
\midrule
 $KO^{-n} (K3)$ & $\mathbb{Z} \oplus e( 22\mathbb{Z}_2,\mathbb{Z})$ & $\mathbb{Z}_2$ & $\mathbb{Z}_2 \oplus 22\mathbb{Z}$ &0
\end{tabular}

\vspace{0.5cm}
\begin{tabular}{ c | c c c c }
n &4 & 5 & 6 & 7 \\
\midrule
 $KO^{-n} (K3)$ & $2\mathbb{Z}$& $\mathbb{Z}_2$ & $\mathbb{Z}_2 \oplus 22\mathbb{Z}$ & $22\mathbb{Z}_2$ 
\end{tabular}

\end{center}
\caption{KO-groups $KO^{-n} (K3)$, $n=0,\ldots,7$, up to extensions. The result can be extrapolated to $n\geq 8$ by Bott periodicity.}
\label{tableKOK3}
\end{table}

\section{Physical interpretation}
\label{sec_interpretation}

In this section, we show that the cobordism and K-theory groups of $X$ previously calculated with the AHSS can be interpreted in terms of the dimensional reduction of global symmetries, thus making contact with section \ref{sec_dimredsymm}. 
For $X\in\{S^k,T^k,K3,CY_3\}$, the analysis turns out to be particularly simple, since all differentials in the AHSS vanish and extensions are trivial, as we explicitly showed. 
For more complicated backgrounds, these simplifications might not occur, but the AHSS should still give the correct  answer.

First, we expect the story to become substantially more involved if one turns on fluxes. 
For instance, allowing for non-trivial NS-NS three-form flux $H$ leads to the computation of $H$-twisted K-theory groups $K^{-n}_H(X)$ and the corresponding cobordism groups $\Omega^{{\rm Spin}^c,H}(X)$. 
In this case where $W_3= 0$, i.e.~we have a ${\rm Spin}^c$-structure, the absence of Freed--Witten anomalies implies that  the $H$-flux through a $D$-brane must vanish. 
This will result in non-trivial maps $d_r:E^{p,q}_r\to E^{p+r,q-r+1}$ in the
evaluation of the AHSS.

Second, even in the purely geometric case (no fluxes), when computing say $KO^{-n}(X)$ there could be non-trivial differentials, indicating e.g.~that certain cycles are not Spin.
An explicitly verification of these expectations is left for future work.

\subsection{General aspects}
\label{sec_genasp}

All of the analysed examples have in common that the final results of the AHSS can be expressed in a convenient, compact manner. 
For the K-theory groups $K^{-n}(X)$  of a $k$-dimensional manifold $X\in\{S^k,T^k,K3,CY_3\}$, we have in fact
\begin{equation}
\label{finalKtheoryX}
K^{-n}(X)= \bigoplus_{m=0}^k\,  b_{k-m}(X)\,   K^{-n-m}({\rm pt})\, ,
\end{equation}
with $n\ge 0$.
The interpretation of this result in terms of $D$-branes is as follows.
Say we are in $d=10$ dimensions and compactify the theory on the $k$-dimensional manifold $X$, so that the total space is $\mathbb{R}^{1,d-k-1}\times X$. 
Then, $K^{-n}(X)$ classifies all $D$-branes that are of codimension $n$ in the flat space $\mathbb{R}^{1,d-k-1}$. 
From the $d$-dimensional point of view, these are given by the set of all codimension $n+m$ branes wrapping $(k-m)$-cycles on the compact space $X$.
Hence, the result \eqref{finalKtheoryX} just reflects that the dimensional reduction performed following this perhaps naive geometrical reasoning is already the correct answer on these manifolds.
Nevertheless, thanks to the AHSS we also learn that none of the  wrapped $D$-branes experiences a Freed--Witten anomaly nor that there is an instantonic decay-channel.

The relation \eqref{finalKtheoryX} has a nice connection to the completeness hypothesis. 
The right hand side of \eqref{finalKtheoryX} is indeed a lattice of charges $({\bf q_1}, \dots, {\bf q_k})$, where each entry ${\bf q}_m$ is a charge vector with $b_{k-m}$ components.\footnote{In general this would be slightly inaccurate, since the groups of the point might be actually direct sums, so one of them could correspond to more sites in the lattice. Here, we neglect this complication for the sake of simplicity. The analysis can be directly adapted.} 
The fact that they can and indeed they are populated independently of one another means that in general the full spectrum of charges (or rather stable states with that given charge) is complete. 
To understand the point, consider the simple two-dimensional situation in which the lattice is just $\mathbb{Z}\oplus \mathbb{Z}$. 
In this case, one not only has stable bound states of branes associated to say $(1,0)$ and $(0,1)$, but also to $(1,1)$. 
Thus, what the relation \eqref{finalKtheoryX} is telling us is that to any non-vanishing element $({\bf q_1}, \dots, {\bf q_k})$ must be associated a stable object and, in this sense, the spectrum is complete. 
In general, especially in the presence of multicharged or non-BPS branes, the situation might become highly involved, but K-theory should give the correct answer.

For cobordism groups we found an analogous result, namely that for $n\ge 0$  they can
also be expressed as
\begin{equation}
\label{finalcobordtheoryX}
\Omega^{{\rm Spin}^c}_{n+k}(X)= \bigoplus_{m=0}^k\, b_{k-m}(X)\,   \Omega^{{\rm Spin}^c}_{n+m}({\rm pt})\, .
\end{equation}
The case $-k\le n<0$ will be discussed later. 
We propose the following intuitive interpretation of this result.
First, recall that in the definition of $\Omega_n(X)$ one introduces continuous maps $f :M\to X$, for every $n$-dimensional compact manifold $M$, such that $[M,f] \in \Omega_n(X)$. 
A non-vanishing term labelled by $m$ in the sum on the right hand side indicates that the map $f:M\to X$ from the $(n+k)$-dimensional manifold $M$ into the $k$-dimensional manifold $X$ is such that it wraps $M$ around a non-trivial $(k-m)$-cycle of $X$, while no other obstruction is introduced by the map in the remaining $(n+m)$ directions of $M$.
Since there are $b_{k-m}$ different $(k-m)$-cycles on $X$, we get $b_{k-m}$ factors of $\Omega^{{\rm Spin}^c}_{n+m}({\rm pt})$ in the total cobordism group $\Omega^{{\rm Spin}^c}_{n+k}(X)$.

Taking into account that the objects charged under the cobordism groups $\Omega_n({\rm pt})$ are the $(d-n)$-dimensional gravitational solitons mentioned in section \ref{sec:cobgroupsrev},
one can provide a similar interpretation as for the K-theory groups.  
Accordingly, $\Omega^{{\rm Spin}^c}_{n+k}(X)$ classifies all gravitational solitons that are of codimension $n$ in the flat space $\mathbb{R}^{1,d-k-1}$. 
From the $d$-dimensional point of view, they are given by the set of all codimension $(n+m)$ objects wrapping $(k-m)$-cycles on the compact space $X$.

Concretely, defining a basis $\{\Sigma_m^a\}$ of $m$-cycles on $X$, with $a=1,\ldots,b_m(X)$, and taking into account that $\Omega^{{\rm Spin}^c}_{\rm even}({\rm pt})=\mathbb{Z}$, for a given $m$-charge vector
\begin{equation}
\label{chargevectorm}
{\bf q_m}=(q_m^1,\ldots,q_m^{b_m})\in \mathbb Z^{b_m},
\end{equation}
the map $f$ is such the $(n+k)$-dimensional manifold $M_{n+k}$ is wrapped $q_m^a$ times around the $m$-cycle $\Sigma_m^a$ of $X$.
Hence, one can think of such an $m$-cycle to be shared between $M$ and $X$.

For all values of the index $n+k$, our goal is to explain how to organise the information contained in K-theory and cobordism groups of $X$ and then reconstruct tadpole cancellation conditions known from string theory. 
As we will see, for $n\ge 0$ this is quite straightforward, whereas in the regime $-k\le n<0$ we will encounter some new issues.
We thus assume $n \geq 0$ for the time being.
Given the previous results, we can understand how the Hopkins--Hovey isomorphism applies to cobordism and K-theory groups of manifolds $X$ which are not just a point. 
Via the relation
\begin{equation}
\label{Ktheorydual}
K^{-n}(X)=K_{n+k}(X),
\end{equation}
valid for $X$ a $k$-dimensional Spin$^c$ manifold, the K-theory result \eqref{finalKtheoryX} can be formally brought into the same form as \eqref{finalcobordtheoryX}, namely we can pass from generalised cohomology to homology.\footnote{The relation \eqref{Ktheorydual} and the analogous one for real K-theory, namely
\begin{equation}
\label{KOsupsub}
KO^{-n}(X) = KO_{n+k}(X),
\end{equation}
for $X$ a $k$-dimensional Spin manifold, follow e.g.~from Theorem 2.9 of section $V$ of \cite{Rudyakbook}, after recalling that a manifold is K-oriented (resp.~KO-oriented) iff it is Spin$^c$ (resp.~Spin). See also \cite{Rudyak}.}
Therefore (for $n\geq 0$) the ABS orientation can be extended to a map
\begin{equation}
\alpha^c_X:  \Omega^{{\rm Spin}^c}_{n+k}(X) \to K_{n+k}(X),
\end{equation}
acting as $\alpha^c$ in \eqref{ABSorient} on each term $ \Omega^{{\rm Spin}^c}_{n+k-m}({\rm pt})$.
Dividing by the kernel of this map provides an isomorphism between cobordism and K-theory classes on $X$.
Hence, at least for these simple cases, the latter isomorphism is directly inherited from the isomorphism between $ \Omega^{{\rm
    Spin}^c}_{n}({\rm pt})$ and $ K_{n}({\rm pt})$.
As we have shown, the AHSS gives analogous simple results for the Spin cobordism groups $\Omega^{{\rm Spin}}_{n+k}(X)$ and the real K-theory classes $KO_{n+k}(X)$ for $X\in\{S^k,T^k\}$.
This implies that the above structure carries over to such cases, as well.

We can also give an interpretation in terms of global symmetries.
In our examples, the groups $K_{n+k}(X)$ and $\Omega^{{\rm Spin}^c}_{n+k}(X)$ classify all global $(D-1-n)$-form charges in the non-compact $D=d-k$ dimensions. 
These can be thought of as arising from the dimensional reduction of global $d-1-n, d-2-n,\ldots, d-1-k-n$ form charges along the $k,k-1,\ldots,0$ cycles of $X$.
Due to the simple underlying structure, it is now clear that the fate of these global symmetries will follow the standard rules of the dimensional reduction. 
As already laid out in section \ref{sec_dimredsymm}, if a global symmetry in $D$ dimensions descends from a global symmetry in $d$ dimensions, then its gauging involves the dimensionally reduced gauge field in $d$ dimensions and also the corresponding dimensionally reduced $D$-branes (defects). 
In fact, the whole tadpole cancellation condition in $D$ dimensions arises from the dimensional reduction of the tadpole cancellation condition in $d$ dimensions.
We will provide more concrete examples in section \ref{sec:exCY3}.

\subsection{Example of a Calabi-Yau threefold}
\label{sec:exCY3}

Let us now focus on the case of the ten-dimensional type IIB superstring compactified on a Calabi-Yau threefold $X$.
In the previous section, we computed
\begin{equation}
  K^0(X)=b_6\underbrace{K^0({\rm pt})}_\mathbb{Z}\,\oplus\, b_4\underbrace{K^{-2}({\rm pt})}_{\mathbb{Z}}\, \oplus\, b_2\underbrace{K^{-4}({\rm pt})}_{\mathbb{Z}}\, \oplus\, b_0\underbrace{K^{-6}({\rm pt})}_{\mathbb{Z}},
\end{equation}
with $b_0=b_6=1$. The corresponding $D$-branes are all of codimension zero in the flat $\mathbb{R}^{1,3}$ space.  
In particular, in subsequent order, the four types of (single charged) $D$-branes corresponding to the $K^{-2n}({\rm pt})$ groups are: $D9$-branes wrapping the entire $CY_3$, $D7$-branes wrapping the $b_4$ 4-cycles of the $CY_3$, $D5$-branes wrapping the $b_2$ 2-cycles of the $CY_3$ and finally $D3$-branes being point-like on the $CY_3$. 
At the next level, we found
\begin{equation}
  K^{-1}(X)=b_3\underbrace{K^{-4}({\rm pt})}_\mathbb{Z},
\end{equation}
corresponding to a codimension one brane in $\mathbb{R}^{1,3}$ and given by $D5$-branes wrapping any of the $b_3$ three-cycles on the $CY_3$.
As already explained, for {\it all} multi-charges there should exist corresponding bound states of the  single charged states. 
This is consistent with the completeness hypothesis.

As we have inferred from the AHSS, the corresponding cobordism groups split in a very similar manner
\begin{equation}
  \Omega^{{\rm Spin}^c}_6(X)=b_6\underbrace{\Omega^{{\rm Spin}^c}_0({\rm pt})}_\mathbb{Z}\,\oplus\, b_4\underbrace{\Omega^{{\rm Spin}^c}_2({\rm pt})}_{\mathbb{Z}}\, \oplus\, b_2\underbrace{\Omega^{{\rm Spin}^c}_4({\rm pt})}_{\mathbb{Z}\oplus\mathbb{Z}}\, \oplus\, b_0\underbrace{\Omega^{{\rm Spin}^c}_6({\rm pt})}_{\mathbb{Z}\oplus\mathbb{Z}}
\end{equation}
and
\begin{equation}
  \Omega^{{\rm Spin}^c}_7(X)=b_3\underbrace{\Omega^{{\rm
        Spin}^c}_4({\rm pt})}_{\mathbb{Z}\oplus \mathbb{Z}}\,.
\end{equation}
As mentioned, this pattern is related to dimensional reduction
of global symmetries. Indeed, from 
\eq{
     \Omega^{{\rm Spin}^c}_6(X)=\mathbb{Z}\oplus b_4 \mathbb{Z}
     \oplus b_2 ( \mathbb{Z}\oplus \mathbb{Z}) \oplus  (\mathbb{Z}\oplus \mathbb{Z})   
   }
we infer that there is a $3b_2+3$ dimensional lattice of $\mathbb{Z}$-valued global 3-form charges in $\mathbb{R}^{1,3}$ (recall $b_4=b_2$). 
These are the dimensional reduction of the ten-dimensional 9-form, 7-form, 5-form and 3-form global symmetries along the 6, 4, 2, 0-cycles of the $CY_3$.

We now explain how to organise the information in the groups above and reconstruct tadpole cancellation conditions.
Consider a six-dimensional Spin$^c$-manifold $M_6$ that lies in the contribution  $b_m \Omega^{{\rm Spin}^c}_{6-m}({\rm pt})$ to $ \Omega^{{\rm Spin}^c}_6(X)$ but, contrary to the Calabi-Yau $X$, it is not necessarily a solution to  the string theory equations of motion.
Hence, in this sense, $M_6$ can be off-shell.
Since there must exist a continuous map $f:M_6\to X$, the manifold $M_6$  shares some of the $m$-cycles with the background space $X$. 
Which $m$-cycles are shared depends on the non-zero entries in the charge vector \eqref{chargevectorm}.
Then, the magnetic $(6-m)$-form currents are  obtained from the cobordism invariants \eqref{cobinvspinc}, which we repeat below for convenience 
\begin{equation}
\begin{aligned}
  \tilde J_0(M_6)&={\rm td}_0(M_6)=1\,,\\
  \tilde J_2(M_6)&={\rm td}_2(M_6)={\frac 12} c_1(M_6)\,,\\
  \tilde J_{4,1}(M_6)&={\rm td}_4(M_6)={\frac{1}{12}}\left(c_2(M_6)+c_1^2(M_6)\right)\,,\qquad
  \tilde J_{4,2}(M_6)=c^2_1(M_6)\,,\\
  \tilde J_{6,1}(M_6)&={\rm td}_6(M_6)={\frac{1}{24}} c_2(M_6)\,
  c_1(M_6)\,,\qquad\qquad\ 
  \tilde J_{6,2}(M_6)={\frac 12}c^3_1(M_6) \,.
\end{aligned}
\end{equation}
Concretely, we propose that the magnetic $(6-m)$-form currents are defined by  expanding  the right hand sides into a basis of those $(6-m)$-forms in $H^{6-m}(M_6;\mathbb{Z})$ that also lie in $H^{6-m}(X;\mathbb{Z})$ (again depending on the entries in the charge vector).
For the Poincar\'e dual of the currents, denoted with hat, this means that we expand
\begin{equation}
\label{expandcurrents}
\hat{\tilde J}_{m,i}(M_6)=\sum_{a=1}^{b_m} \alpha^{a}_{m,i}\, q_m^a\, \Sigma_m^a + \ldots
\end{equation}
where the dots indicate more contributions along $m$-cycles of $M_6$ that do not lie in $X$. Note that the (co)homology of $M_6$ can in principle be bigger than that of $X$. Since this expansion is also valid for $M_6\ne X$, it allows us to go slightly off-shell. 
More in general, topological K-theory and cobordism groups classify all global charges that can be present in principle, irrespective of properties like supersymmetry or being on-shell.

Recall that the Todd classes also define the ABS orientation at fixed degree, i.e.~$\alpha_n^c(M_6)={\rm Td}(M_6)$. 
Due to this map and the fact that all K-theory global symmetries are gauged, we can infer that  at fixed $n=0,2,4,6$ always one linear combination of the above currents is gauged. Such a combination is the one entering the tadpole cancellation conditions.
We discuss this below for all four classes of global symmetries in turn.

\begin{itemize}
    \item 

First, we have $\Omega_0^{{\rm Spin}^c}({\rm pt})$ and $K^0({\rm pt})$.
The factor $\Omega_0^{{\rm Spin}^c}({\rm pt})=\mathbb Z$ gives rise to a single global 3-form symmetry in four dimensions, with the trivial magnetic current $\tilde J_0(M_6)={\rm td}_0(M_6)=1$. 
In ten dimensions, the corresponding 9-form symmetry is gauged with the charged objects being  $D9$-branes, classified by $K^0({\rm pt})=\mathbb Z$.
This leads to the tadpole constraint
\begin{equation}
N \,  \delta^{(0)}(M_6)+ a^{(0)}\, {\rm td}_0(M_6)=0
\end{equation}
where $\delta^{(0)}(M_6)$ denotes the 0-form Poincar\'e dual to the 6-cycle $M_6$ wrapped by the stack of $N$ $D9$-branes.\footnote{Formally, this 0-form arises from the ten-dimensional delta $\delta^{(0)}(\mathbb{R}^{1,3}\times M_6) = \delta^{(0)}(\mathbb{R}^{1,3})\wedge \delta^{(0)}(M_6)$.}
For $a^{(0)}=0,-32$, this is the familiar $D9$-tadpole cancellation condition in type IIB/type I string theory. 
Thus, one linear combination of the initial $\Omega_0^{{\rm Spin}^c}({\rm pt})\oplus K^0({\rm pt})=\mathbb{Z}\oplus \mathbb{Z}$ global symmetries is gauged, while the orthogonal one is in general broken.

\item 
Second, we have $\Omega_2^{{\rm Spin}^c}({\rm pt})$ and $K^{-2}({\rm pt})$.
The factor $b_4\Omega_2^{{\rm Spin}^c}({\rm pt})=b_4\mathbb{Z}$ gives rise to $b_4$ global 3-form symmetries in four dimensions, whose preserved magnetic 0-form currents $\tilde j^{(2)a}_0$ (again in $D=4$ and with the notation of section \ref{sec_dimredsymm}) are given by the expansion of the ten-dimensional 2-form current $\tilde J_2(M_6)={\rm td}_2(M_6)$ in a cohomological basis $\omega_{(2)a} \in H^2(X;\mathbb Z)$, namely
\begin{equation}
\tilde J_2(M_6)=\sum_{a=1}^{b_4} \,   \tilde j_{0}^{(2)a}\,\wedge\, \omega_{(2)a}\,.
\end{equation}
Note that $b_4=b^2$, so that this is the Poincar\'e dual to the expansion \eqref{expandcurrents}, where we included the charges $q_4^a$ into the coefficients.
Similarly, for a $D7$-brane classified by $K^{-2}({\rm pt})$ and wrapping 4-cycles $\Sigma_{4}\in H_4(M_6;\mathbb Z)$ that are contained in $X$ (times the flat space $\mathbb R^{1,3}$), we can expand its Poincar\'e dual 2-form as 
\begin{equation}
\delta^{(2)}(\mathbb{R}^{1,3}\times\Sigma_{4})=\sum_{a=1}^{b_4}   \delta^{(0)}(\mathbb{R}^{1,3})^{(2)a}  \, \wedge \omega_{(2)a}\,.
\end{equation}
In ten dimensions, the gauging of the corresponding 7-form global symmetry is associated to a tadpole constraint
\begin{equation}
\label{7branetad}
\sum_{j \in \text{def}} N_j  \,  \delta^{(2)}(\mathbb{R}^{1,3}\times \Sigma_{4,j})+ a^{(2)}\, {c_1(M_6)\over 2}=0
\end{equation}
which upon expansion in a cohomological basis of $H^2(X;\mathbb Z)=b_4 \mathbb Z$ leads to $b_4=b^2$ tadpole cancellation conditions.   
Hence, a subgroup $b_4 \mathbb Z$ of the initially present global symmetry $b_4  \left(\Omega_2^{{\rm Spin}^c}({\rm pt})\oplus K^{-2}({\rm pt})\right)=b_4  \left(\mathbb Z\oplus\mathbb Z\right)$ is gauged while the orthogonal $b_4 \mathbb Z$ group is broken. 
Of course, for $M_6=X$ we have $c_1(X)=0$ and the tadpole cancellation condition simplifies, but the power of K-theory and cobordism is that they allow us to go off-shell and see terms that could appear in principle, even if they are absent for the on-shell configurations. 
We note that, for $a^{(2)}=-24$ and $M_6$ being the base $B_3$ of an elliptically fibered Calabi-Yau fourfold, \eqref{7branetad} is the well known 7-brane tadpole constraint of F-theory.

\item 
Third, we have $\Omega_4^{{\rm Spin}^c}({\rm pt})$ and $K^{-4}({\rm pt})$.
The factor $b_2\Omega_4^{{\rm Spin}^c}({\rm pt})=b_2\left(\mathbb Z\oplus \mathbb Z\right)$ gives rise to $2b_2$ global 3-form symmetries in four dimensions.
Notice that this time the ABS  orientation between  K-theory and cobordism is not an isomorphism. 
The preserved magnetic 0-form currents $\tilde j^{(4)a}_{0,i}$, $i=1,2$, in $D=4$ are  given by the expansion of the ten-dimensional 4-form currents $\tilde J_{4,i}(M_6)$ in a cohomological basis $\hat\omega_{(4)a}$ of $H^4(X;\mathbb Z)$
\begin{equation}
\label{expandcurrents4}
\tilde J_{4,i}(M_6)=\sum_{a=1}^{b_2} \,   \tilde j^{(4)a}_{0,i}\, \wedge \,\hat\omega_{(4)a}\,.
\end{equation}  
The defects classified by $K^{-4}({\rm pt})$ are $D5$-branes wrapping 2-cycles $\hat \Sigma_{2}$ on $M_6$ that are shared with $X$ (times the flat space $\mathbb{R}^{1,3}$).
Again their Poincar\' e duals can be expanded similarly to \eqref{expandcurrents4}. The gauging of the ten-dimensional 5-form symmetry implies a tadpole condition of the form
\begin{equation}
\label{5branetad}
\sum_{j\in \text{def}} N_j  \,  \delta^{(4)}(\mathbb{R}^{1,3}\times \hat \Sigma_{2,j})+ a^{(4)}_1\,
\left({c_2(M_6)+c_1^2(M_6)\over 12}\right)+a^{(4)}_2\, c^2_1(M_6)=0,
\end{equation}
where we followed the general strategy reviewed in section \ref{sec_gaugandtad}, i.e.~both cobordism invariants can in principle appear in the gauging procedure. 
Upon expansion in a cohomological basis of $H^4(X;\mathbb Z)=b_2 \mathbb Z$, one obtains $b_2=b^4$ tadpole cancellation conditions. 
Hence, a subgroup $b_2 \mathbb Z$ of the initially present global symmetry
is gauged while the orthogonal group is broken.
The type I string on $X=K3\times T^2$ leads to such a $D5$-brane tadpole constraint for $a^{(4)}_1=-12$ and $a^{(4)}_2=3/2$. 
Another setup is the $\Omega \sigma$ orientifold of type IIB on $X=K3\times T^2$ presented in \cite{Dabholkar:1996zi, Gimon:1996ay}.
This example has only $O5$-planes and leads to the above  tadpole constraint with $a^{(4)}_1=-24$ and $a^{(4)}_2=0$.
Thus, we see again that more information is needed to completely specify the gauging related to specific string models.

\item
Fourth, we have $\Omega_6^{{\rm Spin}^c}({\rm pt})$ and $K^{-6}({\rm pt})$.
The factor $\Omega_6^{{\rm Spin}^c}({\rm pt})=\mathbb Z\oplus \mathbb Z$ gives rise to 2 global 3-form symmetries in four dimensions, whose preserved magnetic 0-form currents $\tilde j^{(6)}_{0,i}$, $i=1,2$ (again in $D=4$) are  given by the reduction  of the ten-dimensional 6-form currents $\tilde J_{6,i}(M_6)$  along the volume 6-form of $M_6$,
\begin{equation}
\label{expandcurrents6}
\tilde J_{6,i}(M_6)=  \tilde j^{(6)}_{0,i}\; {\rm vol}(M_6)\,.
\end{equation}  
The defects classified by $K^{-6}({\rm pt})$ are $D3$-branes being point-like on $M_6$. Then, the gauging of the ten-dimensional 3-form symmetry implies a tadpole condition of the general form
\begin{equation}
\label{3branetad}
\sum_{j\in \text{def}} N_j  \,  \delta^{(6)}(\mathbb{R}^{1,3}\times {\rm pt}_j)+ a^{(6)}_1\, {c_2(M_6) \,c_1(M_6)\over 24}+a^{(6)}_2\,  {c^3_1(M_6)\over 2}=0\,.
\end{equation}
Again, for a Calabi-Yau manifold, such as $M_6=X$, the two contributions from cobordism are vanishing but the off-shell nature of cobordism itself makes them visible in the general case.
For $a^{(6)}_1=-12$ and  $a^{(6)}_2=-30$, this  tadpole condition is known to be realised in F-theory compactified on a smooth elliptically fibered Calabi-Yau fourfold with base $M_6=B_3$.

\end{itemize}

Finally, let us discuss the four-dimensional global 2-form symmetries related to $K^{-1}(X)=b_3K^{-4}({\rm pt})=b_3 \mathbb{Z}$ and  $\Omega^{{\rm Spin}^c}_7(X)=b_3 \Omega^{{\rm Spin}^c}_4({\rm pt})=b_3 (\mathbb{Z}\oplus \mathbb{Z})$.
From the ten-dimensional perspective, these arise from the reduction of the global 5-form symmetries along the $b_3$ 3-cycles of $X$.
Concerning $\Omega^{{\rm Spin}^c}_7(X)$, the $2b_3$ preserved magnetic 1-form currents $\tilde j^{(3)a}_{1,i}$, with $i=1,2$, in $D=4$ are given by the dimensional reduction of the ten-dimensional 4-form currents $\tilde J_{4,i}(M_6)$ along the basis 3-forms $\omega_{(3)a}\in H^3(X;\mathbb Z)$,
\begin{equation}
\label{expandcurrentsthreee}
\tilde J_{4,i}(M_6)=\sum_{a=1}^{b_3} \,   \tilde j^{(3)a}_{1,i}\wedge \omega_{(3)a} \,.
\end{equation}
Note that this is meant in principle and that the currents $\tilde j^{(3)a}_{1,i}$ can also be vanishing.
The $D5$-brane defects wrapping 3-cycles $\Sigma_{3}$ on $M_6$ shared with $X$ times a three-dimensional submanifold $\Pi_3$ of the flat space $\mathbb R^{1,3}$ can be expanded in a similar fashion 
\begin{equation}
\delta^{(4)}(\Pi_3 \times \Sigma_3)=\sum_{a=1}^{b_3}   \delta^{(1)}(\Pi_3)^{(3)a} \wedge \omega_{(3)a} .
\end{equation}
In ten dimensions, the global symmetry of $K^{-4}({\rm pt})$ is gauged leading to a magnetic Bianchi identity 
\begin{equation}
d\tilde F_3=\sum_{j\in \text{def}} N_j  \,  \delta^{(4)}(\Pi_{3,j}\times \Sigma_{3,j})+ a^{(4)}_1\, \tilde J_{4,1}(M_6)+a^{(4)}_2\,  \tilde J_{4,2}(M_6)\,.
\end{equation}
Expanding now also the magnetic field strength as
\begin{equation}
\tilde F_3=\sum_{a=1}^{b_3} \tilde f_0^{(3)a}  \, \wedge\, \omega_{(3)a} ,
\end{equation}
we arrive at $b_3$ Bianchi identities for the four-dimensional 0-forms
\begin{equation}
d\tilde f_0^{(3)a}=\sum_{j\in \text{def}}  N_j  \,  \delta^{(1)}(\Pi_{3,j})^{(3)a}+ a^{(4)}_1\, \tilde j^{(3)a}_{1,1}+a^{(4)}_2\,  \tilde j^{(3)a}_{1,2}\,.
\end{equation}
Thus, everything fits nicely together once more. 
The discussion for higher groups, such as $\Omega_8^{{\rm Spin}^c}(X)$ and $\Omega_9^{{\rm Spin}^c}(X)$, together with their K-theory counterparts, follows the same logic.

\subsubsection*{Summary of results}

We demonstrated that, for the example of a Calabi-Yau space $X$, the K-theory and cobordism classes on $X$ for $n\ge 0$  are to be interpreted from the point of view of global symmetries and their subsequent gauging. 
In this situation, the AHSS is simple in the sense that no non-trivial maps, i.e.~differentials, appear and the outcomes reproduce the naive expectation from dimensional reduction.
Of course, a more involved task is to compute K-theory and cobordism classes where maps can be non-trivial and D-branes become inconsistent or unstable. 
However, even if in these cases the $D$-brane spectrum for a background space $X$ changes, the map between K-theory and cobordism is proven to be intact, so that the related global symmetries are guaranteed to disappear simultaneously.
Therefore, we expect that an interpretation in terms of gauging will still be very similar to what we discussed above.

We also expect our results to carry over to the correspondence of KO-groups and Spin-cobordisms. 
A new aspect is the appearance of $\mathbb Z_2$ torsion groups related to non-BPS branes on the K-theory side.
As discussed (see e.g.~\eqref{tadS1xS1} or the examples in \cite{Blumenhagen:2021nmi}), the corresponding cobordism groups  can decouple from tadpoles
or, more precisely, charge neutrality conditions, so they would need to be broken by some unknown defects.
At this stage, we cannot exclude that a more thorough analysis reveals some subtle aspects, but this is beyond the  scope of this paper.

\subsection{Fate of low-dimensional $\Omega^{{\rm Spin}^c}_{n}(X)$}
\label{sec_morecontr}

It  remains to discuss what happens in the regime $-k\le n<0$, for which the cobordism groups $\Omega^{{\rm Spin}^c}_{n+k}(X)$ are still non-vanishing. 
To get a better idea on what is different with respect to the regime $n\ge 0$, we start by asking what the corresponding K-theory groups are, namely $K_{n+k}(X)=K^{-n}(X)$ with $-k\le n<0$, and what they physically mean.
For concreteness, consider e.g.~the class $K^2(CY_3)$. 
Extrapolating the relation \eqref{finalKtheoryX} to $n=-2$, we would get
\begin{equation}
\label{Kregbeyond}
K^{2}(X)= \bigoplus_{m=2}^6\,  b_{6-m}(X)\,   K^{2-m}({\rm pt})=b_{4}(X)\,   K^{0}({\rm pt})\oplus\ldots\, ,
\end{equation}
where from the sum we left out the term $K^2({\rm pt})$, associated to $m=0$. 
The latter could be defined via Bott periodicity to be equal to $\mathbb Z$, but it is not clear what it should represent physically. In addition, it does not appear on the cobordism side \eqref{finalcobordtheoryX}.
The term written explicitly on the right hand side of \eqref{Kregbeyond} would correspond to a $D9$-brane wrapped on a 4-cycle of the $CY_3$. 
However, for dimensional reasons the $D9$ should really wrap a 6-chain (which can be thought of as the 4-cycle times a 2-chain) on the $CY_3$ and, as such, it is not a topologically non-trivial configuration.
This argument suggests that the groups $K_{n+k}(X)=K^{-n}(X)$ with $-k\le n<0$ do not admit  a clear physical interpretation in terms of wrapped D-branes.
This is further supported by extrapolating the result from section \ref{sec_genasp}, namely the fact that $K^{-n}(X)$   corresponds to codimension $n$ branes in $D=d-k$ dimensions. 
Clearly, for $n$ negative the branes do not fit.

For the corresponding cobordism groups  $\Omega^{{\rm Spin}^c}_{n+k}(X)$ with  $-k\le n<0$, the reasoning is completely analogous with the D-branes exchanged by gravitational solitons. 
As also claimed in section \ref{sec_genasp}, $\Omega^{{\rm Spin}^c}_{n+k}(X)$ classifies all gravitational solitons that are of codimension $n$ in the flat non-compact space. 
Again, for $n$ negative  the gravitational solitons do not fit.

\section{Conclusion}
\label{sec_conclusion}

The fact that there should be no global symmetries is believed to be a fundamental property of quantum gravity and is one of the best tested swampland conjectures. The cobordism conjecture \cite{McNamara:2019rup} is a recent generalisation of this fact, which interprets a non-vanishing cobordism group as a higher-form global symmetry in an effective field theory of quantum gravity. Demanding its absence, one is either led to breaking or gauging the symmetry.  While breaking can lead to the prediction of new objects (defects), gauging can be performed by exploiting the close relation between cobordism and K-theory. Indeed, K-theory charges are always gauged and this reflects to the fate of cobordism charges, as proposed in \cite{Blumenhagen:2021nmi}.

In this work, we gave further support to the idea that cobordism and K-theory groups are charges in quantum gravity by computing the groups associated to a compact manifold $X$, for typical choices employed in string compactifications. 
In particular, we showed that those groups contain precisely the information on how symmetries (broken or gauged) spread across the dimensional reduction of the effective theory on $X$. 
One of the advantages of the cobordism and K-theory description, rather than standard (co)homology, is that it automatically takes into account quantum mechanical effects, such as cancellation of Freed--Witten anomalies. 
This is most clear when performing the computation through the Atiyah--Hirzebruch spectral sequence, where absence of such anomalies is related to the vanishing of a certain differential. 
We reviewed this technique in detail, discussing and working out explicitly several examples. 
Then, we gave a physical interpretation of the results and show how the information of gaugings and tadpoles can be decoded from the cobordism and K-theory groups of $X$.

The work here presented can be extended along various directions. 
It would be important to find a first principle derivation of the unfixed coefficients $(a_i^{(n)})$ in tadpole cancellation conditions without using any string theory input, as this could give support to the so called String Lamppost Principle. 
It would also be interesting to extend the analysis in order to systematically include torsional groups, both in K-theory \cite{Braun:2000zm} (see also \cite{Marchesano:2006ns}) and cobordism. 
Indeed, in most of the examples we discussed torsion was playing little role. 
However, it is unavoidable in more realistic and involved setups. 
One can introduce torsion in the spectral sequence but also on the manifold $X$ itself, for example by considering Calabi-Yau manifolds with torsion \cite{Brunner:2001eg, Brunner:2001sk}. 
The question is then whether torsion classes are killed by the differentials or if they survive until the end of the sequence, thus giving additional contributions, say, to tadpoles. 
Another possible direction to extend this work is by including more structure on top of the considered manifold, such as gauge fields, which are again unavoidable in realistic settings. 
One should then turn to refined version of the groups here considered, such as twisted and differential K-theory, see e.g.~\cite{Fredenhagen:2000ei,Bunke:2010mq,BunkeLect,FreedLect,GradySati}. 
 We hope to come back to these questions in the future.

\noindent
\paragraph{Acknowledgments.}
It is a pleasure to thank Arun Debray for useful discussions.
The work of N.C.~is supported by the Alexander-von-Humboldt foundation.


\appendix

\section{Mathematical tools and results}
\label{sec_app_math}

In this appendix, we collect mathematical tools and results used through the main part of the work.

\subsection{Short exact sequences, extensions and Ext}
\label{sec_app_ses}
Consider the abelian groups $A$, $B$ and $C$.
A short sequence
\begin{equation}
\label{ses}
0 \longrightarrow B \overset{\beta}{\longrightarrow} C \overset{\alpha}{\longrightarrow} A \longrightarrow 0 
\end{equation}
is exact if the map $\beta$ is injective and the map $\alpha$ surjective, i.e.~if $\ker (\alpha ) = {\rm Im} (\beta)$. In this case, we say that $C$ is an extension of $A$ by $B$ and we denote it as
\begin{equation}
C=e(A,B).
\end{equation}
The Splitting Lemma for abelian groups tells us that the extension is trivial,
\begin{equation}
C=A\oplus B,
\end{equation}
iff there is a left inverse to $\beta$ iff there is a right inverse to $\alpha$. In this case, one says that the short exact sequence is split. In general, the extension might not be unique and there can be more extensions besides the trivial one. Equivalence classes of extensions of $A$ by $B$ are in one-to-one correspondence with elements of the group ${\rm Ext}^1(A,B)$, with the trivial extension corresponding to $0$ (see e.g.~Theorem 3.4.3 of \cite{weibel}). 

The definition and main properties of the groups ${\rm Ext}^n(A,B)$ can be found e.g.~in \cite{weibel}, chapter 3. We recall some of them below. As stated in Lemma 3.3.1, if $A$ and $B$ are abelian (as we assume) we have that ${\rm Ext}^n(A,B)=0$ for $n\geq 2$.  Therefore, only the groups associated to $n=0,1$ are relevant for us. We have that ${\rm Ext}^0(A,B)={\rm Hom}(A,B)$, while ${\rm Ext}^1(A,B)$ classifies extensions of $A$ by $B$, as anticipated above. Two useful properties of these groups are
\begin{align}
\label{Extprop1}
{\rm Ext}^n (\oplus_i A_i, B) &= \Pi_i {\rm Ext}^n(A_i,B),\\
\label{Extprop2}
{\rm Ext}^n (A,\Pi_i B_i) &= \Pi_i {\rm Ext}^n(A,B_i),
\end{align}
and we recall that for abelian groups direct product and direct sum coincide. For cyclic groups, we recall the results
\begin{align}
&{\rm Ext}^1 (\mathbb{Z},\mathbb{Z})=0,\\
&{\rm Ext}^1 (\mathbb{Z},\mathbb{Z}_n)=0,\\
\label{Extprop3}
&{\rm Ext}^1 (\mathbb{Z}_n,\mathbb{Z})=\mathbb{Z}_n,\\
&{\rm Ext}^1 (\mathbb{Z}_m,\mathbb{Z}_n)=\mathbb{Z}_k,
\label{Extprop4}
\end{align}
where $k={\rm GCD}(m,n)$. All of this is used in the calculations of section \ref{sec_AHSS}

Let us give two simple examples to illustrate how everything works in a combined way. Let us consider the short exact sequence
\begin{equation}
0\to\mathbb{Z}_2 \to e(\mathbb{Z}_2,\mathbb{Z}_2) \to \mathbb{Z}_2 \to 0.
\end{equation}
Since ${\rm Ext}^1(\mathbb{Z}_2,\mathbb{Z}_2)=\mathbb{Z}_2$, $e(\mathbb{Z}_2,\mathbb{Z}_2)$ is not split, instead we have two possible extensions. Indeed, it is well-known that there are two short exact sequences
\begin{align}
\label{ses1}
&0\to\mathbb{Z}_2 \to \mathbb{Z}_4 \to \mathbb{Z}_2 \to 0,\\
\label{ses2}
&0\to\mathbb{Z}_2 \to \mathbb{Z}_2\oplus \mathbb{Z}_2\to \mathbb{Z}_2\to 0.
\end{align}
Instead, the short exact sequence
\begin{align}
\label{ses4}
&0\to\mathbb{Z}_3 \to \mathbb{Z}_6\to \mathbb{Z}_2\to 0,
\end{align}
is split, since ${\rm Ext}^1(\mathbb{Z}_2, \mathbb{Z}_3)=0$.

\subsection{Universal Coefficient Theorem}
\label{sec_app_UCT}

The universal coefficient theorem  (see e.g.~\cite{weibel}) can be used to express (co)homology groups of a topological space $X$ with coefficients in a left $\mathbb{Z}$-module $A$ in terms of (co)homology groups with coefficients in $\mathbb{Z}$. It can be formulated both for homology and cohomology.

The version for homology groups states that there is a short (noncanonically) split exact sequence
\begin{equation}
0 \to H_n(X) \otimes A \to H_n(X;A )\to {\rm Tor}_1 (H_{n-1}(X),A) \to 0.
\end{equation}
The definition of the groups  ${\rm Tor}_n (A,B)$ can be found e.g.~in \cite{weibel}, chapter 3. As stated in Proposition 3.1.2 and 3.1.4, if $A$ and $B$ are abelian,  ${\rm Tor}_n (A,B)$ are torsion abelian groups and they vanish for $n\geq 2$; if $A$ is also torsion free, ${\rm Tor}_1 (A,B)=0$.
The version for cohomology groups states that there is a short (noncanonically) split exact sequence 
\begin{equation}
0 \to {\rm Ext}^1 H_{n-1}(X;A) \to H_n(X;A )\to {\rm Hom}(H_{n-1}(X),A) \to 0.
\end{equation}

\subsection{Properties of Steenrod squares}
\label{app_steenrodsq}

In this appendix, we collect some useful facts about Steenrod squares. For a nice, pedagogical review and for more information, we refer the reader to \cite{BeaudryCampbell} and references therein. A standard textbook is \cite{MilnorStasheff}. We will work at prime $2$, but it is possible to generalize the discussion to any prime $p$.

A cohomology operation of degree $i$ is a map
\begin{equation}
H^{n}(X;\mathbb{Z}_2)\to H^{n+i}(X;\mathbb{Z}_2).
\end{equation}
It is said to be stable if it commutes with the suspension isomorphism. Steenrod squares, $Sq^i$, are stable cohomology operations of degree $i$ satisfying the following defining properties, for any $i\geq 0$:
\begin{itemize}
\item[{\bf a)}] $Sq^0={\rm Id}$;
\item[{\bf b)}] $Sq^i(x)=x \cup x$, for $x \in H^i(X;\mathbb{Z}_2)$;
\item[{\bf c)}] $Sq^i(x)=0$, for $x \in H^j(X;\mathbb{Z}_2)$ and $j<i$;
\item[{\bf d)}] $Sq^i(x\cup y) = \displaystyle\sum_{m+n=i} Sq^m(x) \cup Sq^n(y)$ (Cartan formula).
\item[{\bf e)}] $Sq^i \circ Sq^j = \displaystyle \sum_{k=0}^{\lfloor i/2 \rfloor} \left(\begin{array}{c}j-k-1\\i-2k\end{array}\right)_{\text{mod 2}} Sq^{i+j-k}\circ Sq^k$, for $0<i < 2j$\\ (Adem relation).
\end{itemize}

The map $Sq^1 \equiv \tilde \beta$ is an example of a Bockstein homomorphism. It is associated to the short exact sequence
\begin{equation}
    0 \to \mathbb{Z}_2 \overset{\times 2}{\to} \mathbb{Z}_4 \overset{\rho}{\to} \mathbb{Z}_2 \to 0,
\end{equation}
where the first map is multiplication by $2$ and the second ($\rho$) is the reduction modulo 2, which induces the long exact sequence
\begin{equation}
\dots \overset{\tilde \beta}{\to}  H^{n}(X;\mathbb{Z}_2) \overset{\times 2}{\to}H^n(X;\mathbb{Z}_4) \overset{\rho}{\to} H^n(X;\mathbb{Z}_2) \overset{\tilde \beta}{\to} H^{n+1}(X;\mathbb{Z}_2)\to\dots\, .
\end{equation}
Here, $\tilde \beta$ is the connecting homomorphism between cohomology groups of different degree. Another Bockstein homomorphism, called $\beta$ in the main text, can be constructed in association with the short exact sequence
\begin{equation}
\label{sesZZZ2}
    0 \to \mathbb{Z} \overset{\times 2}{\to} \mathbb{Z} \overset{\rho}{\to} \mathbb{Z}_2 \to 0,
\end{equation}
inducing in turn the long exact sequence
\begin{align}
\label{lesZZZ2} 
&  \dots \overset{\beta}{\to} H^{n}(X;\mathbb{Z}) \overset{\times 2}{\to} H^{n}(X;\mathbb{Z})\overset{\rho}{\to} H^{n}(X,\mathbb{Z}_2) \overset{\beta}{\to}  H^{n+1}(X,\mathbb{Z}) \to\dots\, .
\end{align}
The two Bocksteins are related by
\begin{equation}
    \tilde \beta = \rho \circ \beta.
\end{equation}

At odd degree $i=2k+1$, one can define an integral lift of the Steenrod squares,
\begin{equation}
Sq_{\mathbb{Z}}^{2m+1} = \beta \circ Sq^{2m},
\end{equation}
which is such that $\rho \circ  Sq_{\mathbb{Z}}^{2m+1} =  Sq^{2m+1}$ and maps $H^{n}(X;\mathbb{Z}_2) \to H^{n+i}(X;\mathbb{Z})$. One further gets a map between integral cohomology by first reducing modulo 2 and then acting with $Sq_{\mathbb{Z}}^i$,
\begin{equation}
Sq_{\mathbb{Z}}^i \circ \rho:    H^{n}(X;\mathbb{Z}) \to H^{n+i}(X;\mathbb{Z}).
\end{equation}
An integral lift of $Sq^i$ for even $i=2m$ does not exist.\footnote{This can be proven as follows. Suppose it exists an integral lift for the even case, $Sq^{2m} = \rho \circ Sq^{2m}_{\mathbb{Z}}$. Exactness of the sequence \eqref{lesZZZ2} means that $\ker \beta = {\rm Im} \rho$, implying in turn $\beta \circ Sq^{2m} = \beta (\rho (Sq^{2m}_\mathbb{Z}))=0$. However, this is contradiction with the Adem relation $Sq^1 \circ Sq^{2m} = Sq^{2m+1} \neq 0$ (recall $Sq^1 = \rho \circ \beta$). Thus, such an integral lift $Sq^{2m}_{\mathbb{Z}}$ cannot exist.}

Given an element $x \in H^{k-i}(X;\mathbb{Z}_2)$, with $k=\text{dim}(X)$, the action of the Steenrod squares can be defined as
\begin{equation}
\label{sqWuclass}
Sq^i(x) = \nu_i \cup x,
\end{equation}
where $\nu_i \in H^i(X;\mathbb{Z}_2)$ is the $i$-th Wu class of $X$ (more precisely, of a real vector bundle over $X$ of rank $k$, which we generically take to be the tangent bundle), such that $\nu_i=0$ if $i > k-i$. Since the total Wu class is the Steenrod square of the total Stiefel-Whitney class, one can express each of the single Wu classes in terms of Stiefel-Whitney classes. At lower degree, one has
\begin{equation}
\begin{aligned}
\nu_1 &= w_1,\\
\nu_2 &= w_2 + w_1 \cup w_1,\\
\nu_3 & = w_1\cup w_2.
\end{aligned}
\end{equation}

In certain cases, one can give an alternative action of $Sq^i$, namely (see e.g.~\cite{Diaconescu:2000wy, Maldacena:2001xj})
\begin{equation}
Sq^i(y) = \iota_* (w_i(N)) \cup y,
\end{equation}
where $y \in H^{n}(X;\mathbb{Z}_2)$, $N$ is the normal bundle of the submanifold $Y\subset X$ Poincar\'e dual to $y$ and $\iota: Y \to X$ is the inclusion.\footnote{More in general \cite{MilnorStasheff}, one can define an action $Sq^i(u) = \pi^*(w_i(\xi)) \cup u$, with $u \in H^k(E;\mathbb{Z}_2)$ and $w_i(\xi) \in H^i(B;\mathbb{Z}_2)$, for any $k$-plane bundle $\xi: F \to E \overset{\pi}{\to} B$ of which the normal bundle $N(B)$ is a particular case.} This is most convenient for physical purposes, such as checking the absence of Freed--Witten anomalies for branes wrapping $Y$, on which we comment in section \ref{sec_FWanomalycanc} (there, following \cite{Diaconescu:2000wy, Maldacena:2001xj}, we directly employ the integral lift $W_3(N)$ of $w_3(N)$ and omit the pushforward $\iota_*$).

\subsection{Wedge sum, smash product and reduced suspension}
\label{app_smash}

Consider two pointed topological spaces $(X,x_0)$ and $(Y,y_0)$. The wedge sum, $X \vee Y$, is defined as 
\begin{equation}
X \vee Y = X \sqcup Y /  \sim ,
\end{equation}
where the equivalence relation identifies the two base points $x_0$ and $y_0$.
The smash product, $X\wedge Y$, is defined as the quotient of the cartesian product by the wedge sum
\begin{equation}
X \wedge Y = \frac{X \times Y}{X \vee Y}
\end{equation}
It satisfies the properties
\begin{align}
&X \wedge Y \cong Y \wedge X,\\
&(X \wedge Y) \wedge Z \cong X \wedge (Y \wedge X),
\end{align}
where the symbol $\cong$ means homeomorphic as topological spaces.

Consider then the $n$-sphere $S^n$. The reduced suspension of $X$ is defined as
\begin{equation}
\Sigma X \cong S^1 \wedge X.
\end{equation}
The construction can be iterated
\begin{equation}
\Sigma^n X \cong S^n \wedge X.
\end{equation}
An important case is when $X=S^k$, thus giving
\begin{equation}
\Sigma^n S^k \cong S^{n+k}.
\end{equation}
We also recall that
\begin{align}
\Sigma^0 \wedge X \cong S^0 \wedge X  \cong X,
\end{align}
where $S^0 \cong {\rm pt } \sqcup {\rm pt}$. Another useful formula is
\begin{equation}
\label{suspsplit}
\Sigma (X \times Y) \cong \Sigma X \vee \Sigma Y \vee \Sigma (X \wedge Y).
\end{equation}

\subsection{Cobordism groups of spheres and tori}
\label{sec:altstrat}

We can prove that for a generic structure $\xi$ the cobordism groups of spheres ($S^k$) and tori ($T^k$) have a simple decomposition in terms of the respective cobordism groups of the point, namely
\begin{align}
\label{cob_sphere}
&\Omega_n^{\xi}(S^k) =\Omega_n^{\xi}({\rm pt}) \oplus \Omega^{\xi}_{n-k}({\rm pt}),\\
\label{cob_torus_binom}
&\Omega_n^{\xi}(T^k) = \bigoplus_{i = 0}^{k} \binom{k}{i} \Omega_{n-i}^{\xi}({\rm pt}),
\end{align}
where we implicitly assume the groups with negative index to be vanishing.
This explains for example why, when computing Spin and Spin$^c$ cobordism groups of spheres in section \ref{sec_AHSS_cob}, even if we found that in general
\begin{equation}
\Omega^{\rm Spin}_n (S^k) =\left\{
\begin{array}{cc}
\Omega_n^{\rm Spin}({\rm pt}) & n<k,\\
e(\Omega^{\rm Spin}_{n-k} ({\rm pt}), \Omega^{\rm Spin}_n ({\rm pt})) & n \geq k,
\end{array}
\right. 
\end{equation}
(and similarly for Spin$^c$ cobordism), every time the information at our disposal was enough to solve the extension problem, it turned out to be trivial.  
We now prove \eqref{cob_sphere} and \eqref{cob_torus_binom} by induction.

We start from the cobordism groups of spheres, $S^k$. For $S^1$, we have
\begin{equation}
\begin{aligned}
\label{cob_sphere_2}
\Omega_n^{\xi}(S^1) &=  \Omega_n^{\xi}({\rm pt}) \oplus \tilde{\Omega}_{n}^{\xi}(S^1) \\
&= \Omega_n^{\xi}({\rm pt}) \oplus \tilde{\Omega}_{n}^{\xi}(\Sigma(S^0)) \\
&= \Omega_n^{\xi}({\rm pt}) \oplus \tilde{\Omega}_{n-1}^{\xi }(S^0) \\
&= \Omega_n^{\xi}({\rm pt}) \oplus \Omega_{n-1}^{\xi}({\rm pt}).
\end{aligned}
\end{equation}
In passing from the second to the third line we used the suspension axiom $\tilde\Omega_n^\xi(\Sigma X) = \tilde\Omega_{n-1}^\xi (X)$ \cite{davis2012lecture}, while in the last step we employed that $\tilde \Omega_n^\xi (S^0) =\Omega_n^\xi({\rm pt})$, which follows from
\begin{equation}
\Omega_n^{\xi}(S^0) = \Omega_n^{\xi}({\rm pt} \sqcup {\rm pt}) = \Omega_n^{\xi}({\rm pt}) \oplus \Omega_n^{\rm *}(pt) = \Omega_n^{\xi}({\rm pt}) \oplus \tilde{\Omega}_n^{\xi}(S^0).
\end{equation}
Then, we assume the formula to hold for $S^k$ and we prove it for $S^{k+1}$. Using again the Splitting Lemma \eqref{SplittLemmaCob} and the suspension axiom, we have
\begin{equation}
\begin{aligned}
\Omega_n^{\xi}(S^{k+1}) &=  \Omega_n^{\xi}({\rm pt}) \oplus \tilde{\Omega}_{n}^{\xi}(S^{k+1}) \\
&=  \Omega_n^{\xi}({\rm pt}) \oplus \tilde{\Omega}_{n}^{\xi}(\Sigma(S^{k})) \\
&=  \Omega_n^{\xi}({\rm pt}) \oplus \tilde{\Omega}_{n-1}^{\xi}(S^{k}) \\
&=  \Omega_n^{\xi}({\rm pt}) \oplus \Omega_{n-k-1}^{\xi}({\rm pt}).
\end{aligned}
\end{equation}
This proves \eqref{cob_sphere} by induction.

Then, we look at the cobordism groups of tori, $T^k$. The result for $T^1 = S^1$ is already proven in \eqref{cob_sphere_2}. We thus assume the formula to hold for $T^k$ and we prove it for $T^{k+1}$. To this purpose, using \eqref{suspsplit} we can write
\begin{equation}
\Sigma(T^k \times S^1) = \Sigma(T^k) \vee \Sigma(S^1) \vee \Sigma(T^k \wedge S^1)
\end{equation}
and therefore
\begin{equation}
\begin{aligned}
\Omega_n^{\xi}(T^{k+1}) &= \Omega_{n}^{\xi}({\rm pt}) \, \oplus \,\tilde{\Omega}_{n+1}^{\xi}(\Sigma(T^{k+1})) \\
&= \Omega_{n}^{\xi}({\rm pt}) \, \oplus \, \tilde{\Omega}_{n+1}^{\xi}(\Sigma(T^k \times S^1)) \\
&= \Omega_{n}^{\xi}({\rm pt}) \, \oplus \, \tilde{\Omega}_{n+1}^{\xi}(\Sigma(T^{k})) \, \oplus \, \tilde{\Omega}_{n+1}^{\xi}(\Sigma(S^{1})) \, \oplus \, \tilde{\Omega}_{n+1}^{\xi}(\Sigma^2(T^{k})) \\
&= \Omega_{n}^{\xi}({\rm pt}) \, \oplus \,\tilde{\Omega}_{n}^{\xi}(T^{k}) \, \oplus \, \tilde{\Omega}_{n}^{\xi}(S^{1}) \, \oplus \, \tilde{\Omega}_{n-1}^{\xi}(T^{k}) \\
&= \Omega_{n}^{\xi}(T^{k}) \, \oplus \, \Omega_{n-1}^{\xi}(T^{k}),
\end{aligned}
\end{equation}
where we used $\tilde \Omega(X \vee Y) = \tilde \Omega (X) \oplus \tilde \Omega (Y)$, valid for reduced generalized homology theories \cite{davis2012lecture}. We can finally demonstrate that
\begin{equation}
\begin{aligned}
\Omega_n^{\xi}(T^{k+1})    &= \Omega_{n-1}^{\xi}(T^{k}) \, \oplus \, \Omega_{n}^{\xi}(T^{k})\\
&= \bigoplus_{i = 0}^{k} \binom{k}{i} \Omega_{n-1-i}^{\xi}({\rm pt}) \oplus \bigoplus_{i = 0}^{k} \binom{k}{i} \Omega_{n-i}^{\xi}({\rm pt}) \\
&= \bigoplus_{i = 1}^{k+1} \binom{k}{i-1} \Omega_{n-i}^{\xi}({\rm pt}) \oplus \bigoplus_{i = 0}^{k} \binom{k}{i} \Omega_{n-i}^{\xi}({\rm pt}) \\
&=\bigoplus_{i = 0}^{k+1} \binom{k}{i-1} \Omega_{n-i}^{\xi}({\rm pt}) \oplus \bigoplus_{i = 0}^{k+1} \binom{k}{i} \Omega_{n-i}^{\xi}({\rm pt})\\
&= \bigoplus_{i = 0}^{k+1} \binom{k+1}{i} \Omega_{n-i}^{\xi}({\rm pt}).
\end{aligned}
\end{equation}
In passing from the third to the fourth line we just added zero to both terms, while in the last step we used Pascal's formula. This concludes our proof of \eqref{cob_torus_binom} by induction.

An alternative proof can be given by exploiting some more advanced mathematical constructions. 
In particular, one can use that Spin and Spin$^c$ cobordism are generalised homology theories classified by Thom spectra $MSpin$ and $MSpin^c$ respectively. 
One can thus write\footnote{For any spectrum $G$, the reduced generalised homology is defined as $\tilde G_n(X) = [\mathbb{S},G\wedge X]_n$, where $\mathbb{S}=\Sigma^\infty$ is the sphere spectrum. 
The unreduced generalised homology is instead $G_n(X) = [\mathbb{S},G\wedge X_+]_n$, where $X_+ = X \sqcup {\rm pt}$. We refer e.g.~to \cite{davis2012lecture} for more details.}
\begin{equation}
\tilde{\Omega}^{\rm Spin}_n(X) := [\mathbb{S}, MSpin \wedge X]_n,
\end{equation}
where $X$ is a generic topological space, and similarly for Spin$^c$. 
Considering for example $X=S^k$, by exploiting the properties of the smash product and the suspension given in appendix \ref{app_smash}, we have
\begin{equation}
\begin{aligned}
\label{altproof}
\tilde{\Omega}^{\rm Spin}_n (S^k) &:= [\mathbb{S}, MSpin \wedge S^k]_n \\
&= [\mathbb{S}, MSpin]_{n-k}\\
&=\pi_{n-k}(MSpin)\\
&=\Omega_{n-k}^{\rm Spin} ({\rm pt}).
\end{aligned}
\end{equation}
In passing from the first to the second line we used that $[\Sigma X, Y] =[ X, \Omega Y]$ and then that $\Omega \Sigma X=X$, with $\Omega X$ the loop space, while in the last step we used the Pontrjagin-Thom isomorphism. Combining this with the Splitting Lemma \eqref{SplittLemmaCob}, one gets \eqref{cob_sphere}.

\clearpage
\section{Tables of Cobordism Groups}
\label{sec_app_cobordism}

\begin{table}[h!]

\begin{center}
\begin{tabular}{ c | c c c c c }
\toprule
 $n$ & $\Omega_n^{\rm Spin}(S^1)$ & $\Omega_n^{\rm Spin}(S^2)$  & $\Omega_n^{\rm Spin}(S^3)$ & $\Omega_n^{\rm Spin}(S^4)$ & $\Omega_n^{\rm Spin}(S^5)$\\
 \midrule
 0 & $\mathbb{Z}$ & $\mathbb{Z}$ & $\mathbb{Z}$ & $\mathbb{Z}$ & $\mathbb{Z}$ \\
 1 & $\mathbb{Z}_2 \oplus \mathbb{Z}$ & $\mathbb{Z}_2$  & $\mathbb{Z}_2$ & $\mathbb{Z}_2$ & $\mathbb{Z}_2$\\
 2 & $2\mathbb{Z}_2$ & $\mathbb{Z}_2 \oplus \mathbb{Z}$  & $\mathbb{Z}_2$ & $\mathbb{Z}_2$ & $\mathbb{Z}_2$\\
 3 & $\mathbb{Z}_2$ & $\mathbb{Z}_2$ & $\mathbb{Z}$ & 0 & 0\\
 4 & $\mathbb{Z}$ & $\mathbb{Z}_2 \oplus \mathbb{Z}$  & $\mathbb{Z}_2 \oplus \mathbb{Z}$ & $2\mathbb{Z}$ & $\mathbb{Z}$\\
 5 & $\mathbb{Z}$ & 0 & $\mathbb{Z}_2$ & $\mathbb{Z}_2$ & $\mathbb{Z}$\\
 6 & 0 & $\mathbb{Z}$ & 0 & $\mathbb{Z}_2$ & $\mathbb{Z}_2$\\
 7 & 0 & 0 & $\mathbb{Z}$ & $0$ & $\mathbb{Z}_2$\\
 8 & $2\mathbb{Z}$ & $2\mathbb{Z}$ & $2\mathbb{Z}$ & $3\mathbb{Z}$ & $2\mathbb{Z}$\\
 9 & $2\mathbb{Z}_2 \oplus 2\mathbb{Z}$ & $2\mathbb{Z}_2$ & $2\mathbb{Z}_2$ & $2\mathbb{Z}_2$ & $2\mathbb{Z}_2 \oplus \mathbb{Z}$\\
 10 & $5\mathbb{Z}_2$ & $3\mathbb{Z}_2 \oplus 2\mathbb{Z}$ & $3\mathbb{Z}_2$ & $3\mathbb{Z}_2$ & $3\mathbb{Z}_2$\\
 \bottomrule
\end{tabular}
\captionof{table}{Spin cobordism groups of spheres $\Omega_n^{\rm Spin}(S^k)$, $k=1,\ldots, 5$.} 
\end{center}
\end{table}

\begin{table}[h!]
\begin{center}
\begin{tabular}{ c | c c c c c }
\toprule
 $n$ & $\Omega_n^{\rm Spin}(S^6)$ & $\Omega_n^{\rm Spin}(S^7)$ & $\Omega_n^{\rm Spin}(S^8)$ & $\Omega_n^{\rm Spin}(S^9)$ & $\Omega_n^{\rm Spin}(S^{10})$\\
 \hline
 0 & $\mathbb{Z}$ & $\mathbb{Z}$ & $\mathbb{Z}$ & $\mathbb{Z}$ & $\mathbb{Z}$\\
 1 & $\mathbb{Z}_2$ & $\mathbb{Z}_2$ & $\mathbb{Z}_2$ & $\mathbb{Z}_2$ & $\mathbb{Z}_2$\\
 2 & $\mathbb{Z}_2$ & $\mathbb{Z}_2$ & $\mathbb{Z}_2$ & $\mathbb{Z}_2$ & $\mathbb{Z}_2$\\
 3 & 0 & 0 & 0 & 0 & 0\\
 4 & $\mathbb{Z}$ & $\mathbb{Z}$ & $\mathbb{Z}$ & $\mathbb{Z}$ & $\mathbb{Z}$\\
 5 & 0 & 0 & 0 & 0 & 0\\
 6 & $\mathbb{Z}$ & 0 & 0 & 0 & 0\\
 7 & $\mathbb{Z}_2$ & $\mathbb{Z}$ & 0 & 0 & 0\\
 8 & $\mathbb{Z}_2 \oplus 2\mathbb{Z}$ & $\mathbb{Z}_2 \oplus 2\mathbb{Z}$ & $3\mathbb{Z}$ & $2\mathbb{Z}$ & $2\mathbb{Z}$\\
 9 & $2\mathbb{Z}_2$ & $3\mathbb{Z}_2$ & $3\mathbb{Z}_2$ & $2\mathbb{Z}_2 \oplus \mathbb{Z}$ & $2\mathbb{Z}_2$\\
 10 & $3\mathbb{Z}_2 \oplus \mathbb{Z}$ & $3\mathbb{Z}_2$ & $4\mathbb{Z}_2$ & $4\mathbb{Z}_2$ & $3\mathbb{Z}_2 \oplus \mathbb{Z}$\\
 \bottomrule
\end{tabular}
\captionof{table}{Spin cobordism groups of spheres $\Omega_n^{\rm Spin}(S^k)$, $k=6,\ldots, 10$.}
\end{center}
\end{table}

\begin{table}[h!]
\begin{center}
\begin{tabular}{ c | c c c c c }
\toprule
 $n$ & $\Omega_n^{\rm Spin^c}(S^1)$ & $\Omega_n^{\rm Spin^c}(S^2)$  & $\Omega_n^{\rm Spin^c}(S^3)$ & $\Omega_n^{\rm Spin^c}(S^4)$ & $\Omega_n^{\rm Spin^c}(S^5)$\\
 \midrule
 0 & $\mathbb{Z}$ & $\mathbb{Z}$ & $\mathbb{Z}$ & $\mathbb{Z}$ & $\mathbb{Z}$ \\
 1 & $\mathbb{Z}$ & 0  & 0 & 0 & 0\\
 2 & $\mathbb{Z}$ & $2\mathbb{Z}$ & $\mathbb{Z}$ & $\mathbb{Z}$ & $\mathbb{Z}$\\
 3 & $\mathbb{Z}$ & 0 & $\mathbb{Z}$ & 0 & 0\\
 4 & $2\mathbb{Z}$ & $3\mathbb{Z}$  & $2\mathbb{Z}$ & $3\mathbb{Z}$ & $2\mathbb{Z}$\\
 5 & $2\mathbb{Z}$ & 0 & $\mathbb{Z}$ & 0 & $\mathbb{Z}$\\
 6 & $2\mathbb{Z}$ & $4\mathbb{Z}$ & $2\mathbb{Z}$ & $3\mathbb{Z}$ & $2\mathbb{Z}$\\
 7 & $2\mathbb{Z}$ & 0 & $2\mathbb{Z}$ & 0 & $\mathbb{Z}$\\
 8 & $4\mathbb{Z}$ & $6\mathbb{Z}$ & $4\mathbb{Z}$ & $6\mathbb{Z}$ & $4\mathbb{Z}$\\
 9 & $4\mathbb{Z}$ & 0 & $2\mathbb{Z}$ & 0 & $2\mathbb{Z}$\\
 10 & $4\mathbb{Z} \oplus \mathbb{Z}_2$ & $8\mathbb{Z} \oplus \mathbb{Z}_2$ & $4\mathbb{Z} \oplus \mathbb{Z}_2$ & $6\mathbb{Z} \oplus \mathbb{Z}_2$ & $4\mathbb{Z} \oplus \mathbb{Z}_2$\\
 \bottomrule
\end{tabular}
\captionof{table}{Spin$^c$ cobordism groups of spheres $\Omega_n^{{\rm Spin}^c}(S^k)$, $k=1,\ldots,5$.} 
\end{center}
\end{table}

\begin{table}[h!]
\begin{center}
\begin{tabular}{ c | c c c c c }
\toprule
 $n$ & $\Omega_n^{\rm Spin^c}(S^6)$ & $\Omega_n^{\rm Spin^c}(S^7)$ & $\Omega_n^{\rm Spin^c}(S^8)$ & $\Omega_n^{\rm Spin^c}(S^9)$ & $\Omega_n^{\rm Spin^c}(S^{10})$\\
 \midrule
 0 & $\mathbb{Z}$ & $\mathbb{Z}$ & $\mathbb{Z}$ & $\mathbb{Z}$ & $\mathbb{Z}$\\
 1 & 0 & 0 & 0 & 0 & 0\\
 2 & $\mathbb{Z}$ & $\mathbb{Z}$ & $\mathbb{Z}$ & $\mathbb{Z}$ & $\mathbb{Z}$\\
 3 & 0 & 0 & 0 & 0 & 0\\
 4 & $2\mathbb{Z}$ & $2\mathbb{Z}$ & $2\mathbb{Z}$ & $2\mathbb{Z}$ & $2\mathbb{Z}$\\
 5 & 0 & 0 & 0 & 0 & 0\\
 6 & $3\mathbb{Z}$ & $2\mathbb{Z}$ & $2\mathbb{Z}$ & $2\mathbb{Z}$ & $2\mathbb{Z}$\\
 7 & 0 & $\mathbb{Z}$ & 0 & 0 & 0\\
 8 & $5\mathbb{Z}$ & $4\mathbb{Z}$ & $5\mathbb{Z}$ & $4\mathbb{Z}$ & $4\mathbb{Z}$\\
 9 & 0 & $\mathbb{Z}$ & 0 & $\mathbb{Z}$ & 0\\
 10 & $6\mathbb{Z} \oplus \mathbb{Z}_2$ & $4\mathbb{Z} \oplus \mathbb{Z}_2$ & $5\mathbb{Z} \oplus \mathbb{Z}_2$ & $4\mathbb{Z} \oplus \mathbb{Z}_2$ & $5\mathbb{Z} \oplus \mathbb{Z}_2$\\
 \bottomrule
\end{tabular}
\captionof{table}{Spin$^c$ cobordism groups of spheres $\Omega_n^{{\rm Spin}^c}(S^k)$, $k=6,\ldots,10$.}
\end{center}
\end{table}

\begin{table}[h!]
\begin{center}
\begin{tabular}{ c | c c }
\toprule
$n$ & $\Omega^{\rm Spin}_n (T^2)$ & $\Omega^{{\rm Spin}^c}_n (T^2)$\\
\midrule
0 & $\mathbb{Z}$ & $\mathbb{Z}$\\
1 & 2$\mathbb{Z}\oplus \mathbb{Z}_2$ &  $2\mathbb{Z}$\\
2 & $\mathbb{Z}\oplus 3\mathbb{Z}_2$  & $2\mathbb{Z}$\\
3 &  $3\mathbb{Z}_2$  & $2\mathbb{Z}$\\
4 & $\mathbb{Z}_2\oplus\mathbb{Z}$  & $3\mathbb{Z}$\\
5 & $2\mathbb{Z}$ & $4\mathbb{Z}$ \\
6 & $\mathbb{Z}$ & $4\mathbb{Z}$ \\
7 & 0 & $4\mathbb{Z}$ \\
8 & $2\mathbb{Z}$ & $6\mathbb{Z}$ \\
9 & $2\mathbb{Z}_2\oplus 4\mathbb{Z}$ & $8\mathbb{Z}$ \\
10 & $7\mathbb{Z}_2\oplus 2\mathbb{Z}$ & $8\mathbb{Z}\oplus \mathbb{Z}_2$ \\
\bottomrule
\end{tabular}
\captionof{table}{Cobordism groups of 2-torus,  $\Omega_n^{\rm Spin}(T^2)$, $\Omega_n^{{\rm Spin}^c}(T^2)$.}
\end{center}
\end{table}

\begin{table}[h!]
\begin{center}
\begin{tabular}{ c | c c }
\toprule
$n$ & $\Omega^{{\rm Spin}^c}_n (K3)$ & $\Omega^{{\rm Spin}^c}_n (CY_3)$\\
\midrule
0 & $\mathbb{Z}$ & $\mathbb{Z}$\\
1 & 0 &  0\\
2 & $23\mathbb{Z}$  & $(b_2+1)\mathbb{Z}$\\
3 & 0  & $b_3\mathbb{Z}$\\
4 & $25\mathbb{Z}$  & $(2+2b_2)\mathbb{Z}$\\
5 & 0 & $b_3\mathbb{Z}$ \\
6 & $47\mathbb{Z}$ & $(3+3b_2)\mathbb{Z}$ \\
7 & 0 & $2b_3\mathbb{Z}$ \\
8 & $50\mathbb{Z}$ & $(5+4b_2)\mathbb{Z}$ \\
9 & 0 & $2b_3\mathbb{Z}$\\
10 &$94 \mathbb{Z} \oplus \mathbb{Z}_2$  & $(6+6 b_2)\mathbb{Z} \oplus \mathbb{Z}_2$\\
\bottomrule

\end{tabular}
\captionof{table}{Spin$^c$ cobordism groups of CY manifolds, $\Omega_n^{{\rm Spin}^c}(K3)$, $\Omega_n^{{\rm Spin}^c}(CY_3)$.}
\end{center}
\end{table}
\FloatBarrier

\clearpage
\section{Tables of K- and KO-theory groups}
\label{sec_app_Ktheory}


\begin{center}
\resizebox{\textwidth}{!}{
\begin{tabular}{ c | c c c c c c c c c }
\toprule
$n$ & $K^{-n}(S^1)$ & $K^{-n}(S^2)$ & $K^{-n}(S^3)$ &  $K^{-n}(S^4)$ &$K^{-n}(S^5)$ & $K^{-n}(S^6)$ & $K^{-n}(S^7)$ & $K^{-n}(S^8)$\\
 \midrule
0 &$\mathbb{Z}$ &$2\mathbb{Z}$ &$\mathbb{Z}$ &$2\mathbb{Z}$ & $\mathbb{Z}$ &$2\mathbb{Z}$ & $\mathbb{Z}$ &$2\mathbb{Z}$  \\
  
1 &$\mathbb{Z}$ &0 &$\mathbb{Z}$ &0 & $\mathbb{Z}$ &0 & $\mathbb{Z}$ &0 \\
\bottomrule
\end{tabular}
 }
\captionof{table}{K-groups of spheres $K^n(S^k)$, $k=1,\ldots, 8$. } 
 \end{center}

\vspace{.3cm}

\begin{center}
\begin{tabular}{ c | c c c c c c c }
\toprule
$n$ & $KO^{-n}(S^1)$ & $KO^{-n}(S^2)$ & $KO^{-n}(S^3)$ &  $KO^{-n}(S^4)$\\
\midrule
0 &$\mathbb{Z}\oplus\mathbb{Z}_2$ &  $\mathbb{Z}\oplus\mathbb{Z}_2$ & $\mathbb{Z}$ & $2\mathbb{Z}$ \\
  
1 & $2\mathbb{Z}_2$ & $\mathbb{Z}_2$ &$\mathbb{Z}_2\oplus\mathbb{Z}$ & $\mathbb{Z}_2$ \\
  
2 & $\mathbb{Z}_2$ &$\mathbb{Z}_2\oplus\mathbb{Z}$ & $\mathbb{Z}_2$ & $\mathbb{Z}_2$ \\  
  
3 & $\mathbb{Z}$ &0 &0 &0\\

4 & $\mathbb{Z}$ &$\mathbb{Z}$ & $\mathbb{Z}$ &$2\mathbb{Z}$ \\

5 &0 &0 &$\mathbb{Z}$ &$\mathbb{Z}_2$\\

6 &0 &$\mathbb{Z}$ & $\mathbb{Z}_2$ &$\mathbb{Z}_2$ \\

7 &$\mathbb{Z}$ &$\mathbb{Z}_2$ & $\mathbb{Z}_2$ &0  \\

8 &$\mathbb{Z}\oplus\mathbb{Z}_2$ & $\mathbb{Z}\oplus\mathbb{Z}_2$ & $\mathbb{Z}$  & $2\mathbb{Z}$ \\
\bottomrule
\end{tabular}
\captionof{table}{KO-groups of spheres $KO^{-n}(S^k)$, $k=1,\ldots,4$.}
\end{center}

\vspace{.1cm}

\begin{center}
\begin{tabular}{ c | c c c c c c c  }
\toprule
$n$ & $KO^{-n}(S^5)$ & $KO^{-n}(S^6)$ & $KO^{-n}(S^7)$ & $KO^{-n}(S^8)$ \\
\midrule
0 &$\mathbb{Z}$ & $\mathbb{Z}$ & $\mathbb{Z}$  & $2\mathbb{Z}$\\

1 &$\mathbb{Z}_2$ & $\mathbb{Z}_2$ & $\mathbb{Z}_2\oplus \mathbb{Z}$ & $2\mathbb{Z}_2$ \\
  
2 & $\mathbb{Z}_2$ &$\mathbb{Z}_2\oplus\mathbb{Z}$ & $2\mathbb{Z}_2$ &$2\mathbb{Z}_2$ \\
  
3 & $\mathbb{Z}$ &$\mathbb{Z}_2$ & $\mathbb{Z}_2$ &0 & \\

4 & $\mathbb{Z}\oplus \mathbb{Z}_2$ &$\mathbb{Z}\oplus \mathbb{Z}_2$ & $\mathbb{Z}$ &$2\mathbb{Z}$ \\

5 & $\mathbb{Z}_2$ &0 &$\mathbb{Z}$ & 0 \\

6 &0 & $\mathbb{Z}$ &0 &0 \\

7 &$\mathbb{Z}$ &0 &0 &0 \\

8 &$\mathbb{Z}$ & $\mathbb{Z}$ & $\mathbb{Z}$  & $2\mathbb{Z}$   \\

\bottomrule
\end{tabular}
\captionof{table}{KO-groups of spheres $KO^{-n}(S^k)$, $k=6,\ldots,8$. Note that Bott periodicity is respected in a two-fold way, i.e. $KO^{-n}(S^k)=KO^{-n\pm 8}(S^k)=KO^{-n}(S^{k+8})$.}
\end{center}

\begin{center}
\resizebox{\textwidth}{!}{
\begin{tabular}{ c | c c c c c c c }
\toprule
$n$  & $K^{-n}(T^2)$ & $K^{-n}(T^3)$ &  $K^{-n}(T^4)$ &$K^{-n}(T^5)$ & $K^{-n}(T^6)$ & $K^{-n}(T^7)$ & $K^{-n}(T^8)$\\
 \midrule
0 &$2\mathbb{Z}$ &$4\mathbb{Z}$ &$8\mathbb{Z}$ &$16\mathbb{Z}$ & $32\mathbb{Z}$ &$64\mathbb{Z}$ & $128\mathbb{Z}$  \\
  
1 &$2\mathbb{Z}$ &$4\mathbb{Z}$ &$8\mathbb{Z}$ &$16\mathbb{Z}$ & $32\mathbb{Z}$ &$64\mathbb{Z}$ & $128\mathbb{Z}$  \\
\bottomrule
\end{tabular}
 }
\captionof{table}{K-groups of tori  $K^n(T^k)$, $k=2,\ldots, 8$. } 
 \end{center}

\vspace{.4cm}

\begin{center}
\resizebox{.8\textwidth}{!}{
\begin{tabular}{ c | c c c c c c c }
\toprule
$n$ & $KO^{-n}(T^2)$ & $KO^{-n}(T^3)$ & $KO^{-n}(T^4)$ & $KO^{-n}(T^5)$\\
\midrule
0 &$\mathbb{Z}\oplus 3 \mathbb{Z}_2$ &$\mathbb{Z}\oplus 6 \mathbb{Z}_2$& $2\mathbb{Z}\oplus 10\mathbb{Z}_2$ & $6\mathbb{Z}\oplus 15\mathbb{Z}_2$\\
  
1 & $3\mathbb{Z}_2$ & $\mathbb{Z}\oplus 4\mathbb{Z}_2$ &$4\mathbb{Z}\oplus 5\mathbb{Z}_2$ & $10\mathbb{Z}\oplus 6\mathbb{Z}_2$ \\
  
2 & $\mathbb{Z}\oplus \mathbb{Z}_2$ &$3\mathbb{Z}\oplus \mathbb{Z}_2$ & $6\mathbb{Z}\oplus \mathbb{Z}_2$&  $10\mathbb{Z}\oplus \mathbb{Z}_2$ \\  
  
3 & $2\mathbb{Z}$ &$3\mathbb{Z}$ & $4\mathbb{Z}$ & $6\mathbb{Z}$\\

4 & $\mathbb{Z}$ &$\mathbb{Z}$ & $2\mathbb{Z}$ & $6\mathbb{Z}\oplus \mathbb{Z}_2$ \\

5 & 0  &$\mathbb{Z}$ &$4\mathbb{Z} \oplus \mathbb{Z}_2$ & $10\mathbb{Z}\oplus 6 \mathbb{Z}_2$\\

6 &$\mathbb{Z}$ &$3\mathbb{Z}\oplus \mathbb{Z}_2$ & $6\mathbb{Z}\oplus 5\mathbb{Z}_2$ & $10\mathbb{Z}\oplus 15\mathbb{Z}_2$  \\

7 &$2\mathbb{Z}\oplus \mathbb{Z}_2$ &$3\mathbb{Z}\oplus 4 \mathbb{Z}_2$ & $4\mathbb{Z}\oplus 10\mathbb{Z}_2$  & $6\mathbb{Z}\oplus 20\mathbb{Z}_2$  \\
\bottomrule
\end{tabular}
}
\captionof{table}{KO-groups of tori $KO^{-n}(T^k)$, $k=2,\ldots, 5$.}
\end{center}

\vspace{.4cm}

\begin{center}
\begin{tabular}{ c | c c c c c c c }
\toprule
$n$ & $KO^{-n}(T^6)$ & $KO^{-n}(T^7)$ & $KO^{-n}(T^8)$ \\
\midrule
0 &$16\mathbb{Z}\oplus 21\mathbb{Z}_2$ &$36\mathbb{Z}\oplus 28 \mathbb{Z}_2$& $72\mathbb{Z}\oplus 36\mathbb{Z}_2$ \\
  
1 & $20\mathbb{Z}\oplus 7\mathbb{Z}_2$ & $36\mathbb{Z}\oplus 8\mathbb{Z}_2$ &$64\mathbb{Z}\oplus 10\mathbb{Z}_2$  \\
  
2 & $16\mathbb{Z}\oplus \mathbb{Z}_2$ &$28\mathbb{Z}\oplus 2\mathbb{Z}_2$ & $56\mathbb{Z}\oplus 10\mathbb{Z}_2$ \\  
  
3 & $12\mathbb{Z}\oplus \mathbb{Z}_2$ &$28\mathbb{Z}\oplus 8\mathbb{Z}_2$ & $64\mathbb{Z}\oplus 36 \mathbb{Z}_2$ \\

4 & $16\mathbb{Z}\oplus 7\mathbb{Z}_2$ & $36\mathbb{Z}\oplus 28\mathbb{Z}_2$ & $72\mathbb{Z}\oplus 84\mathbb{Z}_2$ \\

5 &$20\mathbb{Z}\oplus 21\mathbb{Z}_2$ &$36\mathbb{Z}\oplus 56 \mathbb{Z}_2$ & $64\mathbb{Z}\oplus 126\mathbb{Z}_2$ \\

6 &$16\mathbb{Z}\oplus 35\mathbb{Z}_2$ &$28\mathbb{Z}\oplus 70 \mathbb{Z}_2$ & $56\mathbb{Z}\oplus 126\mathbb{Z}_2$ \\

7 &$12\mathbb{Z}\oplus 35\mathbb{Z}_2$ &$28\mathbb{Z}\oplus 56 \mathbb{Z}_2$ & $64\mathbb{Z}\oplus 84\mathbb{Z}_2$ \\
\bottomrule
\end{tabular}
\captionof{table}{KO-groups of tori $KO^{-n}(T^k)$, $k=6,7,8$.}
\end{center}

\clearpage

\begin{table}[h!]
\begin{center}
\begin{tabular}{ c | c c }
\toprule
$n$ & $K^{-n} (K3)$ & $K^{-n}(CY_3)$\\
\midrule
0 & $24\mathbb{Z}$ & $(2+2b_2)\mathbb{Z}$\\
1 & 0 &  $b_3 \mathbb{Z}$\\
\bottomrule

\end{tabular}
\captionof{table}{K-groups of CY manifolds, $K^{-n} (K3)$ and $K^{-n}(CY_3)$.}
\end{center}
\end{table}

\begin{table}[h]
\begin{center}
\hspace{1.5cm}\begin{tabular}{ c | c c c c c c c c}
\toprule
n & 0 & 1 & 2 &  3 &4 & 5 & 6 & 7 \\
\midrule
 $KO^{-n} (K3)$ & $\mathbb{Z} \oplus e( 22\mathbb{Z}_2,\mathbb{Z})$ & $\mathbb{Z}_2$ & $\mathbb{Z}_2 \oplus 22\mathbb{Z}$ &0 & $2\mathbb{Z}$& $\mathbb{Z}_2$ & $\mathbb{Z}_2 \oplus 22\mathbb{Z}$ & $22\mathbb{Z}_2$ \\
\bottomrule
\end{tabular}
\end{center}
\caption{KO-groups of K3, $KO^{-n} (K3)$.}
\end{table}


\bibliography{references}  
\bibliographystyle{utphys}


\end{document}